\documentclass[12pt,oneside,a4paper,onecolumn]{elsarticle}
\usepackage{times}
\usepackage{fullpage}

\usepackage{xcolor}
\usepackage{lineno,hyperref}
\usepackage{amsmath}
\usepackage{amssymb}
\usepackage{graphicx}
\usepackage{caption}
\usepackage{subcaption}
\usepackage{tikz}
\usepackage{slashed}
\usetikzlibrary{patterns}
\usetikzlibrary{shapes.arrows}
\usepackage{pgfplots}
\usepackage{placeins}
\pgfplotsset{compat=newest}
\usepackage{mathtools} 
\usepackage{diagbox}
\usetikzlibrary{arrows.meta}
\modulolinenumbers[5]
\usepackage{adjustbox}

\journal{Journal of \LaTeX\ Templates}

\begin{document}
\begin{frontmatter}

\title{A variational interface-preserving and conservative phase-field method for the surface tension effect in two-phase flows}

\author[UBC]{Xiaoyu Mao}
\author[UBC]{Vaibhav Joshi}

\author[UBC]{Rajeev Jaiman\corref{cor1}}
\ead{rjaiman@mech.ubc.ca}
\cortext[cor1]{Corresponding author}
\address[UBC]{Department of Mechanical Engineering, University of British Columbia, Vancouver, Canada}




\begin{abstract}
We present a finite element based variational interface-preserving and conservative phase-field formulation for the  modeling  of  incompressible two-phase flows with surface tension dynamics. The preservation of the hyperbolic tangent interface profile of the convective Allen-Cahn phase-field formulation relies on a novel time-dependent mobility model. The mobility coefficient is adjusted adaptively as a function of gradients of the velocity and the order parameter in the diffuse interface region in such a way that the free energy minimization properly opposes the convective distortion. The ratio of the convective distortion to the free energy minimization is termed as the convective distortion parameter, which characterizes the deviation of the diffuse interface profile from the hyperbolic tangent shape due to the convection effect. In the phase-field formulation, the mass conservation is achieved by enforcing a Lagrange multiplier with both temporal and spatial dependence on the phase-field function. We integrate the interface-preserving and conservative phase-field formulation with the incompressible Navier-Stokes equations and the continuum surface tension force model for the simulation of incompressible two-phase flows. A positivity preserving scheme designed for the boundedness and stability of the solution is employed for the variational discretization using unstructured finite elements. We examine the convergence and accuracy of the Allen-Cahn phase-field solver through a generic one-dimensional bistable diffusion-reaction system in a stretching flow. We quantify and systematically assess the relative interface thickness error and the relative surface tension force error with respect to the convective distortion parameter. Two- and three-dimensional rising bubble cases are further simulated to examine the effectiveness of the proposed model on the volume-preserving mean curvature flow and the interface-preserving capability.  Finally, we demonstrate the applicability of the proposed model for a complex case of two bubbles rising and merging with a free surface, which includes complex topological changes and the surface tension dynamics using unstructured finite elements.
\end{abstract}


\end{frontmatter}

\linenumbers

\section{Introduction}
Two-phase flow of immiscible fluids is ubiquitous in many natural phenomena and engineering applications. Examples include bubbly cavitating flows around marine propellers \cite{brennen2014cavitation}, wake bubbles behind ships \cite{smirnov2005bubble}, and the bubble sweep-down problem of oceanographic vessels \cite{mallat20183d}. The accurate representation and evolution of the fluid-fluid interface are crucial in the numerical simulation of immiscible two-phase flows. When the surface tension force plays a significant role, the handling of mutual dependency between the representation and the evolution of interfaces becomes considerably challenging. The representation of the interface geometry has a direct impact on the surface tension dynamics, which is one of the important driving forces during the interface evolution.  The quality of two-phase flow solutions is very sensitive to the approximation of surface tension force in the capillary dominated regime. The modeling of immiscible two-phase flows poses other well-known difficulties regarding high density and viscosity ratios, the mass conservation, the discontinuity of properties across the interface, and the topological changes during simulating realistic flows. In the present work, we particularly focus on the accurate representation of interfaces for the surface tension force calculation in incompressible two-phase flow problems, while retaining the mass conservation during complex topological changes.
 
The interface between immiscible two-phase fluids can be represented by a sharp or diffuse interface approach. In the sharp interface approach, the fluid-fluid interface is treated as a sharp boundary separating the domains of the two phases. The boundary is explicitly parameterized by specifying its location and geometry. The parameterization can be accomplished by tracking the interface with a conforming mesh, which is the approach of the arbitrary Lagrangian-Eulerian method \cite{hirt1997arbitrary} and the front-tracking method \cite{unverdi1992front}. However, the mesh operation during complex topological changes, such as the merging and the breaking-up of the interface, poses significant difficulties. An alternative approach for the parameterization is to reconstruct the sharp interface according to a function representing the volume fraction \cite{rider1998reconstructing}, which is employed in the volume of fluid (VOF) method \cite{hirt1981volume}. Although the mesh operation is avoided, the volume fraction function suffers from difficulties in the calculation of the normal and curvature from the reconstructed interface. This issue is resolved in the diffuse-interface approach by eliminating the requirement of interface parameterization and by introducing a smooth phase-field function.
In the diffuse-interface approach, the change of physical properties at the interface is considered as a gradual variation within a transitional region with a finite thickness. The smooth variation can be modeled by a continuous scalar-valued function serving as a phase indicator for the two-phase mixture. The function is chosen as the signed distance function to the interface $d$ in the level set method \cite{malladi1995shape}, or a hyperbolic tangent function $\tanh(d/\sqrt{2}\varepsilon)$ in the phase-field method, where $\varepsilon$ is an interface thickness parameter controlling the thickness of the diffuse interface region. With the decrease in the interface thickness parameter, the phase-field function converges to the Heaviside function description in the sharp-interface approach. The gradients of the functions acquire non-zero values only in the diffuse interface region, which can be utilized to reformulate the surface tension force as a volumetric source term. 
Two popular models to impose the surface tension force are: (i) continuum surface force (CSF) model, which distributes the sharply defined force with a Dirac Delta function  to the diffuse interface region \cite{brackbill1992continuum}, (ii) free energy based surface tension force model, which imposes the energy balance in the context of the phase-field method \cite{van1979thermodynamic,jacqmin1999calculation}. In this paper, we consider the phase-field method based on the transient Allen-Cahn equation concerning the computational efficiency and simplicity. The CSF model is employed for the surface tension dynamics to circumvent the chemical potential calculation in the free energy based surface tension model, which is not required in the Allen-Cahn phase-field equation. 

The hyperbolic tangent shape and the thickness of the interface in the phase-field method \cite{jacqmin1999calculation} or in the conservative level set method \cite{olsson2005conservative,desjardins2008accurate} will not necessarily remain the same during the interface evolution. The diffuse interface region may undergo stretching from the convection, which leads to the development of nonuniformity. We refer to this phenomenon as the convective distortion. While the thinning of the interface causes the numerical difficulty of resolving a high gradient with less computational elements, the thickening of the interface results in the loss of accuracy. Furthermore, the surface tension force models are subjected to inaccuracies due to the convective distortion, since their prerequisite interface profiles are no longer maintained. In the level set method, this issue is resolved by solving a reinitialization equation until the interface profiles are recovered.
This additional procedure can be computationally expensive than solving the convection equation \cite{hartmann2010accuracy}. The reinitialization process may also lead to poor mass conservation, which entails further corrections in the level set method \cite{olsson2005conservative,tornberg2000}. 
While the phase-field methods resemble the conservative level-set methods, there are some fundamental differences.
For example, the property of maintaining the interface profile close to the hyperbolic tangent function, which minimizes the free energy, is embedded in the phase-field method. 
This liberates the phase-field method from the interface reinitialization required in the level set method, thereby reducing the computational cost \cite{chiu2011conservative} 
and providing robustness for any kind of geometric manipulation to compute the interfacial curvature. Likewise the level-set method, the phase-field methods have the advantage to handle any topological changes in the interface due to their Eulerian description.   Furthermore, the mass conservation can be enforced in a relatively simpler manner and the variational foundation of the phase-field method provides provable energy stability and discrete conservation \cite{joshi2018positivity}. For these reasons, the phase-field method has attracted more interest in the modeling of the two-phase flow problems in the recent years.
 
\subsection{Review of the phase-field method}
The phase-field method considers a diffuse representation of the interface geometry and describes the minimization of the free energy functional \cite{cahn1958free}. The diffuse interface between the two phases is described as a region in which the phases are mixed and store the free energy.  The free energy functional $\mathcal{E}$  can be written as:
\begin{align}\label{free energy}
\mathcal{E}:H^1(\Omega)\cap L^4(\Omega)\to\mathbb{R}_{\geqslant 0},\  \mathcal{E}(\phi(\boldsymbol{x},t)) = 
    \int_{\Omega} \left( F(\phi(\boldsymbol{x},t))+\frac{\varepsilon^2}{2}|\nabla\phi(\boldsymbol{x},t)|^2\right)d\Omega,
\end{align}
where $\Omega$ is a bounded fluid domain consisting of spatial points $\boldsymbol{x}$ at time $t$, $H^{1}(\Omega)$ denotes the space of square-integrable real-valued functions with square-integrable derivatives on $\Omega$, $L^4(\Omega)$ denotes the function space in which the fourth power of the function is integrable, $\mathbb{R}_{\geqslant 0}$ represents the set of non-negative real numbers, $\phi(\boldsymbol{x},t)$ is referred to as the order parameter or the phase-field function which indicates the components of the two-phase mixture.
The first term $F(\phi(\boldsymbol{x},t))$ in Eq.~(\ref{free energy}), which is called the bulk or mixing energy, depends on the local composition of the two phases mixture indicated by $\phi$.  To invoke the phase separation, a double-well potential is employed: $ F(\phi)=\frac{1}{4}\left(\phi^2-1\right)^2$,
in which the minimum bulk energy is attained with separated pure phases $\phi=1$ and $\phi=-1$. The second term, which is called interfacial or gradient energy, depends on the composition of the immediate environment indicated by $\nabla\phi$. The gradient energy dictates the interaction between the two phases.
The ratio between the bulk energy and the interfacial energy is determined by the interface thickness parameter $\varepsilon$.  In the free energy minimization, the bulk energy minimization prefers pure components and separated phases, while the interfacial energy minimization prefers a mixed uniform phase. Using the interplay between these two effects, the interface thickness parameter $\varepsilon$ controls the diffuse interface geometry.

The phase-field methods are generally based on the Cahn-Hilliard and the Allen-Cahn phase-field equations \cite{cahn1961spinodal,allen1979microscopic}, which considers the gradient flow minimizing the free energy functional as the cause of the phase-field function evolution. The Cahn-Hilliard equation satisfies the mass conservation naturally \cite{kim2016basic}. However, the equation is a fourth-order partial differential equation (PDE), which is cumbersome during the numerical discretization. In contrast, the Allen-Cahn equation is a second-order convection-diffusion-reaction PDE which has attractive numerical properties from the implementation standpoint. Although the original Allen-Cahn equation is not mass-conservative, the conservation property can be realized by adding a Lagrange multiplier \cite{rubinstein1992nonlocal,brassel2011modified} or employing an anti-curvature term \cite{sun2007sharp}. The former is more stable, while the latter is more accurate \cite{chai2018comparative}. Concerning the computational efficiency and the stability, we employ the Allen-Cahn phase-field equation with a Lagrange multiplier for solving two-phase flow problems in the current study.  In the original Allen-Cahn equation, the evolution of the phase-field function seeks the minimum of the free energy functional:
 \begin{equation}\label{ac}
    \frac{\partial \phi}{\partial t}=-\gamma\left(\frac{\delta \mathcal{E}(\phi)}{\delta \phi}\right),
\end{equation}
 where $\gamma$ is the mobility coefficient and $\frac{\delta\mathcal{E}(\phi)}{\delta \phi} =(F'(\phi)-\varepsilon^2\nabla^2\phi)$ represents  the variational derivative of the free energy functional. Eq.~(\ref{ac}) can be formulated as the gradient flow of  the free energy functional in $L^2$ space \cite{shen2010numerical}:
 $\frac{\partial}{\partial t}\mathcal{E}(\phi)=-\gamma\int_{\Omega}\left|\frac{\delta\mathcal{E}(\phi)}{\delta \phi}\right|^2d\Omega$.
 The mobility coefficient determines the intensity of the gradient flow of the free energy functional and controls the speed at which the interface geometry relaxes to the equilibrium profile and shape with minimum free energy.  For a planar interface in equilibrium, which can be considered as a one-dimensional case, the equilibrium interface profile can be solved as:
 $ \phi_{\text{eq}}(n)=\tanh\left(\frac{n}{\sqrt{2}\varepsilon}\right)$,
 where $n$ is the coordinate normal to the interface.
 The equilibrium interface profile is shown in Fig.~\ref{sche} (a). The thickness of the diffuse interface is $\mathcal{O}(\varepsilon)$.
When the phase-field method is used for the fluid-fluid interface evolution in the two-phase flow problems, the convection of the flow field and the volume conservation must be considered, which leads to the convective form of the conservative Allen-Cahn phase-field equation:
\begin{align}\label{AC}
    \frac{\partial \phi}{\partial t}+\boldsymbol{u}\cdot\nabla\phi=-\gamma\left(\frac{\delta \mathcal{E}(\phi)}{\delta \phi}-\beta(t)\sqrt{F(\phi)}\right),
\end{align}
where $\boldsymbol{u}$ represents the convective velocity of the fluid flow, $\beta(t)$ is the time-dependent part of the Lagrange multiplier for mass conservation \cite{brassel2011modified}, which is given by $\beta(t) = \frac{\int_{\Omega} F'(\phi)d\Omega}{\int_{\Omega}\sqrt{F(\phi)}d\Omega}$.

\subsection{Dynamics of the conservative Allen-Cahn equation}
In the convective form of the conservative Allen-Cahn equation, the evolution of the order parameter is driven by both the convection and the free energy minimization. 
As the interface profile evolves, instead of the equilibrium interface profile, an actual interface profile is formed as a consequence of the interplay between the convective distortion and the free energy minimization. When the finite thickness interface region is subjected to a positive or negative velocity gradient in its normal direction, which represents an extensional or compressional velocity field, the interface will be extended or compressed and deviate from the equilibrium profile. On the other hand, the deviation from the equilibrium profile increases the free energy. The free energy minimization starts to drive the order parameter back to the equilibrium profile. Consequently, an actual profile $\phi_a$ different from the equilibrium interface profile $\phi_{\mathrm{eq}}$ is formed due to the interplay, as illustrated in Fig.~\ref{sche} (a).  As mentioned earlier, the intensity of the gradient flow minimizing the free energy functional is controlled by $\gamma$. If the free energy minimization dominates the competition over the convective distortion, the actual interface profile will be kept close to the equilibrium profile, thus achieving interface preservation.

%
During the interface evolution, the volume-preserving mean curvature flow induced by the free energy minimization disturbs the convection according to the fluid flow velocity. The free energy of the interface is closely related to the perimeter or area of the interface \cite{modica1987gradient}. Under the constraint of the volume conservation, which is the equivalence of the mass conservation with a constant density of each phase, the free energy minimization will drive the actual interface contour $\phi_a=0$ towards the contour with minimum perimeter or area. Consequently, the interface contour convected by the fluid flow velocity relaxes towards a circular or spherical shape in two or three dimensions, which are denoted as $\phi_{\text{eq}}=0$ \cite{rubinstein1992nonlocal,brassel2011modified,modica1987gradient}. The flow induced by this process is referred to as the volume-preserving mean curvature flow and illustrated in Fig.~\ref{sche} (b). The velocity of the flow can be derived from the asymptotic analysis \cite{rubinstein1992nonlocal,brassel2011modified,rubinstein1989fast}. For the current formulation, consider the interface $\Gamma_I^{\phi}(t)=\{\boldsymbol{x}\in\Omega|\phi(\boldsymbol{x},t)=0\}$, the velocity of the volume-preserving mean curvature flow at the interface is given by:
 \begin{align}\label{mc}
    \boldsymbol{v}(\boldsymbol{x},t)=\gamma\varepsilon^2\left(\kappa(\boldsymbol{x},t)-\frac{1}{|\Gamma_I^{\phi}(t)|}\int_{\Gamma_I^{\phi}}\kappa(\boldsymbol{x},t) ds\right)\boldsymbol{n}^{\phi}_L(\boldsymbol{x},t), \boldsymbol{x}\in\Gamma_I^{\phi}(t),
\end{align}
where $\boldsymbol{v}(\boldsymbol{x},t)$ is the velocity of the volume-preserving mean curvature flow, $\kappa(\boldsymbol{x},t)$ is defined as $\kappa(\boldsymbol{x},t)=\sum\limits_{i=1}^{n_{sd}-1}\kappa_i(\boldsymbol{x},t)$, $n_{sd}$ being the number of dimensions and $\kappa_i(\boldsymbol{x},t)$ being the principle curvatures of the interface, $|\Gamma_I^{\phi}(t)|$ is the perimeter or area of the interface, $\boldsymbol{n}^{\phi}_L
(\boldsymbol{x},t)=\nabla\phi/|\nabla\phi|$ is the unit normal vector of the level sets of $\phi$, $\kappa(\boldsymbol{x},t)$ is defined as positive when $\boldsymbol{n}^{\phi}_L(\boldsymbol{x},t)$ is pointing to the concave side of the interface.
 
 \begin{figure}
 	\begin{minipage}[b]{0.5\textwidth}
 		\trimbox{0 0 0 70}{
 			\begin{tikzpicture}[scale=1.1]
 			\begin{axis}[ 
 			ymin=-1.45,
 			ymax=1.3,
 			xmax=6,
 			xmin=-6,
 			xticklabel=\empty,
 			yticklabel=\empty,
 			ytick style={draw=none},
 			xtick style={draw=none},
 			axis line style = thick,
 			minor tick num=0,
 			axis lines = middle,
 			xlabel=$x$,
 			ylabel=$\phi$,
 			label style = {at={(ticklabel cs:1.1)}}
 			]
 			\shade[left color=white,right color=gray!45] (axis cs: -0.7361,-1.4) rectangle (axis cs:0.7361,1.3);
 			\draw[fill=gray!45,draw=gray!45] (axis cs: 0.7361,-1.4) rectangle (axis cs:6,1.3);
 			
 			\end{axis}
 			
 			\begin{axis}[ 
 			ymin=-1.45,
 			ymax=1.3,
 			xmax=6,
 			xmin=-6,
 			xticklabel=\empty,
 			yticklabel=\empty,
 			ytick style={draw=none},
 			xtick style={draw=none},
 			axis line style = thick,
 			minor tick num=0,
 			axis lines = middle,
 			xlabel=$x$,
 			ylabel=$\phi$,
 			label style = {at={(ticklabel cs:1.1)}}
 			]
 			\addplot [mark=none,draw=black, thick, domain=-6:6, samples=60 ,dashed] {tanh(\x*2)};
 			\addplot [mark=none,draw=black, thick, domain=-6:6, samples=60] {tanh((\x)*0.5)};

 			\node at (axis cs:3.5,1.18) {$\phi_{\text{eq}}$};
 			\draw [dashed, thick] (axis cs:3.5,1)--(axis cs: 3.5 ,1.1);
 			\node at (axis cs:3.5,0.7) {$\phi_a$};
 			\draw [ thick] (axis cs:3.5,0.94)--(axis cs: 3.5 ,0.8);
 			
 			\draw [-stealth, thick, blue] (axis cs: 1.42,0.6)--(axis cs: 1.42,0.3);
 			
 			\draw [-stealth, thick, blue] (axis cs: 1.42,0.6)--(axis cs: 1.42,0.35);
 			\node [blue,align=center] at (axis cs:-3.6,-0.3) {Convective};
 			\node [blue,align=center] at (axis cs:-3.6,-0.47) {distortion};
 			\draw [-stealth, thick, red] (axis cs: 1.42,0.6)--(axis cs: 1.42,0.9);
 			\node [red,align=center] at (axis cs:-3.4,-1.2) {Free energy};
 			\node [red,align=center] at (axis cs:-3.4,-1.37) {minimization};
 			\draw [-stealth, thick, blue] (axis cs: -1.42,-0.6)--(axis cs: -1.42,-0.3);
 			\draw [-stealth, thick, blue] (axis cs: -1.42,-0.6)--(axis cs: -1.42,-0.35);
 			\draw [-stealth, thick, red] (axis cs: -1.42,-0.6)--(axis cs: -1.42,-0.9);

 			\draw [densely dashdotted,semithick](axis cs: 0.7361,-1.4)--(axis cs:0.7361,1.3);
 			\draw [densely dashdotted,semithick](axis cs: -0.7361,-1.4)--(axis cs:-0.7361,1.3);

 			\draw[-stealth,thick](axis cs:-1.2361,-0.13)--(axis cs:-0.7361,-0.13);
 			\draw[-stealth,thick](axis cs:1.2361,-0.13)--(axis cs:0.7361,-0.13);
 			\draw[thick](axis cs:1.2361,-0.13)--(axis cs:1.5361,-0.2);
 			\node[below ] at (axis cs:2.8,-0.1) {$\sim\mathcal{O}(\varepsilon)$};
 			
 			
 			\node at (axis cs:-3.5,0.4) {$\Omega_2$};
 			\node at (axis cs:-3.5,0.2) {$(\phi=-1)$};
 			
 			\node at (axis cs:3.5,0.4) {$\Omega_1$};
 			\node at (axis cs:3.5,0.2) {$(\phi=1)$};
 			\clip (axis cs:-5,-1) rectangle (5, 1);
 			\end{axis}
 			
 			\end{tikzpicture}}
 		
 		\caption*{\hspace{-5 pt}(a)}
 	\end{minipage}
 	\hfill
 	\begin{minipage}[b]{0.43\textwidth}
 		\trimbox{45 10 0 70}{
 			\begin{tikzpicture}[scale=1.3]
 			
 			\begin{axis}[ 
 			ymin=-6,
 			ymax=6,
 			xmax=8,
 			xmin=-8,
 			xticklabel=\empty,
 			yticklabel=\empty,
 			ytick style={draw=none},
 			xtick style={draw=none},
 			axis line style = {draw=none},
 			minor tick num=0,
 			axis lines = middle,
 			label style = {at={(ticklabel cs:1.1)}},
 			axis equal
 			]
 			\draw  [dashed,thick,black] (axis cs: 0,0) circle [radius=4];
 			\draw  [thick,black](0,0) ellipse (5 and 3);
 			\draw [thick,dashed] (axis cs: 2.8,2.8)--(axis cs:4,2.8);
 			\draw [thick] (axis cs: 4.1,1.75)--(axis cs:5,1.75);
 			\node [right] at (axis cs: 4,2.8) {$\phi_{\text{eq}}=0$};
 			\node [right] at (axis cs: 5,1.75) {$\phi=0$};

 			\draw [-stealth, thick, red] (axis cs: 0,3)--(axis cs: 0,3.8);
 			\draw [-stealth, thick, red] (axis cs: 0,-3)--(axis cs: 0,-3.8);
 			\draw [-stealth, thick, red] (axis cs: 5,0)--(axis cs: 4.2,0);
 			\draw [-stealth, thick, red] (axis cs: -5,0)--(axis cs: -4.2,0);
 			\node [red] at (axis cs:0,5) {$\boldsymbol{v}(\boldsymbol{x},t)$};
 			
 			\draw [-stealth, thick, blue] (axis cs: 0,3)--(axis cs: 0,1);
 			\draw [-stealth, thick, blue] (axis cs: 0,-3)--(axis cs: 0,-1);
 			\draw [-stealth, thick, blue] (axis cs: 5,0)--(axis cs: 5.8,0);
 			\draw [-stealth, thick, blue] (axis cs: -5,0)--(axis cs: -5.8,0);
 			\draw [-stealth, thick, blue] (axis cs: 0,3)--(axis cs: 0,1.3);
 			\draw [-stealth, thick, blue] (axis cs: 0,-3)--(axis cs: 0,-1.3);
 			\draw [-stealth, thick, blue] (axis cs: 5,0)--(axis cs: 5.5,0);
 			\draw [-stealth, thick, blue] (axis cs: -5,0)--(axis cs: -5.5,0);
 			\node [blue] at (axis cs:1.8,1.5) {$\boldsymbol{u}(\boldsymbol{x},t)$};
 			\end{axis}
 			\end{tikzpicture}}
 		\caption*{\hspace{-20 pt}(b)}
 	\end{minipage}
 	\caption{Illustrations of the interface dynamics of the convective form of the conservative Allen-Cahn equation: (a) one-dimensional equilibrium interface profile $\phi_{\mathrm{eq}}$ and the actual interface profile $\phi_a$ subjected to an extensional velocity field, and (b) volume-conserved mean curvature flow velocity $\boldsymbol{v}(\boldsymbol{x},t)$ and the convective velocity $\boldsymbol{u}(\boldsymbol{x},t)$ of the interface $\phi(\boldsymbol{x},t)=0$. The free energy minimization described by the equation balances the convective distortion with $\phi_a$ in (a), and induces $\boldsymbol{v}(\boldsymbol{x},t)$ in (b).
} 
 	\label{sche}
 \end{figure}
 


\subsection{Related work and contributions}
The phase-field parameters, namely the mobility coefficient $\gamma$ and the interface thickness $\varepsilon$, play important roles in the interface-preserving capability and the volume-preserving mean curvature flow velocity. These parameters should be judiciously select to produce a physically-consistent interface behavior. Jacqmin \cite{jacqmin1999calculation} suggested that the gradient flow minimizing the free energy should properly oppose the convective distortion, while the gradient flow should converge to zero (i.e., the phase-field equation converges to a pure convection equation) as the diffuse interface converges to the sharp interface. According to the order of magnitude analysis, a mobility coefficient varying between  $\mathcal{O}(\varepsilon)$ and $\mathcal{O}(\varepsilon^2)$ in the Cahn-Hilliard equation was found to be appropriate. 
In \cite{yue2004diffuse}, the magnitude of the total free energy is adjusted dynamically to ensure its consistency with the surface tension force coefficient viewed as free energy density. The explicit calculation of the interface length or area was required during the adjustment.
In \cite{sun2007sharp}, the phase-field propagation equation was formulated for tracking sharp interfaces. The mobility coefficient, the interface thickness parameter, and the maximum convective velocity are combined to construct a non-dimensionalized parameter  via a standard explicit finite difference discretization. The parameter was considered purely as a numerical parameter, the impact of which on the interface profile and evolution was studied for stationary and evolving interfaces. 
It was found that the increase in the mobility coefficient results in better enforcement of the hyperbolic tangent phase-field profile and helps to suppress the instabilities at corners. 
The mobility coefficient was controlled one order of magnitude below its upper limit given by the Courant-Friedrichs-Levy (CFL) condition to avoid significant discretization errors. 
In \cite{li2016phase}, an additional free energy functional punishing the deviation from the hyperbolic tangent profile is designed, which provides a correction term in the Cahn-Hilliard equation via the minimization process. The profile correction term enforces the hyperbolic tangent profile, thereby reducing the interface shrinkage effect, the convective distortion, and improving the surface tension force calculation. This approach was further improved in the profile-flux correction \cite{zhang2017flux} and applied in the turbulent multi-phase flow problems \cite{soligo2019mass}.

In the current study, we propose an interface-preserving and conservative phase-field method for incompressible two-phase flows with a particular emphasis on the accurate surface tension dynamics. A continuum formulation and a systematic approach for determining the phase-field parameters $\gamma$ and $\varepsilon$ are presented. The parameters are formulated by directly considering the associated terms in the convective form of the Allen-Cahn equation in a non-dimensional moving orthogonal curvilinear coordinate system. The term representing the effect of the convective distortion is identified wherein the convective distortion parameter quantifies the ratio between the convective distortion and the free energy minimization. An interface-preserving condition for the parameter enforcing the free energy minimization dominance over the convective distortion is derived. To fulfill the condition, we propose a time-dependent mobility model for controlling the RMS convective distortion parameter in the diffuse interface region. Direct relationships between the RMS convective distortion parameter and relative interface thickness and surface tension force errors are assessed by numerical simulations of the interface convection problems. By establishing a suitable range of mobility coefficient, the excessive gradient flow minimizing the free energy functional and the resulting spurious volume-preserving mean curvature flow are avoided.   

The present study builds upon our previous conservative and energy stable variational scheme for the Allen-Cahn and Navier-Stokes system proposed in \cite{joshi2018positivity}. The scheme was integrated with a mesh adaptivity process in \cite{joshi2018adaptive}, and has been proven to be accurate and stable for a wide range of fluid-structure interaction problems in the inertia dominate regime with high density ratio \cite{joshi2019hybrid}. In the current work, we further improve the accuracy of the scheme in the capillary dominated regime by considering the interface-preserving Allen-Cahn based phase-field model.  We employ the model together with the CSF model, where the surface tension force is transformed into a volume
force spread over a few layers of elements.
We discretize the incompressible Navier-Stokes and Allen-Cahn equations with the finite element method in a fully implicit manner. We maintain the bounded and stable solution of the Allen-Cahn system via the positivity preserving variational (PPV) technique \cite{joshi2017positivity} and the coupling between the Allen-Cahn and the Navier-Stokes systems retains second-order accuracy in time domain \cite{joshi2018positivity,joshi2019hybrid}. With the aid of a generic 1D bistable diffusion-reaction system in a stretching flow,  we first carry out a systematic convergence and verification study of our 1D steady Allen-Cahn solver based on the PPV technique and the implicit discretization. To demonstrate the interface-preserving formulation, we employ the Allen-Cahn  solver for the convection of diffuse interfaces in prescribed incompressible velocity fields for planar and curved situations.
We examine the proposed formulation in the two- and three- dimensional rising bubble benchmark cases through a systematic convergence study.  We compare accuracy and convergence with the sharp interface formulation. Our results show that only when the interface-preserving capability is improved and the volume-preserving mean curvature flow is decreased simultaneously, the simulation results will converge to the accurate solution.  This requires the reduction of the RMS convective distortion parameter and the interface thickness parameter at the same time. 
Finally, we simulate two rising bubbles merging with a free surface with an unstructured mesh to demonstrate the applicability of the proposed model in practical problems, which has complex topological changes of the interface and complex dynamics including bubble-bubble and bubble-free surface interaction. 

 The organization of this paper is as follows: Section 2 presents a mathematical analysis of the diffuse interface profile, wherein the convective distortion parameter and the interface-preserving condition are identified. The time-dependent mobility model is proposed according to the interface-preserving condition. Section 3 describes the implementation of the variational formulation for the interface-preserving conservative Allen-Cahn-Navier-Stokes system with the time-dependent mobility model. Section 4 verifies our implementation through a 1D bistable convection-diffusion-reaction system in a stretching flow and provides a numerical assessment of the errors associated with the convective distortion parameter for a planar and a curved interface. Two- and three-dimensional rising bubble cases are investigated in Section 5 to demonstrate the effect of the proposed model on the volume-preserving mean curvature flow and the interface preservation property.  Section 6 demonstrates the applicability of the model by solving for two bubbles rising and merging with a free surface. The conclusions are summarized in Section 7.
 
\section{Interface-preserving phase-field formulation}
In this section, we present the continuum formulation of the time-dependent mobility model for preserving the hyperbolic tangent profile. The convective form of the Allen-Cahn equation is directly analyzed in a non-dimensional moving orthogonal curvilinear coordinate system. The term representing the influence of the convective distortion is identified in the governing equation. The magnitude of the term depends on a non-dimensional parameter, which we refer to as the convective distortion parameter. An interface-preserving condition is derived for the convective distortion parameter to preserve the interface profile. The time-dependent mobility model is proposed based on the interface-preserving condition.

\subsection{Interface profile in non-dimensional moving orthogonal curvilinear coordinate system}\label{interface profile analysis}
Following the original work of \cite{allen1979microscopic}, we describe the evolution of the two-phase interface in an orthogonal curvilinear coordinate system. This allows us to simplify the governing equation utilizing the property that the level sets of the order parameter are parallel to the interface. Consider a physical domain $\Omega\times]0,T[$ with spatial coordinates $\boldsymbol{x}$ and temporal coordinate $t$. The boundary of the computational domain $\Gamma$ is decomposed as $\Gamma=\Gamma^{\phi}_D\cup\Gamma^{\phi}_H$, where $\Gamma^{\phi}_D$ and $\Gamma^{\phi}_H$ denote the Dirichlet and Neumann boundaries for the order parameter respectively. A diffuse interface that separates the immiscible two-phase fluids defined on $\Omega$ is indicated by the order parameter $\phi(\boldsymbol{x},t)$. The diffuse interface is convected by a velocity field $\boldsymbol{u}(\boldsymbol{x},t)$. In the orthogonal curvilinear coordinate system, the spatial coordinates are given by $\boldsymbol{x}=(n,\tau_1,\tau_2)$, where $n$ is the coordinate of the axis normal to the level sets of $\phi$, and the rest two coordinates $\tau_1$ and $\tau_2$ are the coordinates of the axes which are tangential to the level sets of $\phi$. In this coordinate system, the convection of a diffuse interface is given by the following initial boundary value problem based on the Allen-Cahn equation:
\begin{equation} \label{frame independent}
\left.\begin{aligned}
      \frac{\partial \phi}{\partial t}+\boldsymbol{u}\cdot\nabla\phi&=-\gamma\left(F'(\phi)-\varepsilon^2\nabla^2\phi\right),&&\mathrm{on}\  \Omega ,\\
       \phi&=\phi_D, &&\forall \boldsymbol{x}\in\Gamma^{\phi}_D,\\
        \boldsymbol{n}^{\phi}_{\Gamma}\cdot\nabla\phi&=0,&&\forall \boldsymbol{x}\in\Gamma^{\phi}_H,\\
        \phi\big|_{t=0}&=\phi_0,&&\mathrm{on}\ \Omega,
\end{aligned}\hspace{1cm}\right\}
\end{equation}
where $F'(\phi)=\phi^3-\phi$ is the derivative of the double-well potential with respect to $\phi$, $\boldsymbol{n}^{\phi}_{\Gamma}$ represents the unit vector normal to the boundary of the computational domain and $\phi_0$ represents the initial condition for the order parameter. 
 The velocity in the orthogonal curvilinear coordinate system can be expressed as:
\begin{align}\label{velocity}
    \boldsymbol{u}&=u_n\boldsymbol{n}^{\phi}_L+u_{\tau_1}\boldsymbol{\tau}_1+u_{\tau_2}\boldsymbol{\tau}_2,
\end{align}
where $\boldsymbol{n}^{\phi}_L,\boldsymbol{\tau}_1$ and $\boldsymbol{\tau}_2$ are the unit vectors in the normal and two tangential directions of the level sets of $\phi$ respectively, and $u_n,u_{\tau_1},u_{\tau_2}$ are the corresponding velocity components.
 
\noindent Assume that the normal profile of the interface is almost the same everywhere on the interface. Therefore the derivatives of the order parameter in the tangential directions are negligible. With this assumption, the spatial derivatives of the order parameter can be calculated as:
\begin{align}\label{derivatives}
\nabla\phi=\frac{\partial \phi}{\partial n}\boldsymbol{n}^{\phi}_L,\quad
\nabla^2\phi=\nabla\cdot\nabla\phi=\frac{\partial^2\phi}{\partial n^2}+\frac{\partial\phi}{\partial n}\nabla\cdot\boldsymbol{n}^{\phi}_L.
\end{align}
Notice that $\nabla\cdot\boldsymbol{n}^{\phi}_L=-\kappa$, where $\kappa$ is the summation of principle curvatures of the interface. Substituting Eqs.~(\ref{velocity}) and (\ref{derivatives}) into the first equation of Eq.~(\ref{frame independent}), the convective form of the Allen-Cahn equation in the orthogonal curvilinear coordinate system is given by:
\begin{equation}\label{orthogonal curvilinear}
        \frac{\partial \phi}{\partial t}+u_n\frac{\partial\phi}{\partial n}=-\gamma\left(F'(\phi)-\varepsilon^2\left(\frac{\partial^2\phi}{\partial n^2}-\kappa\frac{\partial\phi}{\partial n}\right)\right).
\end{equation}

 Now we write the convective Allen-Cahn equation in a non-dimensional moving orthogonal curvilinear coordinate system. We non-dimensionalize the coordinate system using the interface thickness parameter. For the convenience of analyzing the interface distortion due to the convective velocity difference in the diffuse interface region, we translate the coordinate system with the interface. As a result, the relative convective velocity to the interface, which leads to the convective distortion, appears explicitly in the governing equation.  To begin with, we carry out the non-dimensionalization of the coordinate system by denoting the dimensionless
coordinates as $\tilde{n},\tilde{\tau_1},\tilde{\tau_2}$, which are non-dimensionalized with respect to $\varepsilon$: 
\begin{equation}\label{non-dimensionalization}
\tilde{n}=n/\varepsilon,\quad \tilde{\tau}_1=\tau_1/\varepsilon,\quad \tilde{\tau}_2=\tau_2/\varepsilon.
\end{equation}
Non-dimensionalizing the spatial coordinates in Eq.~(\ref{orthogonal curvilinear}) accordingly, we obtain: 
\begin{equation}\label{non-dimensionalized}
        \frac{\partial \phi}{\partial t}+u_n\frac{\partial\phi}{\partial \tilde{n}}\frac{1}{\varepsilon}=-\gamma\left(F'(\phi)-\left(\frac{\partial^2\phi}{\partial \tilde{n}^2}-\varepsilon\kappa\frac{\partial\phi}{\partial \tilde{n}}\right)\right).
\end{equation}
Assume that the principal radii of the interface are large compared to the interface thickness, which leads to $\kappa\ll1/\varepsilon$. With this assumption, the last term in Eq.~(\ref{non-dimensionalized}) can be neglected:
\begin{equation}
     \frac{\partial \phi}{\partial t}+u_n\frac{\partial\phi}{\partial \tilde{n}}\frac{1}{\varepsilon}=-\gamma\left(F'(\phi)-\frac{\partial^2\phi}{\partial \tilde{n}^2}\right).\label{curvature flow ignored}
\end{equation}
This completes the non-dimensionalization.
 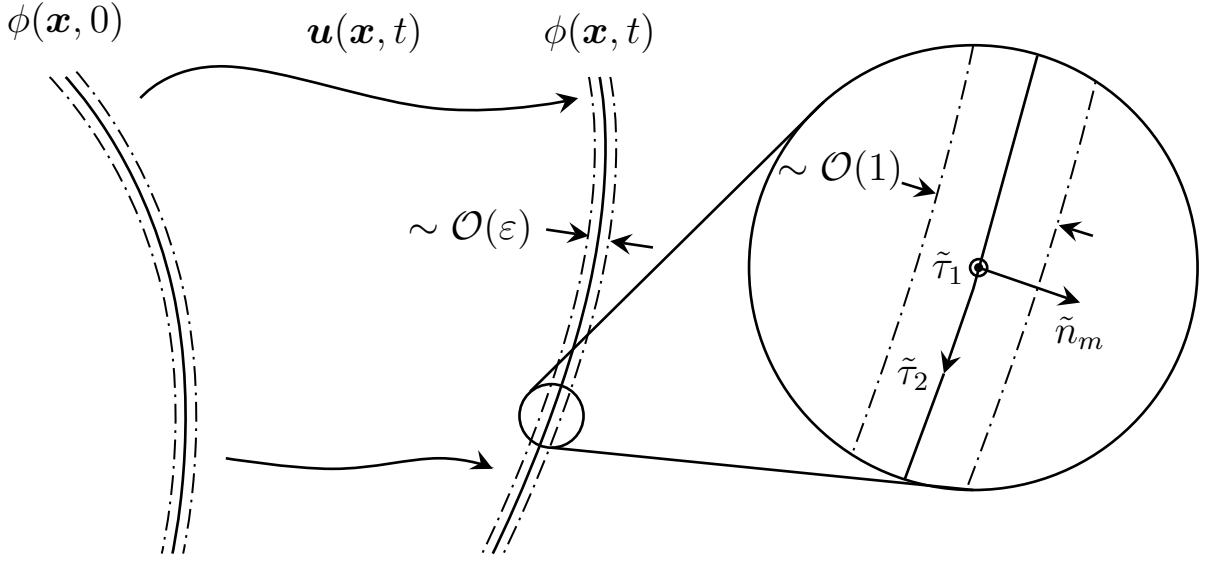
\begin{figure}
     \centering
     \trimbox{20 100 0 100}{
      \begin{tikzpicture}[scale=2.5,every node/.style={scale=0.5}]
     
 \begin{axis}[ 
ymin=-3,
ymax=3,
xmax=6.2,
xmin=-6,
xticklabel=\empty,
yticklabel=\empty,
ytick style={draw=none},
xtick style={draw=none},
axis line style = {draw=none},
minor tick num=0,
axis lines = middle,
label style = {at={(ticklabel cs:1.1)}},
axis equal
]
\draw plot [smooth, tension=1] coordinates { (axis cs: -4.5,2.5) (-3.5,0.5) (-3.5,-2)};
\draw [densely dashdotted,line width=0.3pt] plot [smooth, tension=1] coordinates { (axis cs: -4.4,2.55) (-3.4,0.5) (-3.4,-2)};
\draw [densely dashdotted,line width=0.3pt] plot [smooth, tension=1] coordinates { (axis cs: -4.65,2.5) (-3.6,0.5) (-3.6,-2)};
\node [above] at (-4.5,2.7) {$\phi(\boldsymbol{x},0)$};

\draw plot [smooth, tension=1] coordinates { (axis cs: 0.5,2.5) (0.4,0.5) (-0.5,-2)};
\draw [densely dashdotted,line width=0.3pt] plot [smooth, tension=1] coordinates { (axis cs: 0.4,2.5) (0.3,0.5) (-0.6,-2)};
\draw [densely dashdotted,line width=0.3pt] plot [smooth, tension=1] coordinates { (axis cs: 0.6,2.5) (0.5,0.5) (-0.4,-2)};
\node [above] at (0.5,2.6) {$\phi(\boldsymbol{x},t)$};

\draw [-stealth] plot [smooth,tension=0.7] coordinates {(-3.8,2.3) (-3,2.6)  (-1,2.2)  (0.3,2.3)};
\draw [-stealth] plot [smooth,tension=0.7] coordinates {(-3,-1.1) (-2,-1.2)  (-1,-1.1) (-0.5,-1.2)  };
\node [above] at (-1.7,2.6) {$\boldsymbol{u}(\boldsymbol{x},t)$};

\def\sft{-4}

\draw (0.05,-0.7) circle (0.3);
\draw (4,0.7) circle(2.1);
\draw (0.05,-1)--(0-\sft,-1.4);
\draw (-0.15,-0.47)--(-1.4-\sft,2.265);
\begin{scope}
       \clip(4,0.7) circle(2.1);
       
\draw   [-stealth] plot [smooth, tension=1](axis cs: 0.6-\sft,2.7) to (0-\sft,0.5) to (-0.275-\sft,-0.3) ;
\draw (-0.275-\sft,-0.3)  to (-0.9-\sft,-2);
\draw [-stealth](axis cs: 0.05-\sft,0.7) to (axis cs: 1-\sft,0.37);
\node [below] at (axis cs: 1-\sft,0.37) {$\tilde{n}_m$}; 
\node[left] at (axis cs: -0.275-\sft,-0.3) {$\tilde{\tau}_2$};
\draw [fill=black] (axis cs: 0.05-\sft,0.7) circle [radius=0.03];
\draw  (axis cs: 0.05-\sft,0.7) circle [radius=0.08];
\node[left] at (axis cs: 0.05-\sft,0.7) {$\tilde{\tau}_1$};

\draw [densely dashdotted,line width=0.3pt] plot [smooth, tension=1] coordinates { (axis cs: 0-\sft,2.8) (-0.6-\sft,0.5) (-1.5-\sft,-2)};
\draw [densely dashdotted,line width=0.3pt] plot [smooth, tension=1] coordinates { (axis cs: 1.2-\sft,2.7) (0.6-\sft,0.5) (-0.3-\sft,-2)};
\end{scope}

\draw[-stealth] (0,1.06)--(0.4,1);
\draw[-stealth] (1,0.89)--(0.6,0.95);
\node[left] at (0,1.06) {$\sim\mathcal{O}(\varepsilon)$};

\draw[-stealth] (-0.67-\sft,1.5)--(-0.35-\sft,1.4);
\draw[-stealth] (1.12-\sft,1.0)--(0.8-\sft,1.1);
\node[above] at (-1.25-\sft,1.3){$\sim\mathcal{O}(1)$};

\end{axis}
\end{tikzpicture}}
     \caption{Schematic diagram of the diffuse interface in non-dimensional moving orthogonal curvilinear coordinate system. The coordinate system ($\tilde{n}_m,\ \tilde{\tau}_1$ and $\tilde{\tau}_2$) is attached on the moving interface indicated by $\phi(\boldsymbol{x},t)$ convected in the velocity field $\boldsymbol{u}(\boldsymbol{x},t)$. The thickness of the diffuse interface in the coordinate system is $\sim\mathcal{O}(1)$ due to the non-dimensionalization.}
     \label{transformed coordinates}
 \end{figure}

\noindent Next, we write Eq.~(\ref{curvature flow ignored}) in a coordinate system which translates with the interface. We define the coordinates of the interface as $\Gamma^{\phi}_I(t)=\{(n_0(t),\tau_1,\tau_2)\in\Omega|\phi(n_0,\tau_1,\tau_2,t)=0\}$. The coordinate transformation to the moving coordinate system can be written as:
\begin{equation}
n_m=n-n_0(t),\quad \tilde{n}_m=\tilde{n}-\tilde{n}_0(t),\quad \tilde{n}_0(t)=n_0(t)/\varepsilon,
\end{equation}
where $n_m$ represents the normal coordinate in the moving coordinate system, $\tilde{n}_m$ is the non-dimensionalized $n_m$ with respect to $\varepsilon$, $\tilde{n}_0(t)$ is the non-dimensional normal coordinate of the interface. The coordinate system is illustrated in Fig.~\ref{transformed coordinates}. 

\noindent In the moving coordinate system, the temporal and spatial derivatives of the order parameter become:
\begin{align}
    \frac{\partial\phi(\tilde{n}_m,\tilde{\tau}_1,\tilde{\tau}_2,t)}{\partial t}=\frac{\partial\phi}{\partial t}-\frac{\partial\phi}{\partial \tilde{n}_m}\frac{d \tilde{n}_0(t)}{d t},\quad\frac{\partial\phi}{\partial n}=\frac{\partial\phi}{\partial n_m},\quad
    \frac{\partial\phi}{\partial \tilde{n}}=\frac{\partial\phi}{\partial \tilde{n}_m}, \quad
    \frac{\partial^2\phi}{\partial \tilde{n}^2}=\frac{\partial^2\phi}{\partial \tilde{n}_m^2}.\label{derivatives in moving coordinate}
\end{align}
Replacing the temporal and spatial derivatives in Eq.~(\ref{curvature flow ignored}) with Eq.~\ref{derivatives in moving coordinate}, we have:
\begin{equation}\label{moving coordinate}
    \frac{\partial\phi}{\partial t}-\frac{\partial\phi}{\partial \tilde{n}_m}\frac{d \tilde{n}_0(t)}{d t}+\frac{u_n(\tilde{n}_m,\tilde{\tau}_1,\tilde{\tau}_2,t)}{\varepsilon}\frac{\partial\phi}{\partial \tilde{n}_m}=-\gamma\left(F'(\phi)-\frac{\partial^2\phi}{\partial \tilde{n}_m^2}\right).
\end{equation}
Notice that $\tilde{n}_0(t)$ is the non-dimensional normal coordinate of the interface, the time derivative of which in Eq.~(\ref{moving coordinate}) gives the normal velocity of the interface in the non-dimensional coordinate system. To get the normal velocity at the interface, we substitute $\phi=0$ into Eq.~(\ref{curvature flow ignored}):
\begin{equation}\label{interface velocity}
         \frac{\partial \phi}{\partial t}\bigg|_{\phi=0}+u_n(\tilde{n}_0(t),\tilde{\tau}_1,\tilde{\tau}_2,t)\frac{\partial\phi}{\partial \tilde{n}}\bigg|_{\phi=0}\frac{1}{\varepsilon}=-\gamma\left(-\frac{\partial^2\phi}{\partial \tilde{n}^2}\bigg|_{\phi=0}\right).
\end{equation}
For the hyperbolic tangent profile, $\frac{\partial^2\phi}{\partial \tilde{n}^2}=0$ at $\phi=0$. Assuming that this is approximately satisfied when the convective distortion is not significant, thus the right-hand side of Eq.~(\ref{interface velocity}) is negligible:
\begin{equation}\label{convection equation}
    \frac{\partial \phi}{\partial t}\bigg|_{\phi=0}+\frac{u_n(\tilde{n}_0(t),\tilde{\tau}_1,\tilde{\tau}_2,t)}{\varepsilon}\frac{\partial\phi}{\partial \tilde{n}}\bigg|_{\phi=0}=0.
\end{equation}
The velocity of the interface can be identified from the convection equation ($\ref{convection equation}$) as:
\begin{equation}\label{normal velocity}
    \frac{d \tilde{n}_0(t)}{d t}=\frac{u_n(\tilde{n}_0(t),\tilde{\tau}_1,\tilde{\tau}_2,t)}{\varepsilon}.
\end{equation}
Substituting Eq.~(\ref{normal velocity}) into Eq.~(\ref{moving coordinate}), rewriting the velocity in the moving coordinate system and non-dimensionalize the equation with respect to $\gamma$, we get the governing equation for the interface profile:
\begin{equation}\label{interface profile goverining equation}
    \frac{\partial\phi}{\partial \tilde{t}}+\left(\frac{u_n(\tilde{n}_m,\tilde{\tau}_1,\tilde{\tau}_2,\tilde{t})-u_n(0,\tilde{\tau}_1,\tilde{\tau}_2,\tilde{t})}{\gamma\varepsilon}\right)\frac{\partial\phi}{\partial \tilde{n}_m}=-\left(F'(\phi)-\frac{\partial^2\phi}{\partial \tilde{n}_m^2}\right),
\end{equation}
where $\tilde{t}=t\gamma$ is the non-dimensional time.
As the second term in Eq.~(\ref{interface profile goverining equation}) goes to zero, the equation recovers to the Allen-Cahn equation, which gives the well-known hyperbolic tangent profile. The assumptions used in the derivation that the normal interface profile is almost identical on the interface and $\frac{\partial^2\phi}{\partial \tilde{n}^2}=0$ is approximately satisfied at $\phi=0$  are valid.

\subsection{Interface preserving condition}

As mentioned earlier, a non-zero second term of Eq.~(\ref{interface profile goverining equation}) causes the deviation from the hyperbolic tangent profile due to convection. We refer to the term as the convective distortion term. Multiplying and dividing the term by $\tilde{n}_m$, Eq.~(\ref{interface profile goverining equation}) becomes:
\begin{equation}\label{multi divid}
\frac{\partial\phi}{\partial \tilde{t}}+\left(\frac{u_n(\tilde{n}_m,\tilde{\tau}_1,\tilde{\tau}_2,\tilde{t})-u_n(0,\tilde{\tau}_1,\tilde{\tau}_2,\tilde{t})}{\tilde{n}_m}\right)\frac{1}{\gamma\varepsilon}\tilde{n}_m\frac{\partial\phi}{\partial \tilde{n}_m}=-\left(F'(\phi)-\frac{\partial^2\phi}{\partial \tilde{n}_m^2}\right).
\end{equation}
Suppose that $u_n$ is continuous on $[0,\tilde{n}_m]$ and differentiable on $(0,\tilde{n}_m)$ when $0<\tilde{n}_m$, or $u_n$ is continuous on $[\tilde{n}_m,0]$, and differentiable on $(\tilde{n}_m,0)$ when $\tilde{n}_m<0$, according to the mean value theorem \cite{bartle2000introduction}, there exists $\tilde{n}_s\in(0,\tilde{n}_m)$ or $\tilde{n}_s\in(\tilde{n}_m,0)$ respectively such that:
\begin{equation}\label{mean value theorem}
\left(\frac{u_n(\tilde{n}_m,\tilde{\tau}_1,\tilde{\tau}_2,\tilde{t})-u_n(0,\tilde{\tau}_1,\tilde{\tau}_2,\tilde{t})}{\tilde{n}_m}\right)=\frac{\partial u_n}{\partial \tilde{n}_m}(\tilde{n}_s,\tilde{\tau}_1,\tilde{\tau}_2,\tilde{t}).
\end{equation}
With Eq.~(\ref{mean value theorem}), Eq.~(\ref{multi divid}) can be rewritten as:
\begin{equation}
\frac{\partial\phi}{\partial \tilde{t}}+\left(\frac{\partial u_n}{\partial \tilde{n}_m}(\tilde{n}_s,\tilde{\tau}_1,\tilde{\tau}_2,\tilde{t})\right)\frac{1}{\gamma\varepsilon}\tilde{n}_m\frac{\partial\phi}{\partial \tilde{n}_m}=-\left(F'(\phi)-\frac{\partial^2\phi}{\partial \tilde{n}_m^2}\right),
\end{equation}
where $\tilde{n}_s\in(0,\tilde{n}_m)$ when $0<\tilde{n}_m$ and $\tilde{n}_s\in(\tilde{n}_m,0)$ when $\tilde{n}_m<0$.
Transform the partial derivative of the normal velocity back to the dimensional spatial coordinate system using Eqs.~(\ref{non-dimensionalization}) and (\ref{derivatives in moving coordinate}):
\begin{equation}\label{interface not included}
\frac{\partial\phi}{\partial \tilde{t}}+\left(\frac{\partial u_n}{\partial n}(n_s,\tau_1,\tau_2,\tilde{t})\right)\frac{1}{\gamma}\tilde{n}_m\frac{\partial\phi}{\partial \tilde{n}_m}=-\left(F'(\phi)-\frac{\partial^2\phi}{\partial \tilde{n}_m^2}\right),
\end{equation}
where $n_s\in(0,n_m)$ when $0<n_m$ and $n_s\in(n_m,0)$ when $n_m<0$. 

For the convenience of notation, we replace the notation of the non-dimensional time $\tilde{t}$ with $t$ and denote the normal velocity gradient in the normal direction as:
\begin{equation}\label{normal vel normal grad}
    \zeta(\boldsymbol{x},t)=\frac{\partial u_n}{\partial n}(\boldsymbol{x},t),
\end{equation}
which can be considered as the intensity of the convective distortion (see \ref{appdixA} for detailed explanation).\\

\noindent\textbf{Remark 1.} The magnitude of $\zeta(\boldsymbol{x},t)$ usually increases with the increase in the principal curvatures of the interface. Because high principal curvatures lead to a large surface tension force. When the force is distributed to the diffuse interface region by a Dirac delta function,
the violent variation of the surface tension force term across the diffuse interface gives rise to high velocity gradients, and furthermore causes the large magnitude of $\zeta(\boldsymbol{x},t)$.
\\

\noindent We next define a non-dimensional parameter, which we refer to as the convective distortion parameter: 
\begin{equation}\label{convective distortion parameter def}
 \xi(\boldsymbol{x},t)=\zeta(\boldsymbol{x},t)\big/\gamma. 
\end{equation}
With the notations in Eqs.~(\ref{normal vel normal grad}) and (\ref{convective distortion parameter def}), the governing equation for the interface profile Eq.~(\ref{interface not included}) becomes:
\begin{equation}\label{profile governing equation}
\frac{\partial\phi}{\partial t}+\xi(n_s,\tau_1,\tau_2,t)\tilde{n}_m\frac{\partial\phi}{\partial \tilde{n}_m}=-\left(F'(\phi)-\frac{\partial^2\phi}{\partial \tilde{n}_m^2}\right),
\end{equation}
where $n_s\in(0,n_m)$ when $0<n_m$ and $n_s\in(n_m,0)$ when $n_m<0$.

As the convective distortion term goes to zero, the interface profile approaches the hyperbolic tangent profile. Thus, the interface-preserving condition is given by:
\begin{equation}\label{constrain on the convection term}
\left|\xi(n_s,\tau_1,\tau_2,t)\tilde{n}_m\frac{\partial\phi}{\partial \tilde{n}_m}\right|\leqslant \eta_1,
\end{equation}
where $\eta_1$ is a desired upper bound for the magnitude of the convective distortion term.
Apply the Cauchy–Schwarz inequality for the convective distortion term in the condition (\ref{constrain on the convection term})  as:
\begin{align}\label{Cauchy Swchwarz inequality}
    \left|\xi(n_s,\tau_1,\tau_2,t)\tilde{n}_m\frac{\partial\phi}{\partial \tilde{n}_m}\right|\leqslant\Big|\xi(n_s,\tau_1,\tau_2,t)\Big|\Big|\tilde{n}_m\frac{\partial\phi}{\partial \tilde{n}_m}\Big| \leqslant \eta_1,
\end{align}
where $n_s\in(0,n_m)$ when $n_m > 0$ and $n_s\in(n_m,0)$ when $n_m<0$.
For the hyperbolic tangent profile $\phi({\tilde{n}_m})=\tanh(\tilde{n}_m/\sqrt{2})$, $|\tilde{n}_m\frac{\partial\phi}{\partial \tilde{n}_m}|$ approaches zero outside the diffuse interface region, which satisfies the condition (\ref{Cauchy Swchwarz inequality}). Inside the diffuse interface region, $|\tilde{n}_m\frac{\partial\phi}{\partial \tilde{n}_m}|$ is bounded by a constant. Denoting the constant as $\eta_2$, we have:
\begin{equation}
    \left|\tilde{n}_m\frac{\partial\phi}{\partial \tilde{n}_m}\right|\leqslant \eta_2,\quad \boldsymbol{x}\in\Gamma_{DI}^{\phi}(t),\label{upper bound eta 2}
\end{equation}
where $\Gamma^{\phi}_{DI}(t)$ represents the diffuse interface region at time $t$. We assume that inequality (\ref{upper bound eta 2}) is still valid when the interface profile is close to the hyperbolic tangent profile. Thus, the interface-preserving condition becomes:
\begin{equation}
\boxed{
    |\xi(\boldsymbol{x},t)|\leqslant \eta, \quad \boldsymbol{x}\in \Gamma^{\phi}_{DI}(t)\label{convective distortion parameter}
    }
\end{equation}
where $\eta=\eta_1/\eta_2$ is a desired upper bound for the convective distortion parameter.
Since $\eta_2$ is a constant once the diffuse interface region is defined, $\eta_1\to 0$ as $\eta\to 0$. In other words, the magnitude of the convective distortion term solely depends on $\xi$ and decreases with the reduction of the absolute value of the convective distortion parameter, which leads to the convergence of the interface profile to the hyperbolic tangent profile. 
\\

\noindent\textbf{Remark 2.} The diffuse interface region defines the spatial domain where the convective distortion is considered. A large diffuse interface region includes the distortion away from the interface, which may lead to an overestimation. The overestimation provides better interface-preserving capability but may induce a relatively larger volume-preserving mean curvature flow. In contrast, a small diffuse interface region may underestimate the interface distortion giving rise to a smaller volume-preserving mean curvature flow but weakening the interface-preserving capability. In the current study, we define the diffuse interface region as the region where $90\%$ of the variation of $\phi$ occurs: $\Gamma^{\phi}_{DI}(t)=\{(\boldsymbol{x},t)||\phi(\boldsymbol{x},t)|\leqslant0.9\}$. For the hyperbolic tangent profile, the thickness of the region in the normal direction is $4.164\varepsilon$. 

\subsection{Time-dependent mobility model}
 To satisfy the interface-preserving condition, one needs to adjust the mobility dynamically according to the normal velocity gradient in the normal direction. The interface-preserving condition (\ref{convective distortion parameter}) can be written as:
\begin{equation}\label{interface-preserving condition}
    \gamma\geqslant\frac{|\zeta(\boldsymbol{x},t)|}{\eta},\ \boldsymbol{x}\in\Gamma_{DI}^{\phi}(t).
\end{equation}
We notice that the right-hand side of inequality (\ref{interface-preserving condition}) varies in both space and time. To maintain the inequality, it is natural to consider a mobility model with spatial and temporal dependence. 
However, a mobility model that prohibits the variation in the normal direction while allows variations in tangential directions poses challenges in its construction. Consequently, we consider a time-dependent mobility model in the current study. The mobility coefficient is taken as a constant throughout the computational domain at each time instance, while it is allowed to change as time evolves. This requires a projection at time $t$ from the spatially varying 
$|\zeta(\boldsymbol{x},t)|$ to a real-valued $\gamma(t)$: 
\begin{equation}\label{time dependent mobility}
\mathcal{F}:\Omega\to\mathbb{R}_{\geqslant0},\ \gamma(t)=\frac{1}{\eta}\mathcal{F}(|\zeta(\boldsymbol{x},t)|).
\end{equation}
In the current study, we employ the RMS value in the diffuse interface region for the projection, which relaxes the condition (\ref{interface-preserving condition}) in an average sense:
\begin{equation}
    \mathcal{F}(\varphi(\boldsymbol{x},t))=\sqrt{\frac{\int(\varphi(\boldsymbol{x},t))^2d\Omega}{\int 1 d\Omega}},\ \boldsymbol{x}\in\Gamma^{\phi}_{DI}(t).
\end{equation}
We refer to $\eta$ as the RMS convective distortion parameter in the rest of the paper. As derived in 
Appendix B, the frame independent form of the time-dependent mobility model can be expressed as:
\begin{align}\label{final form}
 \gamma(t)&=\frac{1}{\eta}\mathcal{F}\left(\left|\frac{\nabla\phi\cdot\nabla\boldsymbol{u}\cdot\nabla\phi}{|\nabla\phi|^2}\right|\right).
\end{align}
The frame independent form facilitates its numerical implementation in Cartesian coordinate system introduced in the next section.
\\

\noindent\textbf{Remark 3.} By taking the RMS function to perform the projection, the condition (\ref{interface-preserving condition}) is relaxed in an average sense. Because the mobility is calculated according to the RMS value of $|\zeta(\boldsymbol{x},t)|$, at the location where $|\zeta(\boldsymbol{x},t)|$ exceeds the RMS value, the condition (\ref{interface-preserving condition}) is violated. As mentioned in Remark 1, the violation usually happens in the region with high principle curvatures and singularities of the interface. This allows the merging and breaking-up of the interface, where keeping the hyperbolic tangent profile is no longer required. Other projection methods can be employed for different considerations and requirements.
\\

\section{Variational interface-preserving conservative Allen-Cahn-Navier-Stokes formulation }
\label{formulation}

In this section, we present an variational implementation of the interface-preserving conservative phase-field formulation. For the sake of completeness, we describe the governing equations of the two-phase flow modeling, viz., the incompressible Navier-Stokes equations and the conservative Allen-Cahn equation with the proposed time-dependent mobility model. We begin with the strong form of the equations and then project them into finite element space as the semi-discrete variational form. Specifically, we describe the discretization of the time-dependent mobility model. The section is closed with the coupled linearized matrix form of the variational discretization.

\subsection{Coupling between the Allen-Cahn and incompressible Navier-Stokes Equations}
%
Consider a domain $\Omega\times]0,T[$ consisting of the spatial points $\boldsymbol{x}$ at time $t$. The boundary of the domain, $\Gamma$ can be decomposed in two ways, $\Gamma = \Gamma^{\mathrm{f}}_{D} \cup \Gamma^{\mathrm{f}}_{H}$ and $\Gamma = \Gamma^\mathrm{\phi}_D \cup \Gamma^\mathrm{\phi}_H$, where $\Gamma_D^{\mathrm{f}}$ and $\Gamma_H^{\mathrm{f}}$ denote the Dirichlet and Neumann boundaries for the Navier-Stokes equations respectively, while $\Gamma^{\phi}_D$ and $\Gamma^{\phi}_H$ denote the same for the Allen-Cahn counterpart respectively. The diffuse interface region between the two-phases is denoted as $\Gamma^{\phi}_{DI}(t)$.
The one-fluid formulation for the two-phase incompressible and immiscible fluids system with the boundary conditions is given as:
\begin{align}
\left.
\begin{aligned} \label{NS FULL}
	\rho\frac{\partial {\boldsymbol{u}}}{\partial t} + \rho\boldsymbol{u}\cdot\nabla{\boldsymbol{u}} &= \nabla\cdot {\boldsymbol{\sigma}} + \mathbf{sf} + \boldsymbol{b},&\hspace{4.5cm}&\mathrm{on}\ \Omega,\\
	\nabla\cdot{\boldsymbol{u}} &= 0,&&\mathrm{on}\ \Omega, \\
	\boldsymbol{u} &= \boldsymbol{u}_{D},&&\forall \boldsymbol{x}\in \Gamma^{\mathrm{f}}_{D}, \\
	\boldsymbol{\sigma}\cdot\boldsymbol{n}^{\mathrm{f}} &= \boldsymbol{h},&&\forall \boldsymbol{x}\in \Gamma^{\mathrm{f}}_{H}, \\
	\boldsymbol{u} &= \boldsymbol{u}_{0},&&\mathrm{on}\ \Omega(0),
	\end{aligned}\hspace{5pt}\right\}
	\\ \nonumber \\
	\left.
	\begin{aligned}\label{AC FULL}
	\frac{\partial\phi}{\partial t} + \boldsymbol{u}\cdot\nabla\phi &= -\gamma(t)
\big( F'(\phi) - \varepsilon^2\nabla^2\phi  - \beta(t)\sqrt{F(\phi
)}\big),\ \ &&\mathrm{on}\ \Omega, \\
            \gamma(t)&=\frac{1}{\eta}\mathcal{F}(|\zeta(\boldsymbol{x},t)|)  && \boldsymbol{x}\in\Gamma^{\phi}_{DI}(t)\\
	\phi &= \phi_{D},\ \ &&\forall \boldsymbol{x}\in \Gamma^\mathrm{\phi}_{D},\\
	\nabla\phi\cdot\boldsymbol{n}^\mathrm{\phi}_{\Gamma} &= 0,\ \ &&\forall \boldsymbol{x}\in \Gamma^\mathrm{\phi}_{H},\\
	\phi\big|_{t=0} &= \phi_{0},\ \ &&\mathrm{on}\ \Omega(0), 
\end{aligned}\hspace{5pt}\right\}
\end{align}
where Eq.~(\ref{NS FULL}) and Eq.~(\ref{AC FULL}) represent the Navier-Stokes and Allen-Cahn equations respectively. In the Navier-Stokes equations, $\rho$ is the density of the fluid, $\boldsymbol{u}$ represents the fluid velocity defined for each spatial point $\boldsymbol{x}$ in $\Omega$,  $\boldsymbol{b}$ is the body force on the fluid such as gravity $(\boldsymbol{b}=\rho\boldsymbol{g})$, $\boldsymbol{g}$ being the acceleration due to gravity, $\boldsymbol{u}_{D}$ and $\boldsymbol{h}$ denote the boundary conditions at the Dirichlet and Neumann boundaries respectively, $\boldsymbol{n}^{\mathrm{f}}$ is the unit outward normal to the Neumann boundary and $\boldsymbol{u}_{0}$ represents the initial velocity field at $t=0$. The Cauchy stress tensor for a Newtonian fluid is given as:
\begin{align}
\boldsymbol{\sigma}=-p\boldsymbol{I} + \boldsymbol{T},\quad \boldsymbol{T}=2\mu\boldsymbol{\epsilon}(\boldsymbol{u}),\quad \boldsymbol{\epsilon}(\boldsymbol{u})= \frac{1}{2}\big[ \nabla\boldsymbol{u} + (\nabla\boldsymbol{u})^T \big],
\end{align}
where $p$ is the pressure field,  $\boldsymbol{T}$ and $\boldsymbol{\epsilon}$ represent the shear stress tensor and the fluid strain rate tensor respectively and $\mu$ denotes the dynamic viscosity of the fluid. The physical parameters of the fluid such as $\rho$ and $\mu$ vary with the evolution of the interface indicated by order parameter $\phi$:
\begin{align}
	\rho(\phi) &= \frac{1+\phi}{2}\rho_1 + \frac{1-\phi}{2}\rho_2, \label{dens}\\
	\mu(\phi) &= \frac{1+\phi}{2}\mu_1 + \frac{1-\phi}{2}\mu_2, \label{visc}
\end{align} 
where $\rho_i$ and $\mu_i$ are the density and dynamic viscosity of the $i$th phase of the fluid respectively. The surface tension force $\mathbf{sf}$ is modeled by the continuum surface force (CSF) model \cite{brackbill1992continuum}, in which it is reformulated as a volumetric source term with a Dirac Delta function utilizing the gradient of the phase indicator $\phi$. Several forms of $\mathbf{sf}(\phi)$ have been used in the literature which are reviewed in \cite{kim2012phase,kim2005continuous}. In this study, we employ the following definition \cite{lafaurie1994modelling}:
	\begin{align} \label{surface_tension}
	\mathbf{sf}(\phi) &=\sigma\kappa\boldsymbol{n}^{\phi}_L\delta_S\nonumber\\
	&=\sigma\nabla\cdot\left(\left( \mathbf{I} - \boldsymbol{n}^{\phi}_L\otimes \boldsymbol{n}^{\phi}_L\right)\delta_S \right)\nonumber\\
	&=\sigma\alpha_{\mathrm{sf}}\varepsilon\nabla\cdot\left(|\nabla\phi|^2\mathbf{I}-\nabla\phi\otimes\nabla\phi\right)
	\end{align}
	where $\sigma$ is the surface tension coefficient, $\delta_S=\varepsilon\alpha_\mathrm{sf}|\nabla\phi|^2$ is the Dirac delta function at the interface, $\alpha_\mathrm{sf}=3\sqrt{2}/4$ is a constant derived by the property of the Dirac delta function, $\kappa$ being the summation of the principle curvatures of the interface and $\boldsymbol{n}^{\phi}_L=\nabla\phi/|\nabla\phi|$ denotes the normal vector of the level sets of $\phi$.

On the other hand, in Eq.~(\ref{AC FULL}), $\varepsilon$ is the interface thickness parameter, $\gamma(t)$ is the time-dependent mobility, $F(\phi)$ is the double-well potential, $\eta$ is the RMS convective distortion parameter and $\boldsymbol{n}^{\phi}_L$ is the unit normal vector of the level sets of the order parameter $\phi$. The value of the order parameter at the Dirichlet boundary is denoted by $\phi_{D}$, the initial condition is represented by $\phi_{0}$ and $\boldsymbol{n}^{\phi}_{\Gamma}$ denotes the unit outward normal to the Neumann boundary where a zero flux condition is satisfied. The mass conservation is enforced in the Allen-Cahn equation by a Lagrange multiplier $\beta(t)\sqrt{F(\phi)}$ where $\beta(t) = \int_{\Omega} F'(\phi)\mathrm{d}\Omega / \int_{\Omega}\sqrt{F(\phi)}\mathrm{d}\Omega$, $F'(\phi)$ is the derivative of the energy potential with respect to the order parameter. The Allen-Cahn equation can be transformed into a convection-diffusion-reaction equation as follows:
\begin{align}
	\partial_t\phi + \boldsymbol{u}\cdot\nabla\phi- \gamma(t)(\hat{k}\nabla^2\phi - \hat{s}\phi + \hat{f}) = 0 \ \ &\mathrm{on}\ \Omega^\mathrm{f},
\end{align}
where $\boldsymbol{u}$, $\hat{k}$, $\hat{s}$ and $\hat{f}$ are the convective velocity, modified diffusion coefficient, modified reaction coefficient and the modified source respectively which are defined in \cite{joshi2018positivity}.

\subsection{Semi-discrete Allen-Cahn and Navier-Stokes equations}
In this subsection, we present the semi-discrete variational form of the Navier-Stokes-Allen-Cahn (NS-AC) system, which has been described earlier. We employ the generalized-$\alpha$ technique \cite{Gen_alpha} for the temporal discretization which enables a user-controlled high frequency damping desirable for coarse discretizations in space and time. The following expressions are employed for the temporal discretization of the Navier-Stokes equations:
\begin{align}
	\boldsymbol{u}^\mathrm{n+1} &= \boldsymbol{u}^\mathrm{n} + \Delta t\partial_t\boldsymbol{u}^\mathrm{n} + \varsigma\Delta t (\partial_t\boldsymbol{u}^\mathrm{n+1} - \partial_t\boldsymbol{u}^\mathrm{n}),\\
	\partial_t\boldsymbol{u}^\mathrm{n+\alpha_m} &= \partial_t\boldsymbol{u}^\mathrm{n} + \alpha_\mathrm{m}(\partial_t\boldsymbol{u}^\mathrm{n+1} - \partial_t\boldsymbol{u}^\mathrm{n}),\\
	\boldsymbol{u}^\mathrm{n+\alpha} &= \boldsymbol{u}^\mathrm{n} + \alpha(\boldsymbol{u}^\mathrm{n+1} - \boldsymbol{u}^\mathrm{n}),
\end{align} 
where $\alpha$, $\alpha_\mathrm{m}$ and $\varsigma$ are the generalized-$\alpha$ parameters which are dependent on the user-defined spectral radius $\rho_{\infty}$. The time step size is denoted by $\Delta t$ and $\partial_t$ denotes the partial differentiation with respect to time. Similar expressions can be written for the Allen-Cahn equation as well.

Suppose $\mathcal{S}^\mathrm{h}_{\boldsymbol{u}}$, $\mathcal{S}^\mathrm{h}_{p}$ and $\mathcal{S}^\mathrm{h}_{\phi}$ denote the space of trial solution such that:
\begin{align}
	\mathcal{S}^\mathrm{h}_{\boldsymbol{u}} &= \big\{ \boldsymbol{u}_\mathrm{h}\ |\ \boldsymbol{u}_\mathrm{h} \in (H^1(\Omega))^{d}, \boldsymbol{u}_\mathrm{h} = \boldsymbol{u}_{D}\ \mathrm{on}\ \Gamma^{\mathrm{f}}_{D} \big\},\\
	\mathcal{S}^\mathrm{h}_{p} &= \big\{ p_\mathrm{h}\ |\ p_\mathrm{h} \in L^2(\Omega) \big\},\\
	\mathcal{S}^\mathrm{h}_{\phi} &= \big\{ \phi_\mathrm{h}\ |\ \phi_\mathrm{h} \in H^1(\Omega), \phi_\mathrm{h} = \phi_{D}\ \mathrm{on}\ \Gamma^\mathrm{\phi}_{D} \big\},
\end{align}
where $(H^1(\Omega))^{d}$ denotes the space of square-integrable $\mathbb{R}^{d}$-valued functions with square-integrable derivatives on $\Omega$ and $L^2(\Omega)$ is the space of the scalar-valued functions that are square-integrable on $\Omega$. Similarly, we define $\mathcal{V}^\mathrm{h}_{\boldsymbol{\psi}}$, $\mathcal{V}^\mathrm{h}_{q}$ and $\mathcal{V}^\mathrm{h}_{\phi}$ as the space of test functions such that:
\begin{align}
	\mathcal{V}^\mathrm{h}_{\boldsymbol{\psi}} &= \big\{ \boldsymbol{\psi}_\mathrm{h}\ |\ \boldsymbol{\psi}_\mathrm{h} \in (H^1(\Omega))^{d}, \boldsymbol{\psi}_\mathrm{h} = \boldsymbol{0}\ \mathrm{on}\ \Gamma^{\mathrm{f}}_{D} \big\},\\
	\mathcal{V}^\mathrm{h}_{q} &= \big\{ q_\mathrm{h}\ |\ q_\mathrm{h} \in L^2(\Omega) \big\},\\
	\mathcal{V}^\mathrm{h}_{\phi} &= \big\{ \hat{w}_\mathrm{h}\ |\ \hat{w}_\mathrm{h} \in H^1(\Omega), \hat{w}_\mathrm{h} = 0\ \mathrm{on}\ \Gamma^\mathrm{\phi}_{D} \big\}.
\end{align}
The variational statement of the combined NS-AC system can be written as: \\
 find $[\boldsymbol{u}_\mathrm{h}(t^\mathrm{n+\alpha}),p_\mathrm{h}(t^\mathrm{n+1}),\phi_\mathrm{h}(t^\mathrm{n+\alpha})]\in \mathcal{S}^\mathrm{h}_{\boldsymbol{u}} \times \mathcal{S}^\mathrm{h}_{p} \times \mathcal{S}^\mathrm{h}_{\phi}$ such that $\forall [\boldsymbol{\psi}_\mathrm{h},q_\mathrm{h}, \hat{w}_\mathrm{h}]\in \mathcal{V}^\mathrm{h}_{\boldsymbol{\psi}} \times \mathcal{V}^\mathrm{h}_{q} \times \mathcal{V}^\mathrm{h}_{\phi}$ for the 
incompressible NS equations
\begin{align}
&\int_{\Omega} \rho(\phi) ( \partial_t{\boldsymbol{u}}_\mathrm
{h} + {\boldsymbol{u}}_{\mathrm{h}} \cdot\nabla{\boldsymbol{u}}_{\mathrm{h}})\cdot\boldsymbol{\psi}_{\mathrm{h}} \mathrm{d\Omega} +
\int_{\Omega} {\boldsymbol{\sigma}}_{\mathrm{h}}
: \nabla\boldsymbol{\psi}_{\mathrm{h}} \mathrm{d\Omega} + \underbrace{\int
_{\Omega} \alpha_{\mathrm{sf}}\sigma\varepsilon\left(|\nabla\phi_{\mathrm{h}}|^2\mathbf{I}-\nabla\phi_{\mathrm{h}}\otimes\nabla\phi_{\mathrm{h}}\right):\nabla
\boldsymbol{\psi}_{\mathrm{h}} \mathrm{d\Omega}}_\mathrm{Surface\ tension\ force}  \nonumber \\
+ &\displaystyle\sum_\mathrm{e=1}^\mathrm{n_{el}}\int_{\Omega
^{\mathrm{e}}} \frac{\tau_\mathrm{m}}{\rho(\phi)} (\rho(\phi
){\boldsymbol{u}}_{\mathrm{h}}\cdot\nabla
\boldsymbol{\psi}_{\mathrm{h}}+ \nabla q_{\mathrm{h}} )\cdot
\boldsymbol{\mathcal{R}}_\mathrm{m} \mathrm
{d\Omega^e} \nonumber \\
+ &\int_{\Omega}q_{\mathrm{h}}(\nabla\cdot{\boldsymbol{u}}_{\mathrm{h}}) \mathrm{d\Omega} + \displaystyle\sum_\mathrm{e=1}^\mathrm{n_{el}}\int
_{\Omega^{\mathrm{e}}} \nabla\cdot\boldsymbol{\psi}_{\mathrm{h}}\tau
_\mathrm{c}\rho(\phi) \boldsymbol{\mathcal{R}}_\mathrm
{c} \mathrm{d\Omega^e}\nonumber\\
-&\displaystyle\sum_\mathrm{e=1}^\mathrm{n_{el}}\int_{\Omega
^{\mathrm{e}}} \tau_\mathrm{m} \boldsymbol{\psi}_{\mathrm{h}}\cdot
(\boldsymbol{\mathcal{R}}_\mathrm{m} \cdot
\nabla{\boldsymbol{u}}_{\mathrm{h}}) \mathrm
{d\Omega^e} -\displaystyle\sum_\mathrm{e=1}^\mathrm{n_{el}}\int
_{\Omega^{\mathrm{e}}} \frac{\nabla\boldsymbol{\psi}_\mathrm
{h}}{\rho(\phi)}:(\tau_\mathrm{m}\boldsymbol{\mathcal{R}}_\mathrm
{m} \otimes\tau_\mathrm{m}\boldsymbol
{\mathcal{R}}_\mathrm{m}) \mathrm{d\Omega
^e}\nonumber\\
= &\int_{\Omega} \boldsymbol{b}(t^{\mathrm{n}+\alpha})\cdot
\boldsymbol{\psi}_{\mathrm{h}} \mathrm{d\Omega} + \int_{\Gamma
_{H}} \boldsymbol{h}\cdot\boldsymbol{\psi}_{\mathrm{h}}
\mathrm{d\Gamma}, \label{PG_NS}
\end{align}
and for the Allen-Cahn equation:
\begin{align}
\label{PPV_AC}
&\int_{\Omega}\bigg( \hat{w}_{\mathrm{h}}\partial_t{\phi}_{\mathrm{h}} +
\hat{w}_{\mathrm{h}}\big(\boldsymbol{u}_h\cdot\nabla\phi_{\mathrm{h}}\big) + \underbrace{\gamma\big(t^{n+\alpha} \big)}_\mathrm{Dynamic\ mobility} \big(\nabla
\hat{w}_{\mathrm{h}}\cdot(\hat{k}\nabla\phi_{\mathrm{h}} ) + \hat{w}_{\mathrm{h}}\hat{s}\phi
_{\mathrm{h}} - \hat{w}_{\mathrm{h}}\hat{f}\big) \bigg) \mathrm{d}\Omega\nonumber\\
+& \displaystyle\sum_\mathrm{e=1}^\mathrm{n_{el}}\int_{\Omega
^{\mathrm{e}}}\bigg( \Big(\boldsymbol{u}_h\cdot\nabla \hat{w}_{\mathrm{h}}
\Big)\tau_{\phi}\Big( \partial_t{\phi}_{\mathrm{h}} + \boldsymbol{u}_h\cdot
\nabla\phi_{\mathrm{h}} - \underbrace{\gamma\big(t^{n+\alpha}\big)}\big(\nabla\cdot(\hat{k}\nabla\phi_{\mathrm{h}}) -
\hat{s}\phi_{\mathrm{h}} +\hat{f}\big) \Big) \bigg) \mathrm{d}\Omega^{\mathrm{e}}
\nonumber\\
+& \displaystyle\sum_\mathrm{e=1}^\mathrm{n_{el}}\int_{\Omega
^{\mathrm{e}}} \chi\frac{|\mathcal{R}(\phi_{\mathrm{h}})|}{|\nabla
\phi_{\mathrm{h}}|}k_s^\mathrm{add} \nabla \hat{w}_{\mathrm{h}}\cdot\bigg(
\frac{\boldsymbol{u}_h\otimes\boldsymbol{u}_h}{|\boldsymbol{u}_h|^2}
\bigg) \cdot\nabla\phi_{\mathrm{h}} \mathrm{d}\Omega^{\mathrm{e}} 
\nonumber\\
+&\sum_\mathrm{e=1}^\mathrm{n_{el}} \int_{\Omega^{\mathrm{e}}}\chi
\frac{|\mathcal{R}(\phi_{\mathrm{h}})|}{|\nabla\phi_{\mathrm{h}}|}
k^\mathrm{add}_{c} \nabla \hat{w}_{\mathrm{h}} \cdot\bigg( \mathbf{I} -
\frac{\boldsymbol{u}_h\otimes\boldsymbol{u}_h}{|\boldsymbol{u}_h|^2}
\bigg) \cdot\nabla\phi_{\mathrm{h}} \mathrm{d}\Omega^{\mathrm{e}}=0,
\end{align}
where the terms with under brackets representing the fluid-fluid interface dynamics are central to the current study, $\boldsymbol{\mathcal{R}}_\mathrm{m}$, $\boldsymbol{\mathcal{R}}_\mathrm{c}$ and $\mathcal{R}(\phi_\mathrm{h})$ denote the element-wise residuals for the momentum, continuity and the Allen-Cahn equations, respectively.

In Eq.~(\ref{PG_NS}), the terms in the first line represent the Galerkin projection of the momentum equation in the test function space $\boldsymbol{\psi}_\mathrm{h}$ and the second line comprises of the Petrov-Galerkin stabilization term for the momentum equation. The third line denotes the Galerkin projection and stabilization terms for the continuity equation and the terms in the fourth line are derived via approximation of fine scale velocity on the element interiors based on multi-scale argument \cite{hughes2005conservation, hsu2010improving}. The terms in the final line are the Galerkin projection of the body force and Neumann boundary condition. On the other hand, in Eq.~(\ref{PPV_AC}), the first line is the Galerkin projection of the transient, convection, diffusion, reaction and source terms, the second line represents the Streamline-Upwind Petrov-Galerkin stabilization and the third line depicts the PPV terms that are derived for the multi-dimensional convection-diffusion-reaction equation via satisfaction of the positivity condition at the element matrix level \cite{joshi2017positivity}. Several test cases have been performed to assess the effectiveness of this PPV technique in \cite{joshi2017positivity}. The details of the derivation of the added diffusions $k_s^\mathrm{add}$, $k_c^\mathrm{add}$ and $\chi$ can be found in \cite{joshi2017positivity}, which are given for the present context by \cite{joshi2018positivity}:
\begin{align}
	\chi &= \frac{2}{|\hat{s}|h + 2|\boldsymbol{u}_h|},\\
	k_s^\mathrm{add} &= \mathrm{max} \bigg\{ \frac{||\boldsymbol{u}_h| - \tau_{\phi}|\boldsymbol{u}_h|\hat{s}|h}{2} - (\hat{k} + \tau_{\phi}|\boldsymbol{u}_h|^2) + \frac{\hat{s}h^2}{6}, 0 \bigg\},\\
	k_c^\mathrm{add} &= \mathrm{max} \bigg\{ \frac{|\boldsymbol{u}_h|h}{2} - \hat{k} + \frac{\hat{s}h^2}{6}, 0 \bigg\},
\end{align} 
where $|\boldsymbol{u}_h|$ is the magnitude of the convective velocity and $h$ is the characteristic element length defined in \cite{joshi2017positivity}.  The stabilization parameters $\tau_\mathrm{m}$, $\tau_\mathrm{c}$ and $\tau_{\phi}$ in Eqs.~(\ref{PG_NS}) and (\ref{PPV_AC}) are given by \cite{shakib1991new, brooks1982streamline}:
\begin{align}
	\tau_\mathrm{m} &= \bigg[ \bigg( \frac{2}{\Delta t}\bigg)^2 + \boldsymbol{u}_\mathrm{h}\cdot\boldsymbol{G}\boldsymbol{u}_\mathrm{h} + C_I \bigg( \frac{\mu(\phi)}{\rho(\phi)}\bigg)^2 \boldsymbol{G}:\boldsymbol{G} \bigg]^{-1/2},\qquad \tau_\mathrm{c} = \frac{1}{\mathrm{tr}(\boldsymbol{G})\tau_\mathrm{m}}, \\
	\tau_{\phi} &= \bigg[ \bigg(\frac{2}{\Delta t} \bigg)^2 + \boldsymbol{u}_h\cdot\boldsymbol{G}\boldsymbol{u}_h + 9\hat{k}^2 \boldsymbol{G}:\boldsymbol{G} + \hat{s}^2 \bigg]^{-1/2}.
\end{align}
where $C_I$ is a constant derived from the element-wise inverse estimates \cite{harari1992c}, $\boldsymbol{G}$ is the element contravariant metric tensor and $\mathrm{tr}(\boldsymbol{G})$ is the trace of the contravariant metric tensor. This stabilization in the variational form circumvents the Babu$\mathrm{\check{s}}$ka-Brezzi condition that is required to be satisfied by any standard mixed Galerkin method \cite{Johnson}.

\subsection{Discrete form of the time-dependent mobility model}
 We present the discrete form of the time-dependent mobility model in this subsection. The strong form of the time-dependent mobility model is given by Eq.~(\ref{final form}):
\begin{equation}
    \gamma(t)=\frac{1}{\eta}\mathcal{F}\left(\left|\frac{\nabla\phi\cdot\nabla\boldsymbol{u}\cdot\nabla\phi}{|\nabla\phi|^2}\right|\right), \label{discrete frame independnet}
\end{equation}
where  $\mathcal{F}(\varphi(\boldsymbol{x},t))=\sqrt{\frac{\int(\varphi(\boldsymbol{x},t))^2d\Omega}{\int 1 d\Omega}},\ \boldsymbol{x}\in\Gamma^{\phi}_{DI}(t)$.
In Eq.~(\ref{discrete frame independnet}), $\mathcal{F}\left(\left|\frac{\nabla\phi\cdot\nabla\boldsymbol{u}\cdot\nabla\phi}{|\nabla\phi|^2}\right|\right)$ can be approximated as the RMS of $\left|\nabla\phi\cdot\nabla\boldsymbol{u}\cdot\nabla\phi/|\nabla\phi|^2 \right|$ at all the nodes located inside the diffuse interface region $\Gamma^{\phi}_{DI}(t)$. The nodal value of $\left|\nabla\phi\cdot\nabla\boldsymbol{u}\cdot\nabla\phi/|\nabla\phi|^2 \right|$ is calculated as follows.  

The nodal value of $\boldsymbol{u}_\mathrm{h}(t^{\mathrm{n}+\alpha})$, $\phi_\mathrm{h}(t^{\mathrm{n}+\alpha})$ is used to interpolate the $\nabla\boldsymbol{u}_\mathrm{h}(t^{\mathrm{n}+\alpha})$ and $\nabla\phi_\mathrm{h}(t^{\mathrm{n}+\alpha})$ at the quadrature points, and $L^2$-projection is used to project the value on the quadrature points back to the nodes inside the diffuse interface region \cite{jaiman2016partitioned}. If the node lies outside the diffuse interface region, the value is assigned to be zero. For example, for the node $p$, we have:
\begin{align*}
  \left|\left(\frac{\nabla\phi_\mathrm{h}\cdot\nabla\boldsymbol{u}_\mathrm{h}\cdot\nabla\phi_\mathrm{h}}{|\nabla\phi_\mathrm{h}|^2}\right)_{p}\right|=
  \begin{cases}
    \left|\dfrac{\sum_e\int_{\Omega^e}N_p( \nabla\phi_\mathrm{h}\cdot\nabla\boldsymbol{u}_\mathrm{h}\cdot\nabla\phi_\mathrm{h})/|\nabla\phi_\mathrm{h}|^2 d\Omega^e}{\sum_e\int_{\Omega_e}N_p d\Omega^e}\right|
    & \mathrm{if}\ |\phi_p|\leqslant0.9 ,\\
    \hfil 0                    
    & \mathrm{if}\ |\phi_p|>0.9,
\end{cases}
\end{align*}
where $N_p$ represents the shape function at node $p$. The discrete form of the time-dependent mobility model is given by:
\begin{equation}
    \gamma(t^{\mathrm{n}+\alpha})=\frac{1}{\eta}\left(\sqrt{\frac{1}{n_{DI}}\sum\limits_{p=1}^{n_{DI}} \left|\left(\frac{\nabla\phi_\mathrm{h}\cdot\nabla\boldsymbol{u}_\mathrm{h}\cdot\nabla\phi_\mathrm{h}}{|\nabla\phi_\mathrm{h}|^2}\right)_{p}\right|^2}\right),
\end{equation}
where $n_{DI}$ denotes the number of the nodes lying inside the diffuse interface region.

\subsection{Coupled partitioned matrix formulation}
In this subsection, we present the coupled linearized matrix form of the variationally discretized two-phase flow equations. Employing the Newton-Raphson linearization technique, the coupled two-phase fluid system can be expressed in terms of the solution increments for velocity, pressure and order parameter ($\Delta\boldsymbol{u}$, $\Delta p$ and $\Delta \phi$ respectively) as:
\begin{align} \label{TPFS_LM}
\begin{bmatrix}
\boldsymbol{K}_{\Omega}& & \boldsymbol{G}_{\Omega} & & \boldsymbol{D}_{\Omega}\\ \noalign{\vspace{4pt}}
-\boldsymbol{G}^T_{\Omega}& &\boldsymbol{C}_{\Omega} & & \boldsymbol{0}\\ \noalign{\vspace{4pt}}
\boldsymbol{G}_{AC} & & \boldsymbol{0} & & \boldsymbol{K}_{AC}
\end{bmatrix}
					\begin{Bmatrix} \Delta \boldsymbol{u} \\ \noalign{\vspace{4pt}}
									\Delta p \\ \noalign{\vspace{4pt}}
									\Delta \phi
									 \end{Bmatrix}
				= \begin{Bmatrix} \overline{\boldsymbol{\mathcal{R}}}_\mathrm{m} \\ \noalign{\vspace{4pt}}
								  \overline{\boldsymbol{\mathcal{R}}}_\mathrm{c} \\ \noalign{\vspace{4pt}}
								  \overline{\mathcal{R}}(\phi) \end{Bmatrix}
\end{align}
where $\boldsymbol{K}_{\Omega}$ is the stiffness matrix of the momentum equation consisting of transient, convection, viscous and Petrov-Galerkin stabilization terms, $\boldsymbol{G}_{\Omega}$ is the gradient operator, $\boldsymbol{G}^T_{\Omega}$ is the divergence operator for the continuity equation and $\boldsymbol{C}_{\Omega}$ is the stabilization term for cross-coupling of pressure terms. On the other hand, $\boldsymbol{D}_{\Omega}$ consists of the terms in the momentum equation which depend on the phase-indicator $\phi$, $\boldsymbol{G}_{AC}$ is the velocity coupled term in the Allen-Cahn equation and $\boldsymbol{K}_{AC}$ is the left-hand side stiffness matrix for the Allen-Cahn equation comprising of transient, convection, diffusion, reaction and positivity preserving stabilization terms. Here, $\overline{\boldsymbol{\mathcal{R}}}_\mathrm{m}$, $\overline{\boldsymbol{\mathcal{R}}}_\mathrm{c}$ and $\overline{\mathcal{R}}(\phi)$ represent the weighted residuals of the variational forms in Eqs.~(\ref{PG_NS}-\ref{PPV_AC}).

The two-phase flow system in Eq.~(\ref{TPFS_LM}) is decoupled into two subsystems: Navier-Stokes and Allen-Cahn solves, for which the linear system of equations can be summarized as:
\begin{align} \label{LS_NS}
\begin{bmatrix}
\boldsymbol{K}_{\Omega}& & \boldsymbol{G}_{\Omega}\\ \noalign{\vspace{4pt}}
-\boldsymbol{G}^T_{\Omega}& &\boldsymbol{C}_{\Omega}
\end{bmatrix}
\begin{Bmatrix}
\Delta\boldsymbol{u}\\ \noalign{\vspace{4pt}}
\Delta p
\end{Bmatrix}
&=
\begin{Bmatrix}
\overline{\boldsymbol{\mathcal{R}}}_\mathrm{m} \\ \noalign{\vspace{4pt}}
\overline{\boldsymbol{\mathcal{R}}}_\mathrm{c}
\end{Bmatrix}\\
\begin{bmatrix}
\boldsymbol{K}_{AC}
\end{bmatrix}
\begin{Bmatrix}
\Delta\phi
\end{Bmatrix}
&=
\begin{Bmatrix}
\overline{\mathcal{R}}(\phi)
\end{Bmatrix} \label{LS_AC}
\end{align}
Note that the cross-coupling terms between the Navier-Stokes and the Allen-Cahn equations ($\boldsymbol{D}_{\Omega}$ and $\boldsymbol{G}_{AC}$) are not present in the decoupled form. We solve the decoupled system in a partitioned-block iterative manner which leads to flexibility and ease in its implementation to the existing variational solvers. The linear systems (Eqs.~(\ref{LS_NS}) and (\ref{LS_AC})) are solved by the Generalized Minimal Residual (GMRES) algorithm proposed by \cite{saad1986gmres}. The algorithm relies on Krylov subspace iteration and modified Gram-Schmidt orthogonalization. Instead of construction of the left-hand side matrices explicitly, we only construct the required matrix-vector products of each block matrix in the GMRES solver.   Detailed algorithmic steps for the partitioned coupling of the fully implicit solutions of the conservative Allen-Cahn and the incompressible Navier-Stokes equations can be found in \cite{joshi2018positivity}. The stability and robustness of the partitioned decoupled system has been demonstrated for a broad range of problems involving high-density and viscosity ratios, high Reynolds number and complex topological changes over unstructured meshes \cite{joshi2018positivity,joshi2019hybrid}.

\section{Interface convection problem}

In this section, we first verify the convergence and accuracy of our fully-implicit finite element solver by simulating a bistable steady convection-diffusion-reaction system in a one-dimensional stretching flow. We then turn our attention to the convection of a planar interface and a curved interface in prescribed velocity fields. The convective distortion of the diffuse interface is quantified by the relative interface thickness and surface tension force errors. The dependence of the errors on the convective distortion parameter $\xi$ is assessed systematically. 

\subsection{Verification of steady Allen-Cahn phase-field solver}
\begin{figure}[h]
	\begin{minipage}[b]{0.5\textwidth}
		\captionsetup{skip=1pt}
		\centering
		\includegraphics[scale=0.57]{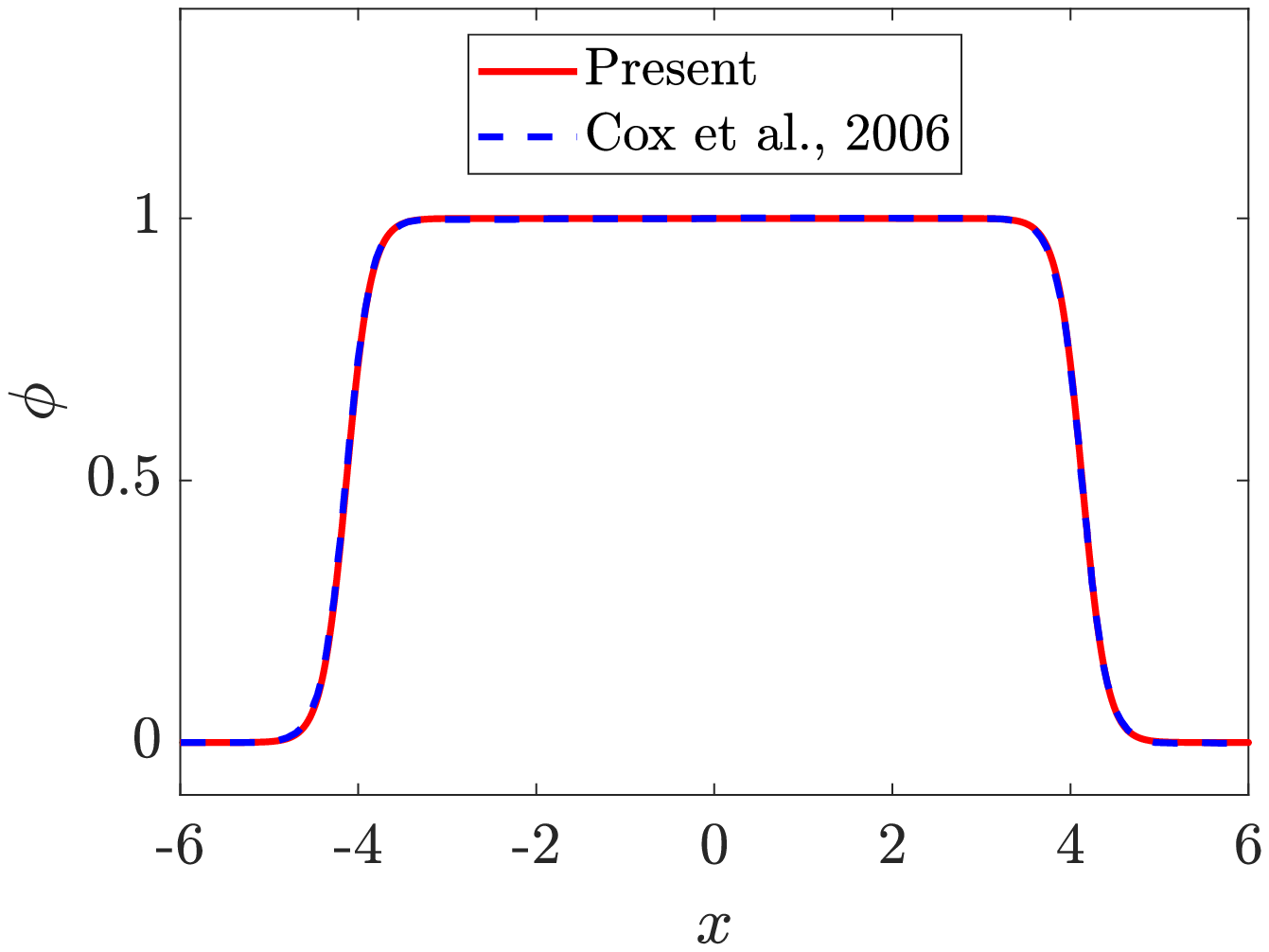}
		\caption*{\hspace{27pt}(a)}
	\end{minipage}
	\hspace{0pt}
	\begin{minipage}[b]{0.5\textwidth}
		\captionsetup{skip=1pt}
		\centering
		\includegraphics[scale=0.57]{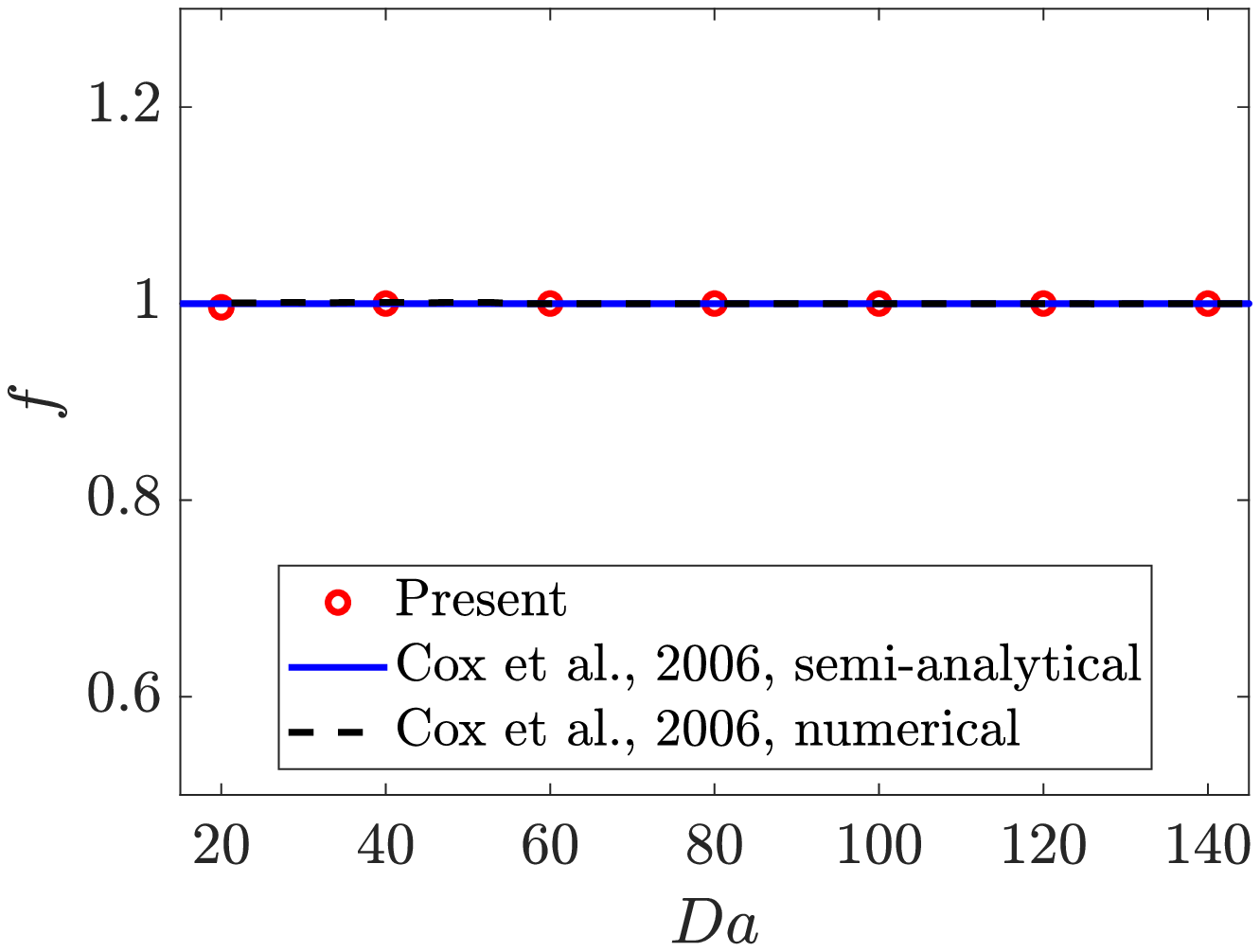}
		\caption*{\hspace{27pt}(b)}
	\end{minipage}
	
	\begin{minipage}[b]{0.5\textwidth}
		\captionsetup{skip=1pt}
		\centering
		\includegraphics[scale=0.57]{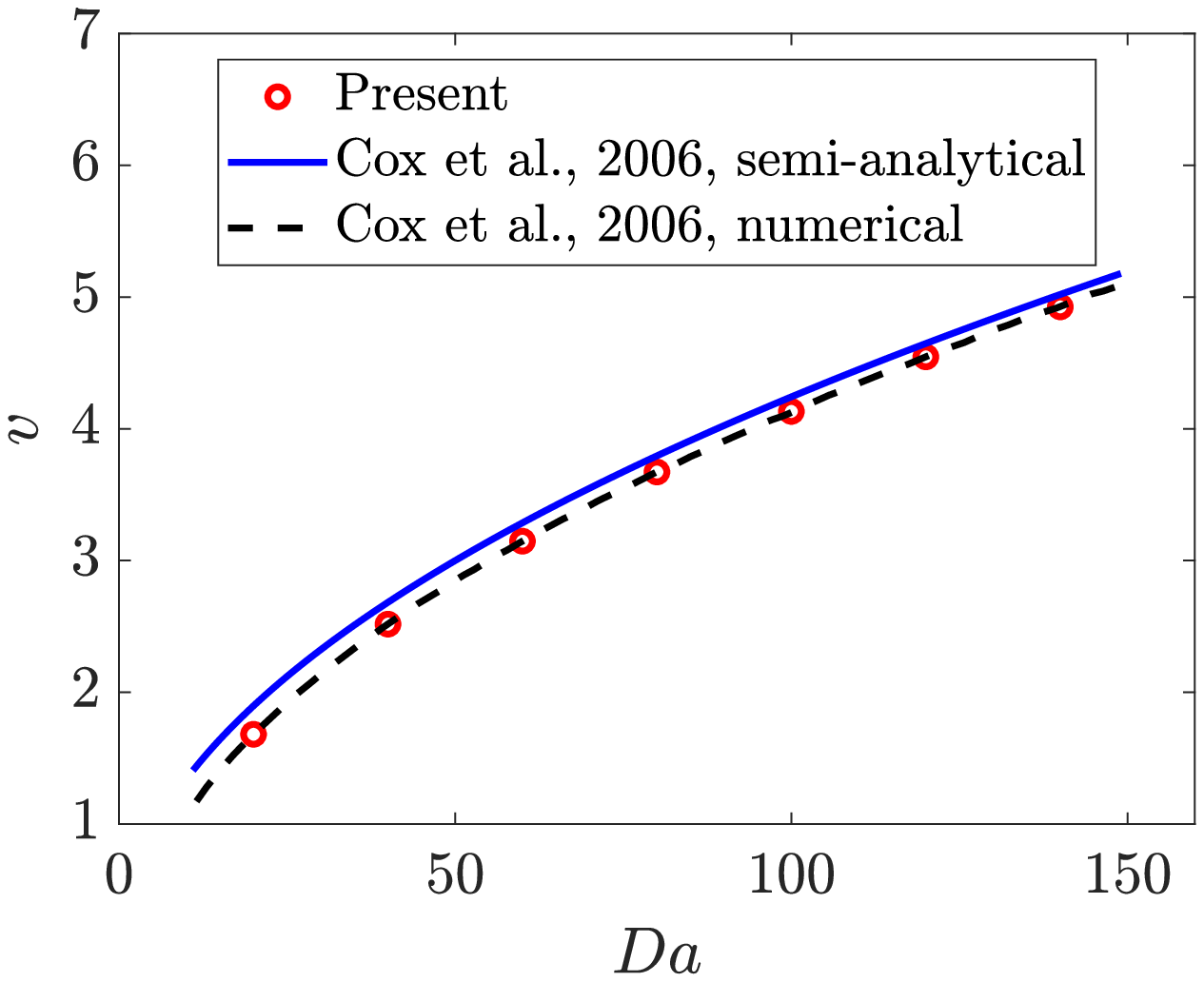}
		\caption*{\hspace{23pt}(c)}
	\end{minipage}
	\hspace{0pt}
	\begin{minipage}[b]{0.5\textwidth}
		\captionsetup{skip=1pt}
		\centering
		\includegraphics[scale=0.57]{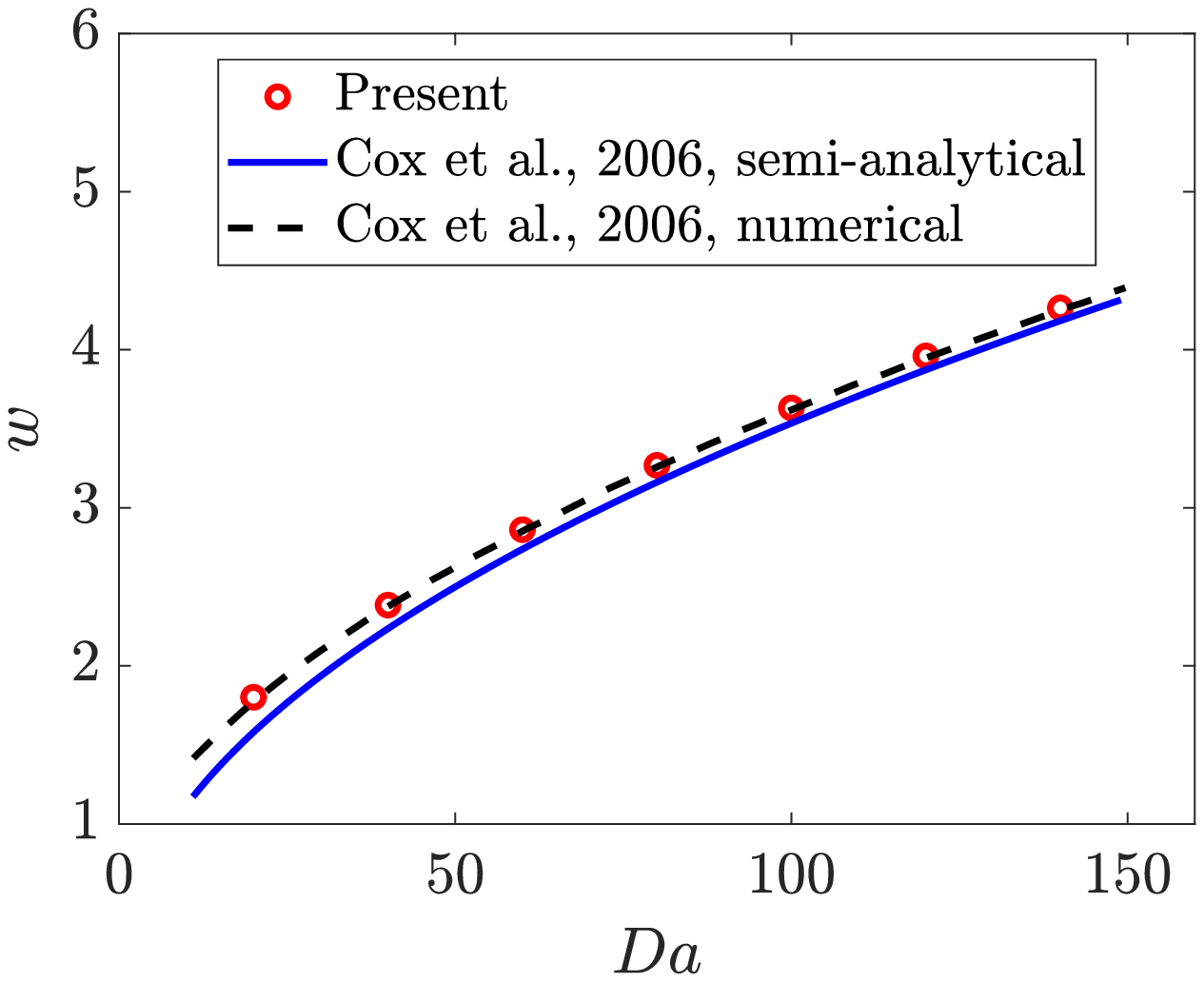}
		\caption*{\hspace{23pt}(d)}
	\end{minipage}
	
	\captionsetup{skip=1pt}
	
	\caption{Accuracy assessment of fully-implicit finite element formulation for a generic bistable convection-diffusion-reaction system:
   (a) steady state plateau-like solution as a function of distance $x$, and (b) the height of the solution $f$, (c) half-width of the solution $v$, (d) inverse width of the diffuse interface $w$ as a function of Damk\"ohler number $Da$.} 
	\label{bivali}
\end{figure}

For simplification, we consider 1D steady-state Allen-Cahn solution of the interface profile Eq.~(\ref{profile governing equation}) for constant convective distortion parameter:
\begin{equation}\label{1D governing equation}
\xi\tilde{x}\frac{d\phi}{d \tilde{x}}=-(\phi^3-\phi-\frac{d^2 \phi}{d\tilde{x}^2}),
\end{equation}
where $\tilde{x}=x/\varepsilon$ is the non-dimensionalized coordinate. The above interface profile equation for a phase-field function can be considered as a special case of a generic bistable steady convection-diffusion-reaction system in a stretching flow, which can be written as:
\begin{equation}\label{bistable system}
-S x\frac{d\phi}{d x}=D\frac{d^2\phi}{d x^2}+R\phi(\phi-1)(A-\phi),
\end{equation}
where $S$ is the stretching rate, $D$ is the diffusion coefficient, $R$ is the reaction rate, and $A$ is a parameter determines the unstable equilibrium phase separating the two stable equilibrium phases with minimum bulk energy. In \cite{cox2006bistable},  Eq.~(\ref{bistable system}) has been solved semi-analytically in the large Damk\"ohler number limit defined as $Da=R/S\gg 1$ with a fixed stretching rate $S=1$. The solutions of Eq.~(\ref{bistable system}) include the plateau-like solution and the pulse-like solution. Since the plateau-like solution can be considered as a solution composed of two diffuse interfaces subjected to the convective distortion (as shown in Fig.~\ref{bivali} (a)), it is used for the verification of our implicit finite element Allen-Cahn solver. The plateau-like solution takes the form:
\begin{equation} \nonumber
\phi(x)=\frac{1}{2}\ f \left[ \tanh\left(w(x+v)\right)-\tanh\left(w(x-v)\right)\right],
\end{equation}
where $f$ represents the height of the ``plateau'', $w$ is referred to as the inverse width of the diffuse interface which is inversely proportional to the width of the diffuse interface, and $v$ is the half-width of the ``plateau''. They are given semi-analytically by:
\begin{equation}\label{semi-ana}
f\sim 1,\ w\sim\sqrt{\frac{Da}{8D}},\ v\sim\sqrt{2DDa}(0.5-A).
\end{equation}

 To form a discrete system which is consistent with Eq.~(\ref{bistable system}), we directly prescribe the velocity as $u=-Sx$ without solving the Navier-Stokes equations. The Lagrange parameter is set to be zero and the reaction term is adjusted accordingly. We consider a one-dimensional computational domain $x\in[-L,L]$. The computational domain is discretized by a uniform mesh of grid size $h$. A zero flux boundary condition is applied on the left and the right boundaries. The initial condition is specified as:
\begin{equation}\phi(x)=\frac{1}{2}\left(\tanh\left(\frac{1}{\sqrt{2}}\left(x+5\right)\right)-\tanh\left(\frac{1}{\sqrt{2}}\left(x-5\right)\right)\right).
\end{equation}
The stretching rate and the diffusion coefficient are taken as $S=1$ and $D=1$ respectively. The reaction rates $R\in[20,140]$ corresponding to $Da\in[20,140]$ are considered for 
the verification purpose. The unstable equilibrium phase is set to be $A=0.2$ and the time step is taken as $\Delta t=0.1$ for the marching of steady state solution.

\begin{figure}[h]
	
	\begin{minipage}[b]{0.5\textwidth}
		\centering
		\includegraphics[scale=0.57,trim= 0 0 0 0,clip]{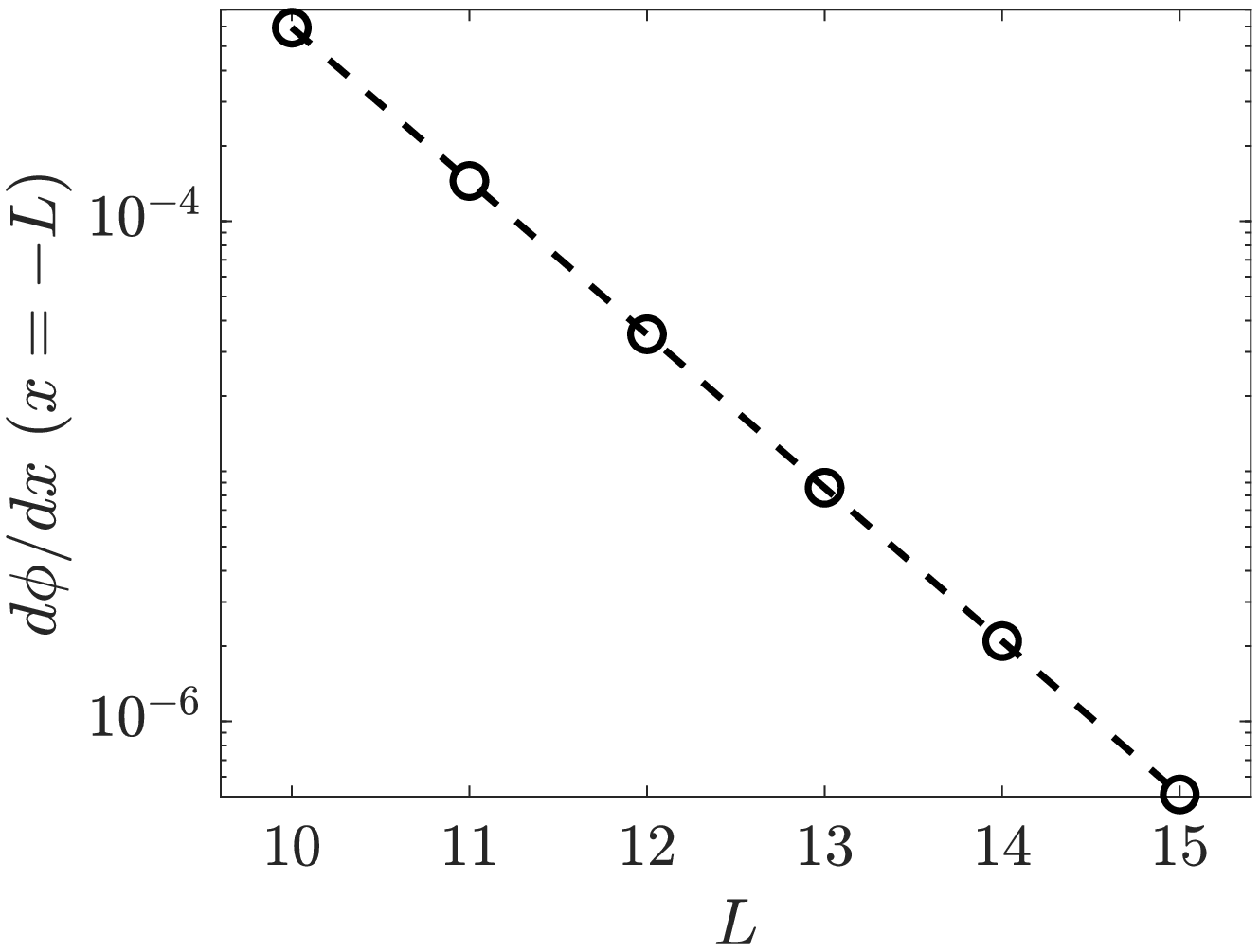}
		\caption*{\hspace{30pt}(a)}
	\end{minipage}
	\hspace{0pt}
	\begin{minipage}[b]{0.5\textwidth}
		\centering
		\includegraphics[scale=0.57,trim= 0 0 0 0,clip]{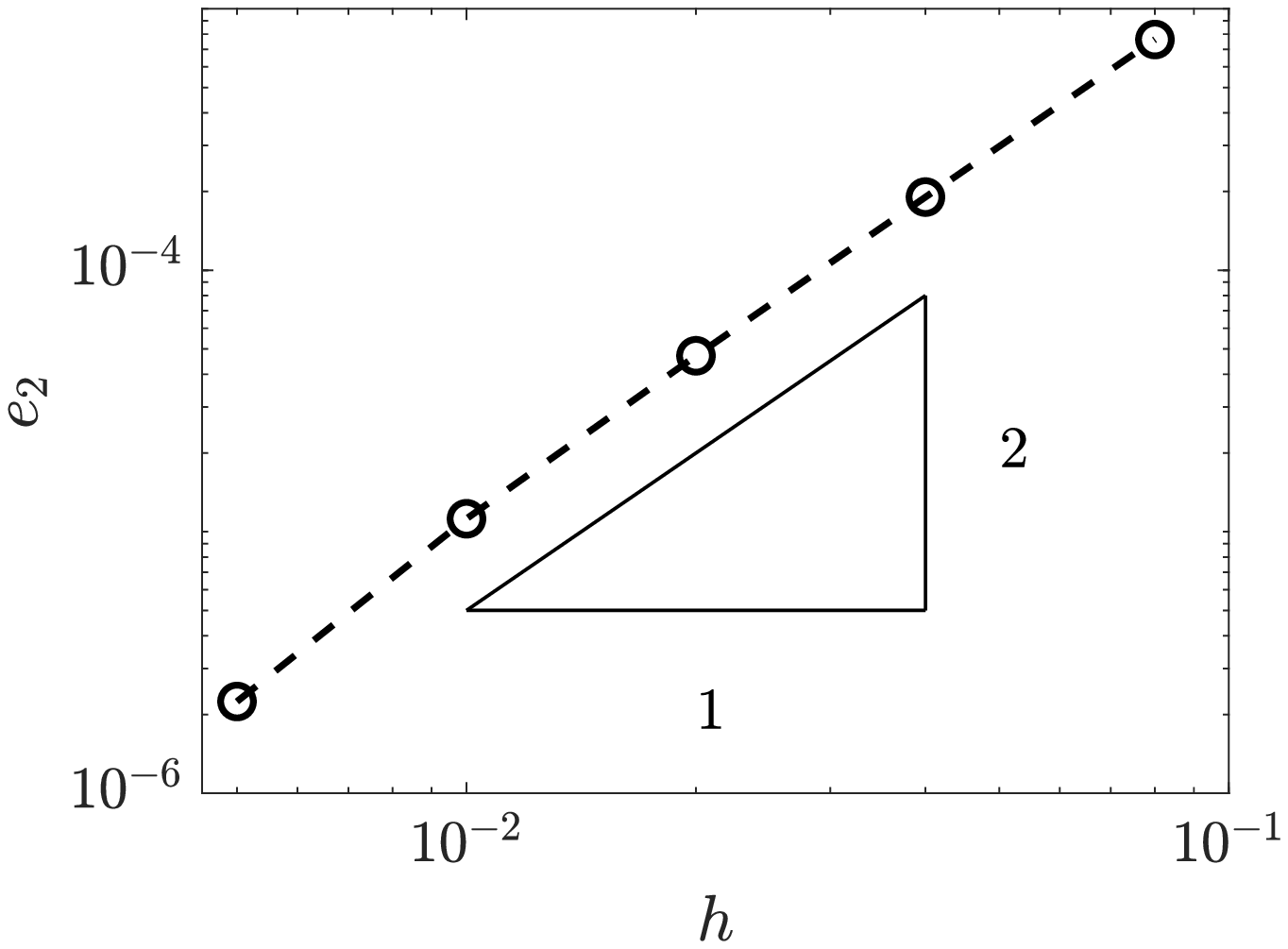}
		\caption*{\hspace{30pt}(b)}
	\end{minipage}
	\caption{Convergence study of phase-field solver for a generic bistable convection-diffusion-reaction system: (a) the variation of the derivative of $\phi$ with respect to domain length $L$, and (b) the relative $L^2$ error ($e_2$) as a function of grid size $h$. } 
	\label{bi}
\end{figure}

We perform a systematic convergence study for the final time when the steady state solution is reached, the length of the computational domain and the grid size. In the convergence study, we set the tolerance to  $10^{-5}$ for both the linear GMRES and the nonlinear Newton solvers. To quantify the error in the convergence study, we define the relative error in $L^2$ norm as
$e_2=||\phi-\phi_{\mathrm{ref}}||_2/||\phi_{\mathrm{ref}}||_2$, where $||\cdot||_2$ denotes the $L^2$ norm of the vector.  We first study the final time for the case $Da=140$, $h=0.01$ and $L=15$. The error level of the steady solution at $t=36$  reaches $e_2=1.3\times10^{-8}$ while considering the solution at $t=40$ as the reference. Hence we consider the solution at $t=40$ as the fully-converged steady state solution. We next investigate the domain length $L$ to ensure that the zero flux boundary condition is far enough so that its influence on the solution is negligible. The solution of $Da=140$, which has the wildest ``plateau'' according to Eq.~(\ref{semi-ana}), is analyzed at $t=40$ with $h=0.01$ and various $L$. The variation of the derivative of $\phi$ at the left boundary with respect to $L$ is shown in Fig.~\ref{bi} (a). When $L=15$, $d\phi/dx\ (x=-L)=5.1\times10^{-7}$. As a result, $L=15$ is considered as the converged domain length. We further investigate the convergence with respect to the grid size $h$. While keeping $t=40$ and $L=15$, $Da=20$ is selected for the mesh convergence study. Because the solution of $Da=20$ has the smallest width of the diffuse interface according to Eq.~(\ref{semi-ana}), which leads to the highest gradient of $\phi$ among all the cases and needs the finest mesh to resolve the gradient effects. By considering the solution at $h=0.0025$ as the reference,  the error $e_2$ is plotted in Fig.~\ref{bi} (b). It shows that our implementation is spatially second order accurate. The relative $L^2$ error reduces to $2.2\times10^{-6}$ when $h=0.005$, which is considered to be converged. To summarize, $t=40$, $L=15$ and $h=0.005$ are taken as the converged  parameters for the numerical simulation used in the verification. 

After establishing the convergence of our phase-field solver, we assess the accuracy against Eq.~(\ref{semi-ana}) and numerical results from \cite{cox2006bistable}. The plateau-like solution at $Da=100$ and the change of $h$, $v$, $w$ with respect to $Da$ are shown in Fig.~\ref{bivali} (a)-(d), respectively. The comparisons clearly show excellent agreements of our numerical results against previously reported analytical and numerical data. The accuracy of our fully implicit variationally discretized solver for the bistable steady convection-reaction-diffusion system in a stretching flow is successfully demonstrated.

\subsection{Convection of a planar interface}
Following the verification of our Allen-Cahn phase-field solver, we now turn our attention to the solution of Eq.~(\ref{1D governing equation}), which can be considered as another scenario of parameter specifications of Eq.~(\ref{bistable system}) describing the convective distortion of the diffuse interface. To form a constant $\xi$ used in Eq.~(\ref{1D governing equation}), the velocity field and the mobility coefficient are explicitly prescribed without solving the Navier-Stokes equations and the time-dependent mobility model.  

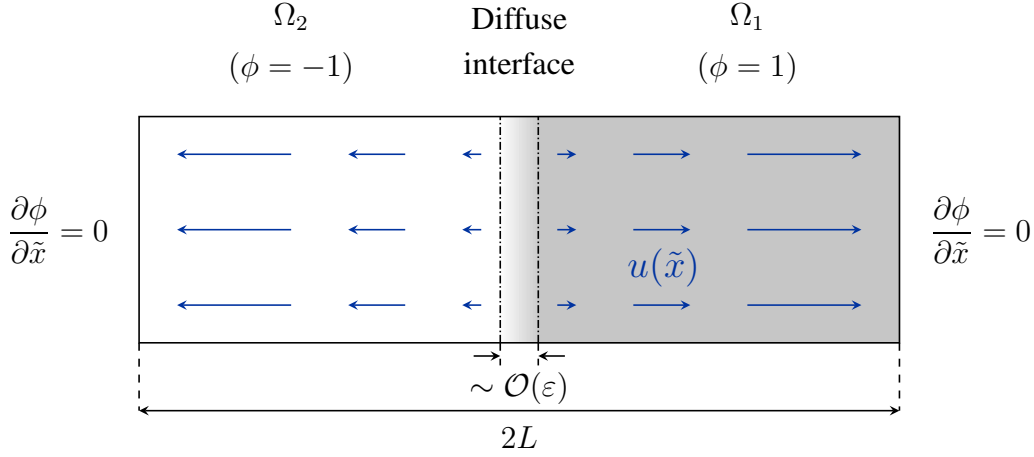
\begin{figure}[h]
	
	\centering
	\begin{tikzpicture}[scale=0.5,line width=0.6]
	\def\le{-0.5};
	\def\re{0.5};
	
	\draw[fill=white,draw=none] ( -10,-3) rectangle (\le,3);
	\shade[left color=white,right color=gray!45] ( \le,-3) rectangle (\re,3);
	\draw[fill=gray!45,draw=gray!45] ( \re,-3) rectangle (10,3);
	\draw (-10,-3) rectangle (10,3);
	\node[above] at (-6,5) {$\Omega_2$};
	\node[below] at (-6,5) {$(\phi=-1)$};
	\node[above] at (6,5) {$\Omega_1$};
	\node[below] at (6,5) {$(\phi=1)$};
	\node[above] at (0,5) {Diffuse};
	\node[below] at (0,5) {interface};
	\definecolor{mycolor}{RGB}{0,51,160}
	\foreach \x in {-6,-3,-1,1,3,6}{
		\foreach \y in {-2,0,2}{
			\draw [-stealth,semithick,mycolor] (\x,\y)--(1.5*\x,\y);
		}
	}
	
	\node[black] at (3.8,-1)[mycolor] {\large $u(\tilde{x})$};
	\draw[densely dashdotted] (\le,-3)--(\le,3);
	\draw[densely dashdotted] (\re,-3)--(\re,3);

	\node at (0,-4.2) {$\sim\mathcal{O}(\varepsilon)$};
	\draw  [dashed](-0.5,-3)--(-0.5,-3.6);
	\draw  [dashed](0.5,-3)--(0.5,-3.6);
	\draw [-stealth] (-1.2,-3.35)--(-0.5,-3.35);
	\draw [-stealth] (1.2,-3.35)--(0.5,-3.35);

	\draw[dashed] (-10,-3)--(-10,-4.8);
	\draw[dashed] (10,-3)--(10,-4.8);
	\draw[stealth-stealth](-10,-4.8)--(10,-4.8);
	\node at (0,-5.5) {$2L$};
	
	\node[left] at (-10.5,0) {$\displaystyle\frac{\partial \phi}{\partial \tilde{x}}=0$};
	\node[right] at (10.5,0) {$\displaystyle\frac{\partial \phi}{\partial \tilde{x}}=0$};
	\end{tikzpicture}
	
	\caption{Schematic diagram showing the computational domain for the convection of a one-dimensional planer interface in a prescribed extensional velocity field. $\Omega_1$ and $\Omega_2$ are domains of the two phases. A zero flux boundary condition for the order parameter is applied on the left and the right boundaries.}
	\label{1Dil}
\end{figure}

In the numerical simulation, the one-dimensional computational domain is taken as $\tilde{x}\in[-L,L]$, where $\tilde{x}=x/\varepsilon$ is the non-dimensional coordinate. The interface thickness parameter is set to be $\varepsilon=1$. The computational domain is discretized by uniform mesh of grid size $h=\Delta\tilde{x}$. A zero flux boundary condition is imposed for the order parameter $\phi$ on the left and the right boundaries. The planar interface is initialized as $\phi(\tilde{x})=\tanh(\tilde{x}/\sqrt{2})$. The mobility coefficient of the Allen-Cahn equation is chosen as $\gamma=1$. The velocity is prescribed as $u(\tilde{x})=a \tilde{x}$, where $a\in[-0.25,0.25]$ is a constant selected according to the desired convective distortion parameter. From the problem setup, the convective distortion parameter can be calculated as $\xi=a$. The time step is taken as $\Delta t=0.1$. 
The problem setup with the illustration of an extensional velocity field is shown in Fig.~\ref{1Dil}.

\begin{figure}[h]
	
	\begin{minipage}[b]{0.5\textwidth}
		\centering
		\includegraphics[scale=0.57,trim= 0 0 0 0,clip]{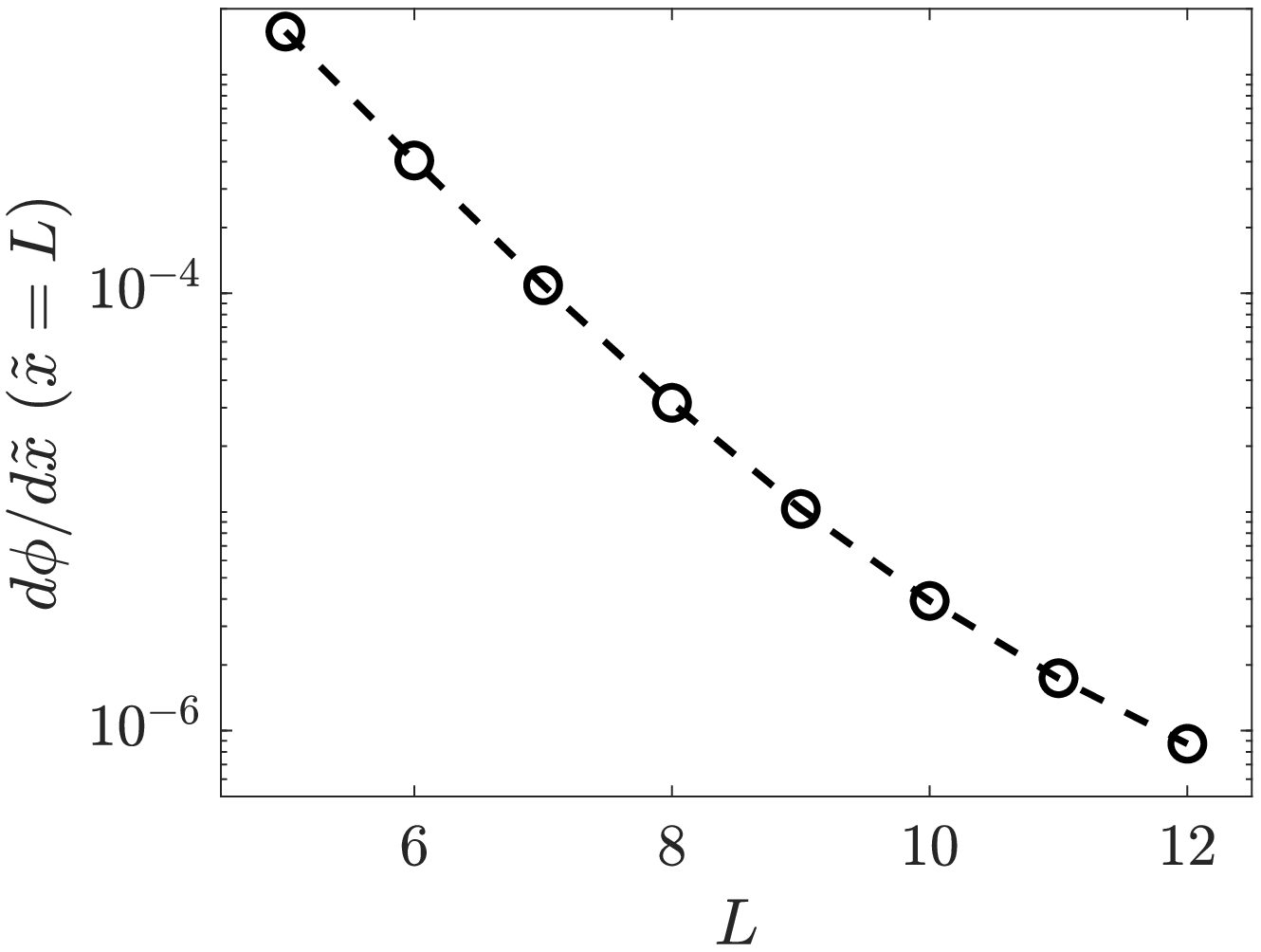}
		\caption*{\hspace{30pt}(a)}
	\end{minipage}
	\hspace{0pt}
	\begin{minipage}[b]{0.5\textwidth}
		\centering
		\includegraphics[scale=0.57,trim= 0 0 0 0,clip]{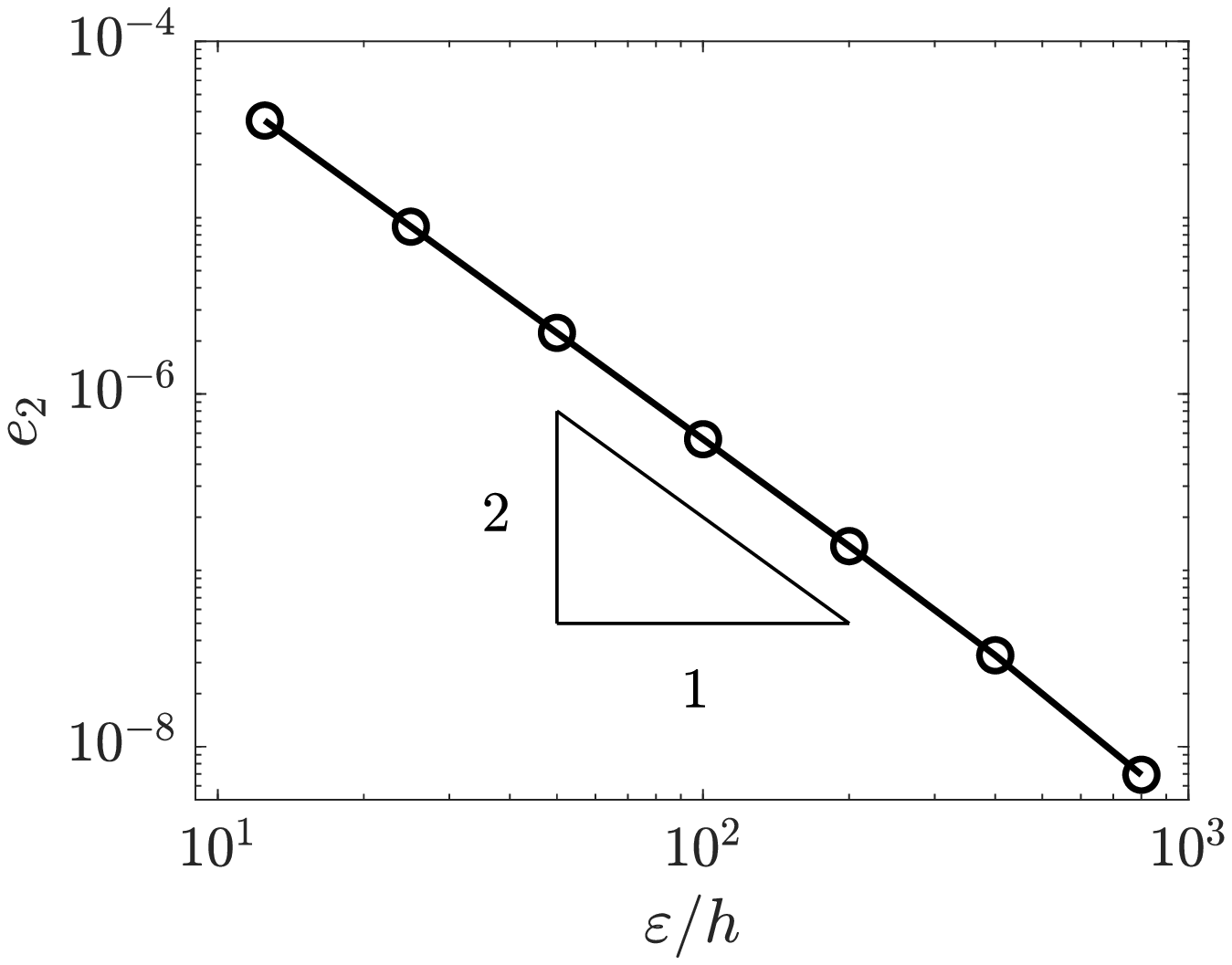}
		\caption*{\hspace{30pt}(b)}
	\end{minipage}
	\caption{Convergence of a convecting planar diffuse interface with a prescribed velocity: (a) the variation of the derivative of $\phi$ with respect to domain length $L$, and (b) the relative $L^2$ error ($e_2$) as a function of mesh resolution $\varepsilon/h$.} 
	\label{xicon}
\end{figure}

We carry out a convergence study to minimize the discretization error so that the effect of $\xi$ can be accurately demonstrated. To ensure that the discretization error is negligible, the convergence of the final time when the steady-state is achieved, the length of the computational domain and the mesh resolution at the diffuse interface are studied with the tolerance of $10^{-8}$ for the linear GMRES and the nonlinear Newton solvers.
The final time is examined for the values of $\xi=-0.25$ and $\xi=0.25$, at which the steady state solution deviates the most from the initial condition among all the cases. The grid size $h=0.01$ and the size of the computational domain $L=10$ are used. By considering $t=20$ as the reference, we check the relative $L^2$ norm of the solution at $t=18$. For $\xi=-0.25$ and $\xi=0.25$, the errors reduce to $e_2=3.4\times10^{-8}$ and $e_2=7.6\times10^{-11}$, respectively. Hence the solution at $t=20$ is considered to be fully-converged.
We proceed to the investigation of $L$ to check whether the zero flux boundary condition is far enough so that its influence on the solution is negligible. The solution at $\xi=0.25$, which leads to the maximum extensional distortion and the widest diffuse interface among all the cases, is analyzed at $t=20$ with $h=0.01$. The derivative of the order parameter on the right boundary $d\phi/d\tilde{x}\ (\tilde{x}=L)$ is plotted as a function of $L$ in Fig.~\ref{xicon} (a). When $L=12$, the derivative reduces to $d\phi/ dx\ (\tilde{x}=L)=8.7\times10^{-7}$, which indicates an error of $\phi$ at the order of $10^{-9}$ (with $h=0.01$). Therefore it is considered as the converged domain length.
The convergence of the mesh resolution in the diffuse interface region is studied at $t=20$ with $L=12$. The solution at $\xi=-0.25$ is investigated, which results in the maximum compressional distortion and the highest gradient among all the cases requiring the finest mesh to resolve. By taking the solution at $\varepsilon/h=1600$ as the reference, the relative $L^2$ error as a function of the mesh resolution $\varepsilon/h$ is plotted in Fig.~\ref{xicon} (b). The plot confirms the second-order spatial accuracy of our implementation. When $\varepsilon/h=800$, the error reduces to $e_2=6.9\times10^{-9}$. The resolution is deemed to be converged. To summarize, $t=20$, $L=12$ and $\varepsilon/h=800$ are employed to minimize the numerical error in the investigation of the effect of $\xi$.  

\begin{figure}[h]
	\centering
	\includegraphics[scale=0.55]{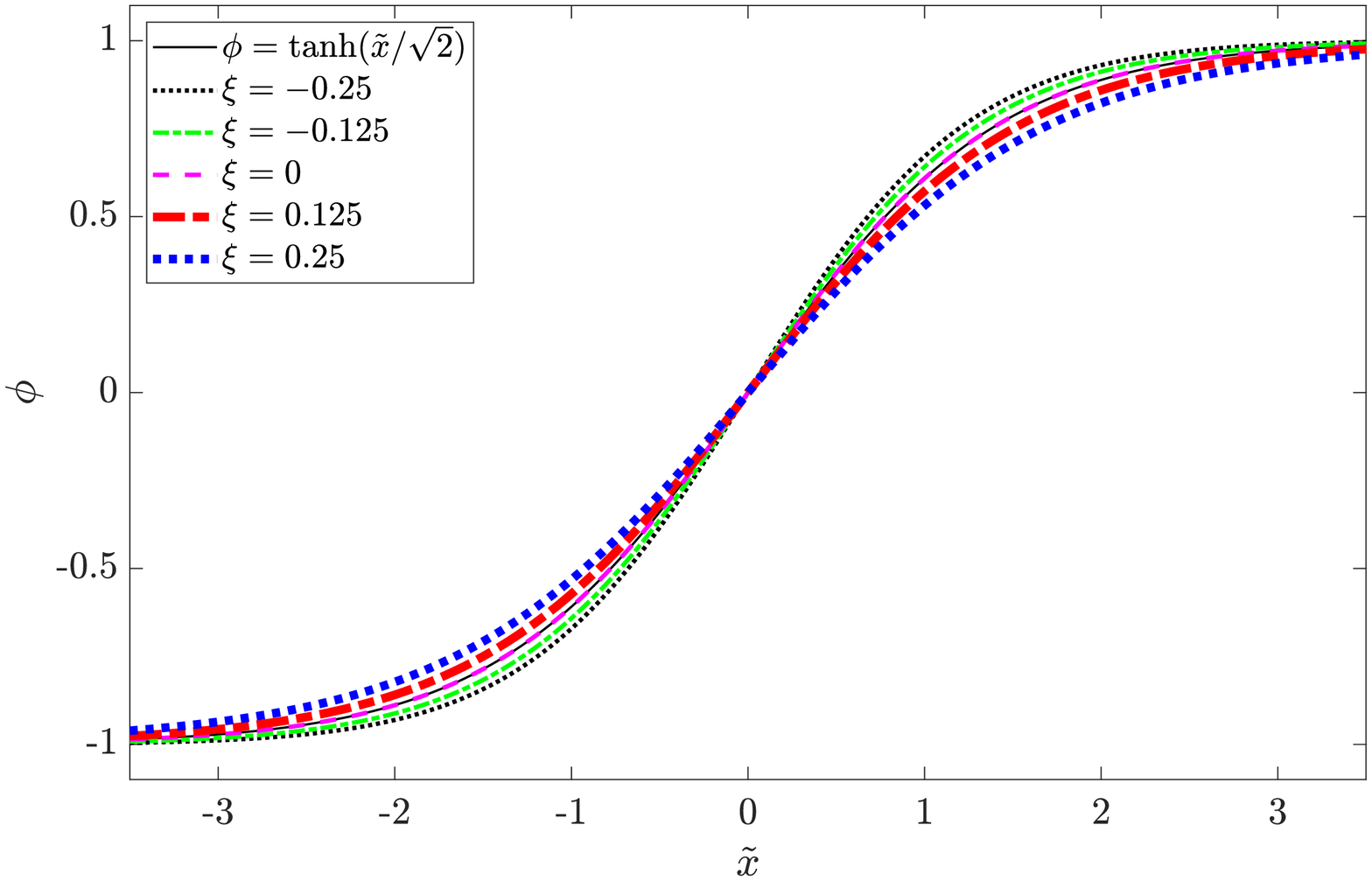}
	\caption{Convection of a planer interface in a prescribed velocity field: the interface profiles corresponding to various distortion parameter $\xi$.}
	\label{xi}
\end{figure}

 With the converged numerical parameters, we first consider the effect of the convective distortion parameter on the interface profile. As shown in Fig.~\ref{xi}, when $\xi=0$, the interface profile from the numerical simulation tends to the hyperbolic tangent profile. When $\xi>0$, an extensional distortion is observed. The extensional distortion increases with an increase in $\xi$. Compressional distortion is noted when $\xi<0$ whereby the compressional distortion increases with the decrease in $\xi$. To quantify the deviation of the interface profile from the hyperbolic tangent profile, we define the relative interface thickness error as:
\begin{equation}\label{thkerror}
    e_{\varepsilon}=\left|\frac{\tilde{\varepsilon}_{d}-\tilde{\varepsilon}_{\mathrm{eq}}}{\tilde{\varepsilon}_{\mathrm{eq}}}\right|,
\end{equation}
where $\tilde{\varepsilon}_d$ and $\tilde{\varepsilon}_{\mathrm{eq}}$ denote the non-dimensionalized distance with respect to $\varepsilon$ from $\phi=-0.9$ to $\phi=0.9$ of the distorted interface, and of the hyperbolic tangent profile respectively. The coordinates of $\phi=0.9$ and $\phi=-0.9$ are linearly interpolated from the numerical solution.The relative interface thickness error as a function of $\xi$ is shown in Fig.~\ref{m1} (a). 

Furthermore, we examine the surface tension force calculation error due to the convective distortion. In the CSF model, the singular surface tension force at the interface is distributed to the diffuse interface region by a Dirac delta function at the interface. In the current model, the function is given by
$\delta_S=\alpha_{\mathrm{sf}}\varepsilon|\nabla\phi|^2$, which should satisfy:
\begin{equation}\label{surftens}
    \int_{-\infty}^{\infty} \alpha_{\mathrm{sf}}\varepsilon\left(\frac{\partial\phi}{\partial n}\right)^2dn=1,
\end{equation}
where $\alpha_{\mathrm{sf}}$ is a constant parameter. For the hyperbolic tangent profile $\phi=\tanh(n/\sqrt{2}\varepsilon)$, the constant parameter can be calculated as $\alpha_{\mathrm{sf}}=3\sqrt{2}/4$.
When the interface profile is affected by the convective distortion, its deviation from the hyperbolic tangent profile leads to the violation of Eq.~(\ref{surftens}). This gives rise to a relative surface tension force error quantified as:
\begin{equation}\label{surftenserror}
    e_{\sigma}=\left|\int_{-\infty}^{\infty}\alpha_{\mathrm{sf}}\left(\frac{\partial\phi}{\partial \tilde{n}}\right)^2d\tilde{n}-1\right|.
\end{equation}
The error is evaluated numerically in the current study. The derivative is calculated by the central difference method at the interior points and the forward or backward difference technique at the boundary points of the computational domain. The integral is calculated by the standard trapezoidal rule. The surface tension force error as a function of $\xi$ is plotted in Fig.~\ref{m1} (b).

\subsection{Convection of a curved interface}
\begin{figure}[h]
 	\centering 
 	\begin{tikzpicture}[xscale=7,yscale=7,line width=0.6]
 	\draw[fill=gray!20]
 	(0:0.5) arc(0:90:0.5) --(90:0.8) arc(90:0:0.8)--cycle;

 	\begin{scope}
 	\clip (0:0.2) arc(0:90:0.2) --(90:0.8) arc(90:0:0.8)--cycle;
 	
 	\foreach \cn in {20,17.5,...,0}
 	{
 		\draw[gray!\cn,fill=gray!\cn](0,0) circle [radius=0.49+0.001*\cn];
 	}
 	\end{scope}
 	\draw
 	(90:0.51)--(90:0.2) arc(90:0:0.2) --(0:0.51);
 	
 	\draw [dashed] (-0.2,0)--(0.2,0);
 	\draw [stealth-stealth] (0,0) -- (0,0.2);
 	\node  [left] at (-0,0.1) {$R_1$};
 	\draw [stealth-stealth] (-0.1,0) -- (-0.1,0.5);
 	\draw [dashed] (0,0.5)--(-0.1,0.5);
 	\node  [left] at (-0.1,0.25) {$R_2$};
 	\draw [stealth-stealth] (-0.2,0) -- (-0.2,0.8);
 	\draw [dashed] (0,0.8)--(-0.2,0.8);
 	\node  [left]at (-0.2,0.4) {$R_3$};
 	\def\cl{black}
 	\begin{scope}
 	\clip (0:0.2) arc(0:90:0.2) --(90:0.8) arc(90:0:0.8)--cycle;
 	
 	\draw[-Latex,domain=0.01:0.075,smooth,variable=\x,\cl,line width=0.5] plot ({\x},{0.03/\x});
 	\draw[-Latex,domain=0.075:0.43,smooth,variable=\x,\cl,line width=0.5] plot ({\x},{0.03/\x});
 	\draw[domain=0.43:0.8,smooth,variable=\x,\cl,line width=0.5] plot ({\x},{0.03/\x});
 	

 	\draw[-Latex,domain=0.01:0.29,smooth,variable=\x,\cl,line width=0.5] plot ({\x},{0.14/\x});
 	\draw[-Latex,domain=0.28:0.54,smooth,variable=\x,\cl,line width=0.5] plot ({\x},{0.14/\x});
 	\draw[domain=0.53:0.8,smooth,variable=\x,\cl,line width=0.5] plot ({\x},{0.14/\x});
 	
 	\draw[-Latex,domain=0.01:0.45,smooth,variable=\x,\cl,line width=0.5] plot ({\x},{0.27/\x});
 	\draw[-Latex,domain=0.44:0.63,smooth,variable=\x,\cl,line width=0.5] plot ({\x},{0.27/\x});
 	\draw[domain=0.63:0.8,smooth,variable=\x,\cl,line width=0.5] plot ({\x},{0.27/\x});
 	\end{scope}
 	\node [above] at (0.469,0.459) {$\Omega_1$};
 	\node [below] at (0.469,0.459) {$(\phi=1)$};
 	\node [above] at (0.28,0.28) {$\Omega_2$};
 	\node [below] at (0.28,0.28) {$(\phi=-1)$};
 	\node [above,\cl] at (0.6,0.1) {\large$\boldsymbol{u}(x,y)$};
 	\draw [line width=0.5,black]
 	(0:0.2) arc(0:90:0.2) --(90:0.8) arc(90:0:0.8)--cycle;
 	
 	\draw [line width=0.5,dashed](0.53,0.6)--(0.6,0.63);
 	
 	\node at (0.7,0.63) {$ \displaystyle \frac{\partial \phi}{\partial n}=0$};
 	\end{tikzpicture}
 	\caption{Schematic diagram showing the computational domain for the convection of a curved interface in a prescribed incompressible velocity field illustrated with streamlines. $\Omega_1$ and $\Omega_2$ are domains of the two phases. A zero flux Neumann boundary conditions for the order parameter is applied on all the boundaries.}
 	\label{2Dil}
 \end{figure}
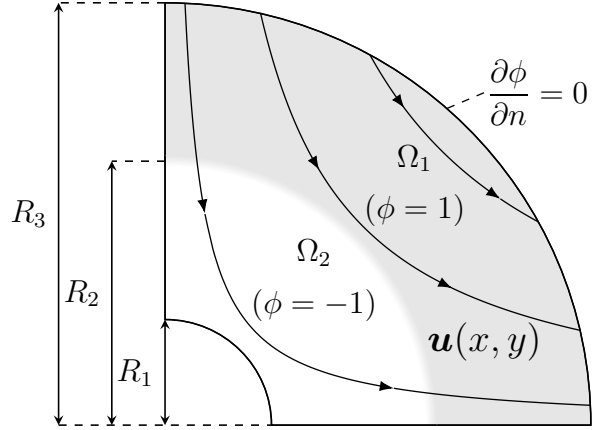

  To demonstrate the extensional and compressional distortion in an incompressible fluid flow, we consider the convection of a curved interface in a prescribed divergence-free 2D velocity field. The computational domain is in the shape of an arch with $r\times\theta\in [R_1,R_3]\times[0,\pi/2]$. A zero flux Neumann boundary condition is imposed for $\phi$ on all the boundaries. The curved interface is initialized as a circular arc centered at $r=0$ with radius $R_2$:
  \begin{align}
  \phi(x,y,0)=-\tanh\left(\frac{R_2-\sqrt{(x^2+y^2)}}{\sqrt{2}\varepsilon}\right).    
  \end{align}
   The curved interface is convected by a divergence-free velocity field $\boldsymbol{u}(x,y)$. The horizontal and vertical components of the velocity are given by $u(x,y)=x$ and $v(x,y)=-y$ respectively. The mobility coefficient in the range of $\gamma=b\in[4,20]$ are investigated, where $b$ is a constant parameter. From the problem setup, the convective distortion parameter can be calculated as $\xi(r,0)=1/b$ at the bottom boundary and $\xi(r,\pi/2)=-1/b$ at the left boundary. In the present cases, we use $R_1=0.1$, $R_2=0.25$ and $R_3=0.55$. A structured mesh of grid size $\Delta r =6.25\times10^{-5}$ and $\Delta \theta=\pi/2400$ is used for the spatial discretization. The interface thickness parameter is chosen as $\varepsilon=6.25\times10^{-4}$. The mesh resolution can be calculated as $\varepsilon/h=10$, which leads to a relative $L^2$ convergence error at the order of $10^{-5}$. The computational domain and the initial condition are illustrated in Fig.~\ref{2Dil}.  The time step size is taken as $\Delta t =1.25\times10^{-4}$. The numerical solutions at $t=0.75$ are analyzed, when the interface thickness at $\theta=0$ and $\theta=\pi/2$ reaches a constant value.

 The volume-preserving mean curvature flow is another source of error in the convection of a curved interface. To isolate the effect of the convective distortion from the effect of the volume-preserving mean curvature flow, the latter is minimized utilizing a small interface thickness parameter $\varepsilon$. From Eq.~(\ref{mc}), the maximum $\boldsymbol{v}(\boldsymbol{x},t)$ occurs at $\gamma=100$, $t=0.75$, which can be calculated as $\max\limits_{\boldsymbol{x}\in\Gamma^{\phi}_I}|\boldsymbol{v}(\boldsymbol{x},t)|=0.0013$. It is negligible compared to the convective velocity in the computational domain which is $\mathcal{O}(0.1)$. On the other hand, the minimum radius of curvature at $t=0.75$ can be calculated as $\min\limits_{\boldsymbol{x}\in\Gamma^{\phi}_I}R=0.028$. Since $\varepsilon \ll R$, the normal velocity gradient in the normal direction introduced by the volume-preserving mean curvature flow being $\mathcal{O}\left((\varepsilon/R)^2\right)$ (by taking the derivative of $\boldsymbol{v}(\boldsymbol{x},t)$ with respect to $R$) is negligible. Consequently, the effect of the volume-preserving mean curvature flow is negligible in these cases.
 
\begin{figure*}
  	\hspace{2cm}
  	\begin{minipage}[b]{0.8\textwidth}
  		\hspace{0cm}
  		\begin{minipage}[b]{0.4\textwidth}
  			\centering
  			\includegraphics[trim=1 1 1 1,clip,height=0.9\textwidth]{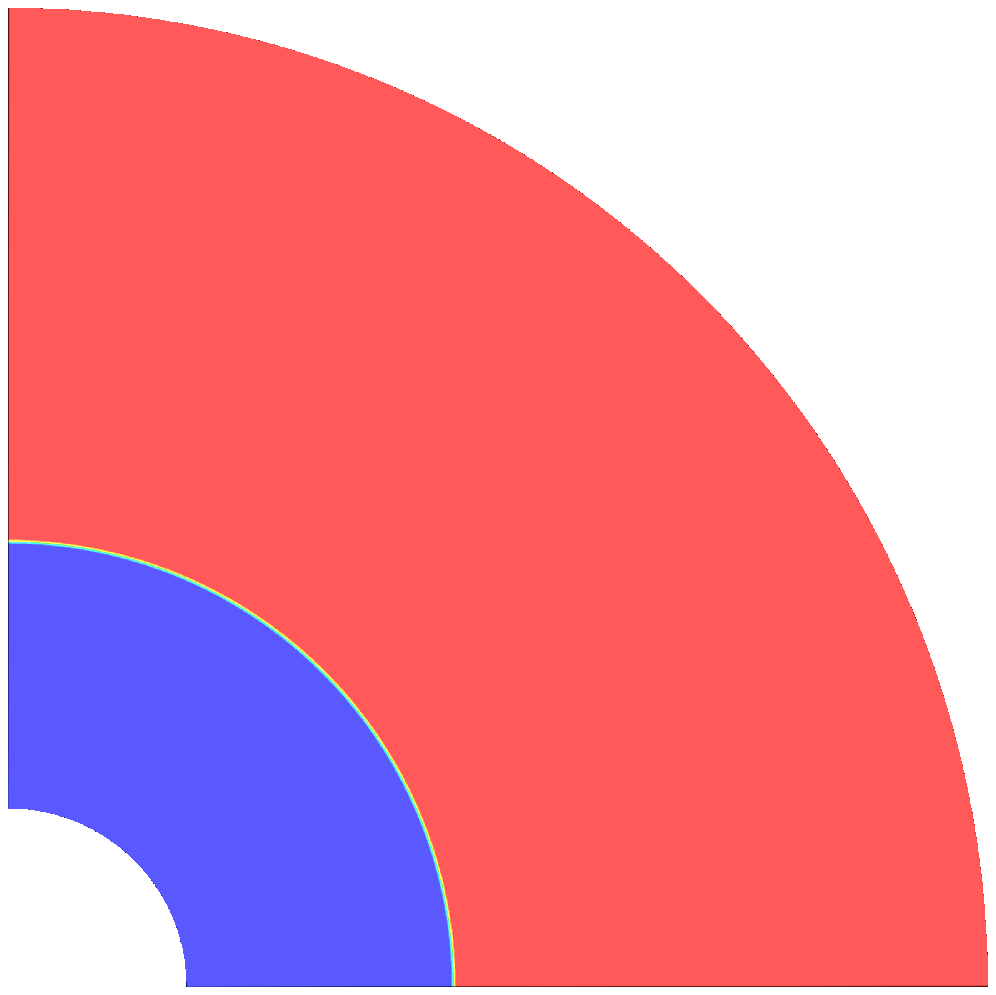}
  			\caption*{\hspace{30pt}(a)}
  		\end{minipage}
  		\hspace{-0.5cm}
  		\begin{minipage}[b]{0.4\textwidth}
  			\centering
  			\includegraphics[trim=1 1 1 1,clip,height=0.9\textwidth]{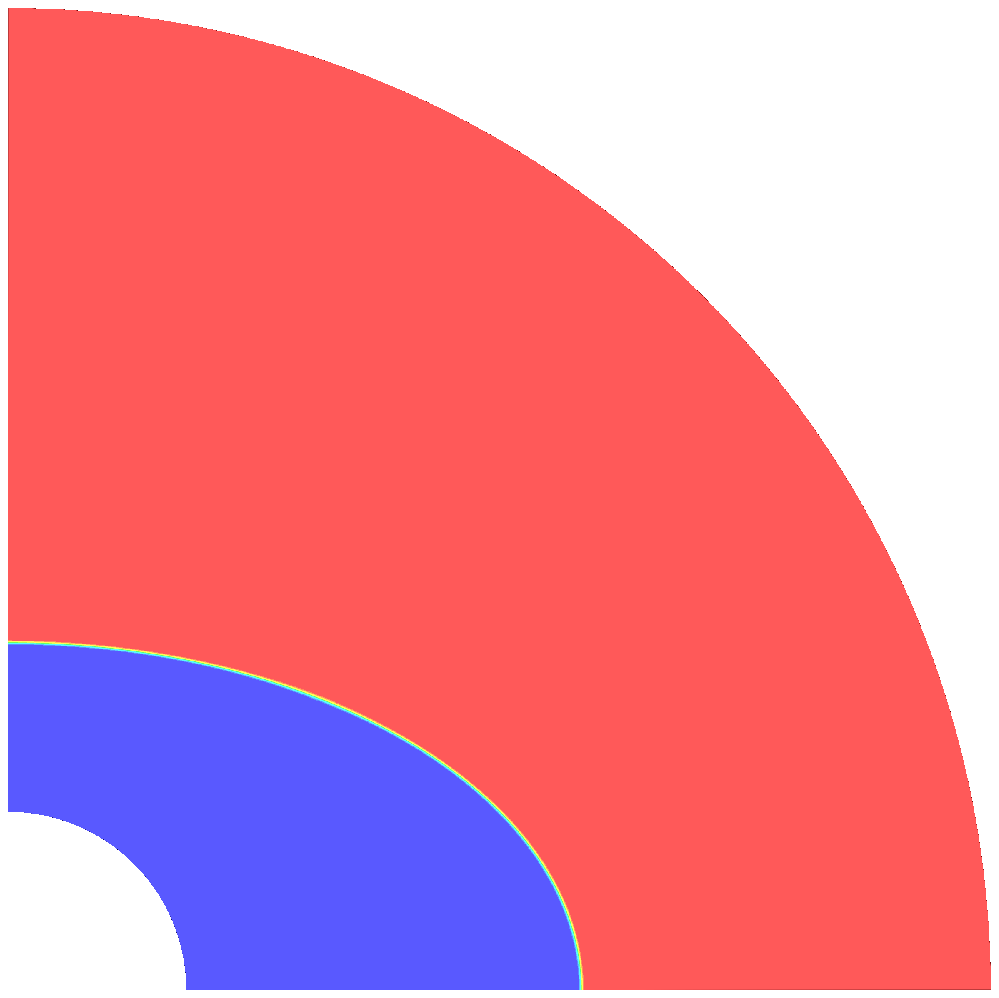}
  			\caption*{\hspace{30pt}(b)}
  		\end{minipage}
  		
  		\hspace{0cm}
  		\begin{minipage}[b]{0.4\textwidth}
  			\centering
  			\includegraphics[trim=1 1 1 1,clip,height=0.9\textwidth]{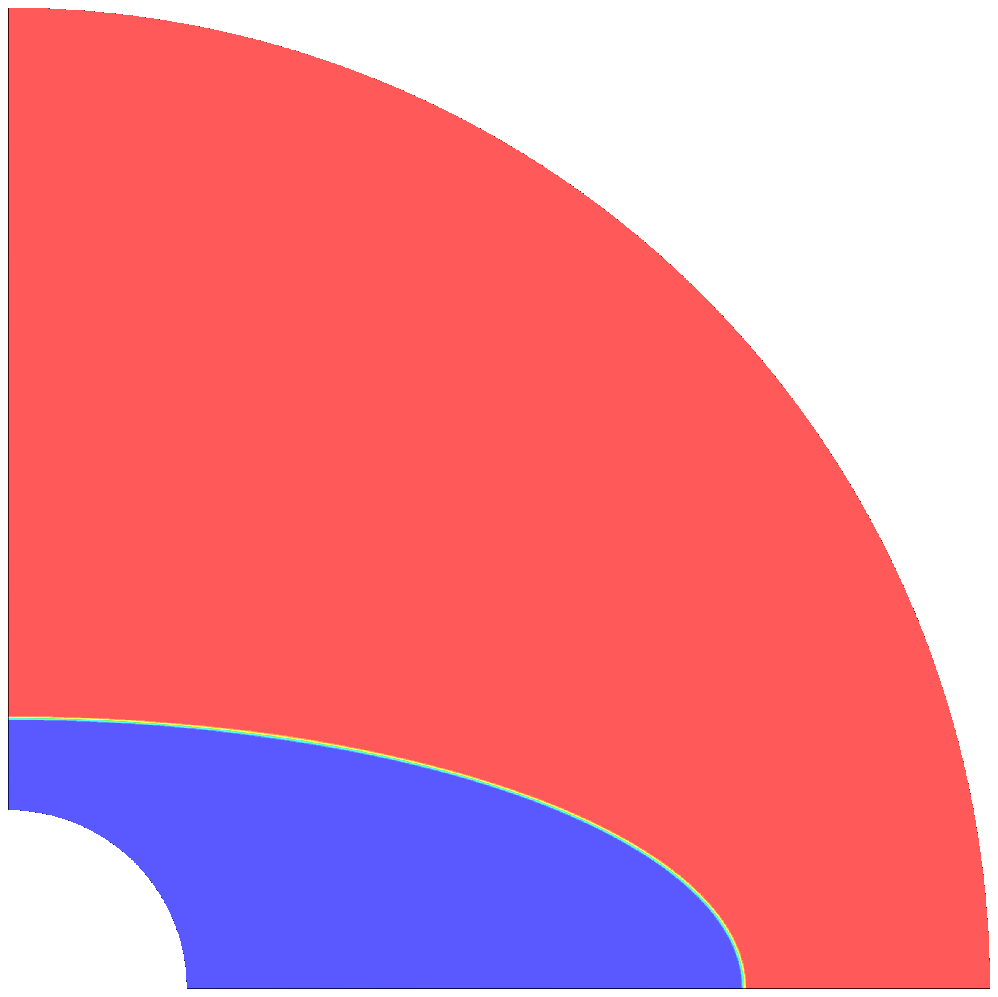}
  			\caption*{\hspace{30pt}(c)}
  		\end{minipage}
  		\hspace{-0.5cm}
  		\begin{minipage}[b]{0.4\textwidth}
  			\centering
  			\includegraphics[trim=1 1 1 1,clip,height=0.9\textwidth]{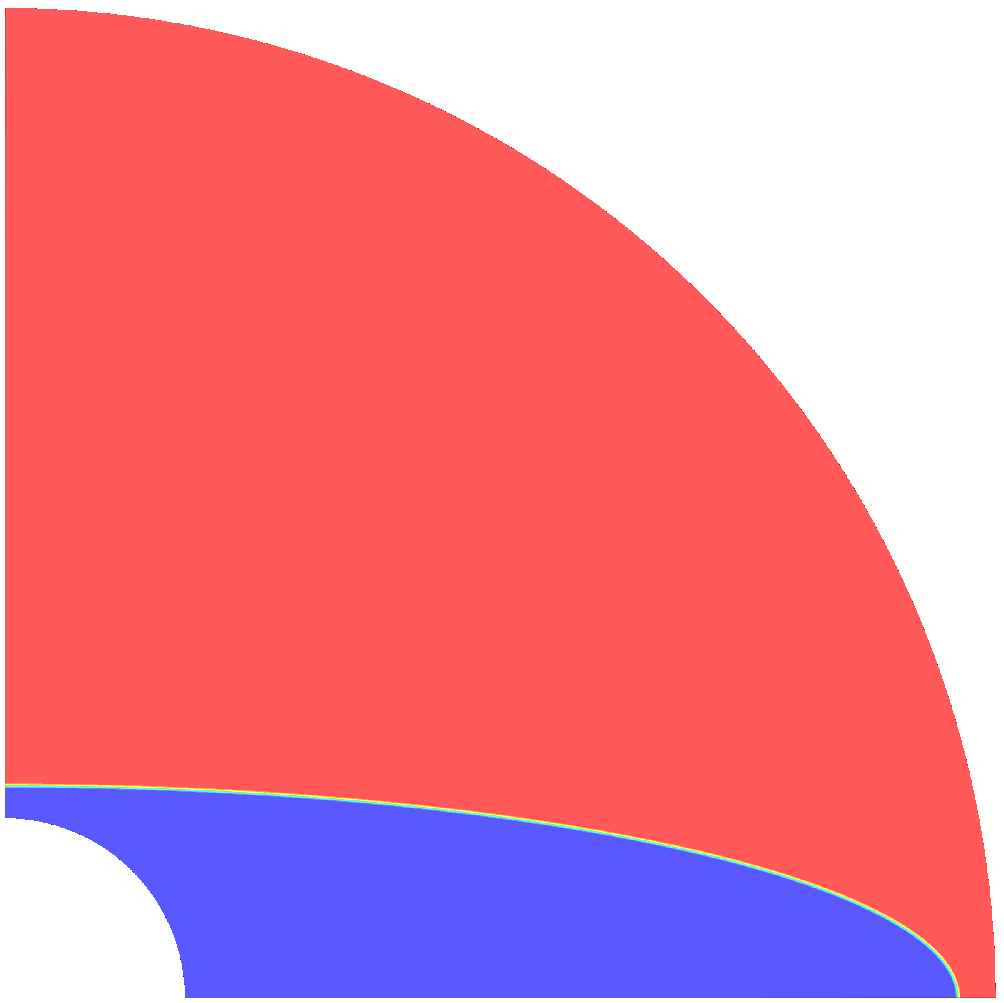}
  			\caption*{\hspace{30pt}(d)}
  		\end{minipage}
  	\end{minipage}
  	\hspace{-3cm}
  	\begin{minipage}[b]{0.1\textwidth}
  		\centering
  		\includegraphics[trim=100 25 100 0,clip,width=4cm,scale=0.42]{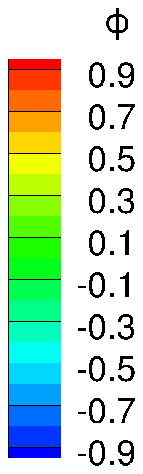}
  	\end{minipage}
  	
  	\begin{minipage}[b]{1\textwidth}
  		\centering
  		\includegraphics[trim=0 0 0 0,clip,scale=0.6]{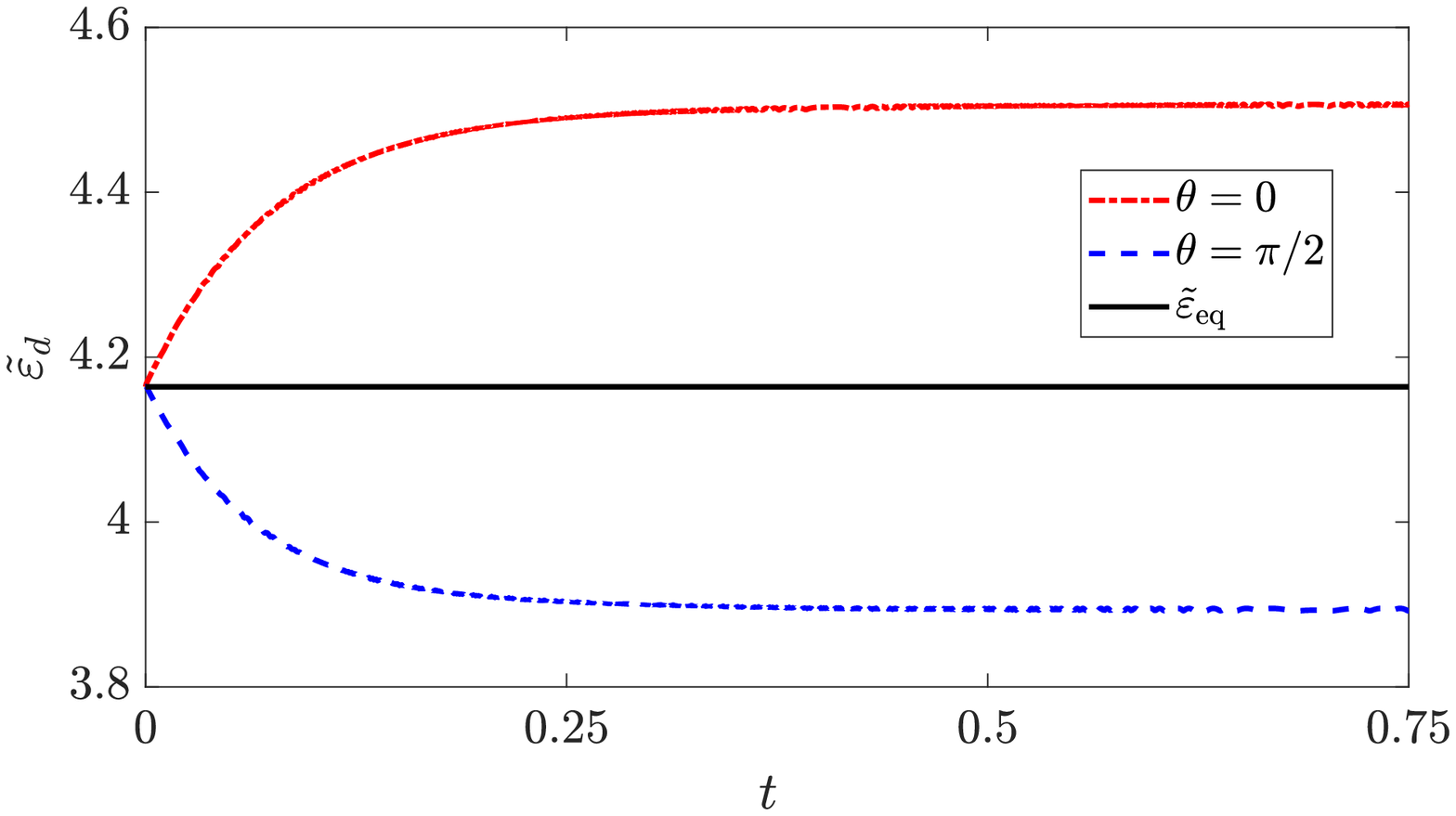}
  		\caption*{\hspace{10pt}(e)}
  	\end{minipage}
  	\caption{Convection of a curved interface with $\xi=0.1$: the contours of $\phi$ at (a) $t=0$, (b) $t=0.25$, (c) $t=0.5$, (d) $t=0.75$, and (e) the time history of the non-dimensional interface thickness $\tilde{\varepsilon}_d$ on the bottom boundary $\theta=0$ and the left boundary $\theta=\pi/2$ with the comparison to the equilibrium interface thickness $\tilde{\varepsilon}_{\mathrm{eq}}$.}
  	\label{ag}
  \end{figure*}

\begin{figure}[h]
    	
    	\begin{minipage}[b]{0.5\textwidth}
    		\centering
    		\includegraphics[scale=0.6,trim= 0 0 0 0,clip]{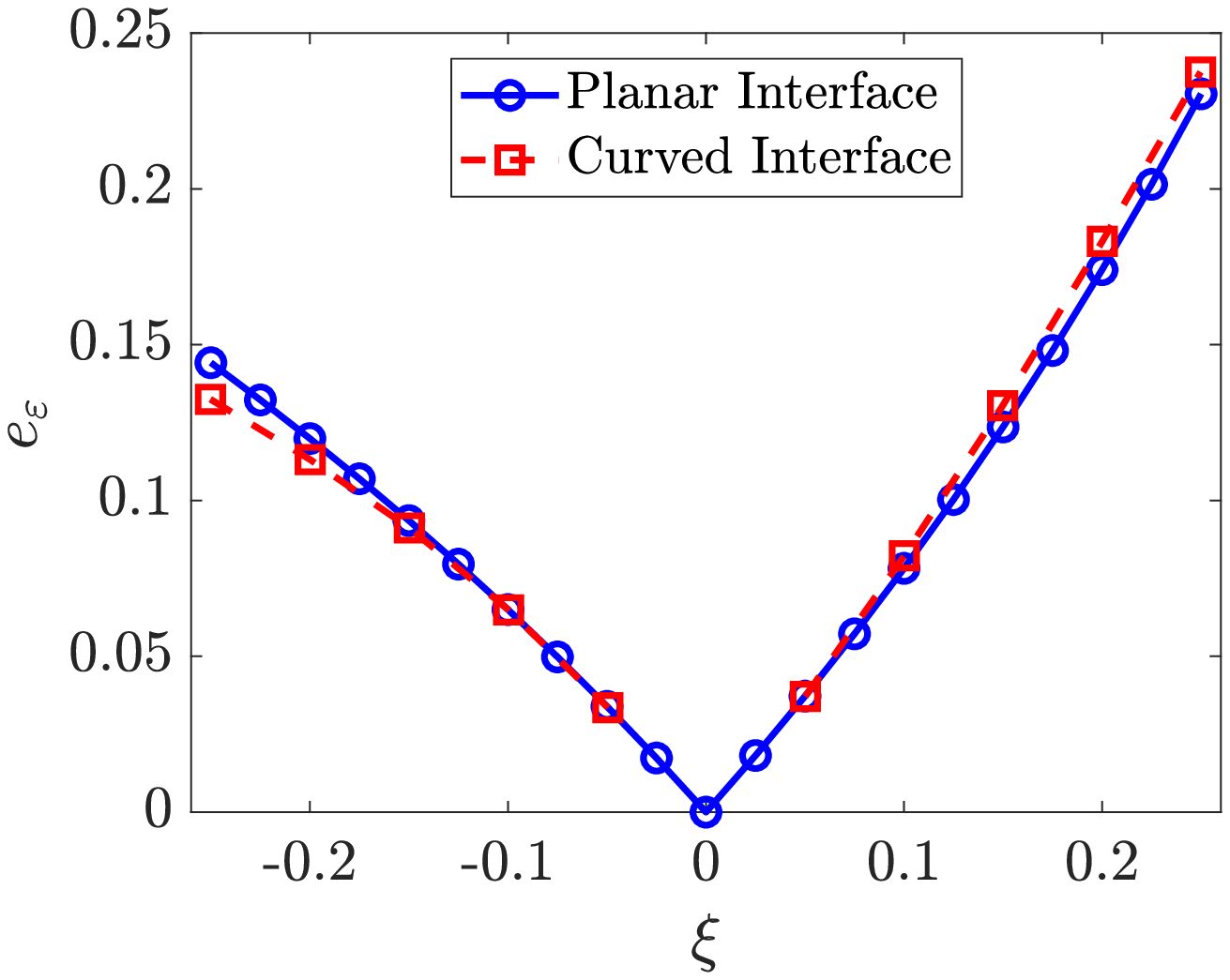}
    		\caption*{\hspace{30pt}(a)}
    	\end{minipage}
    	\hspace{0pt}
    	\begin{minipage}[b]{0.5\textwidth}
    		\centering
    		\includegraphics[scale=0.6,trim= 0 0 0 0,clip]{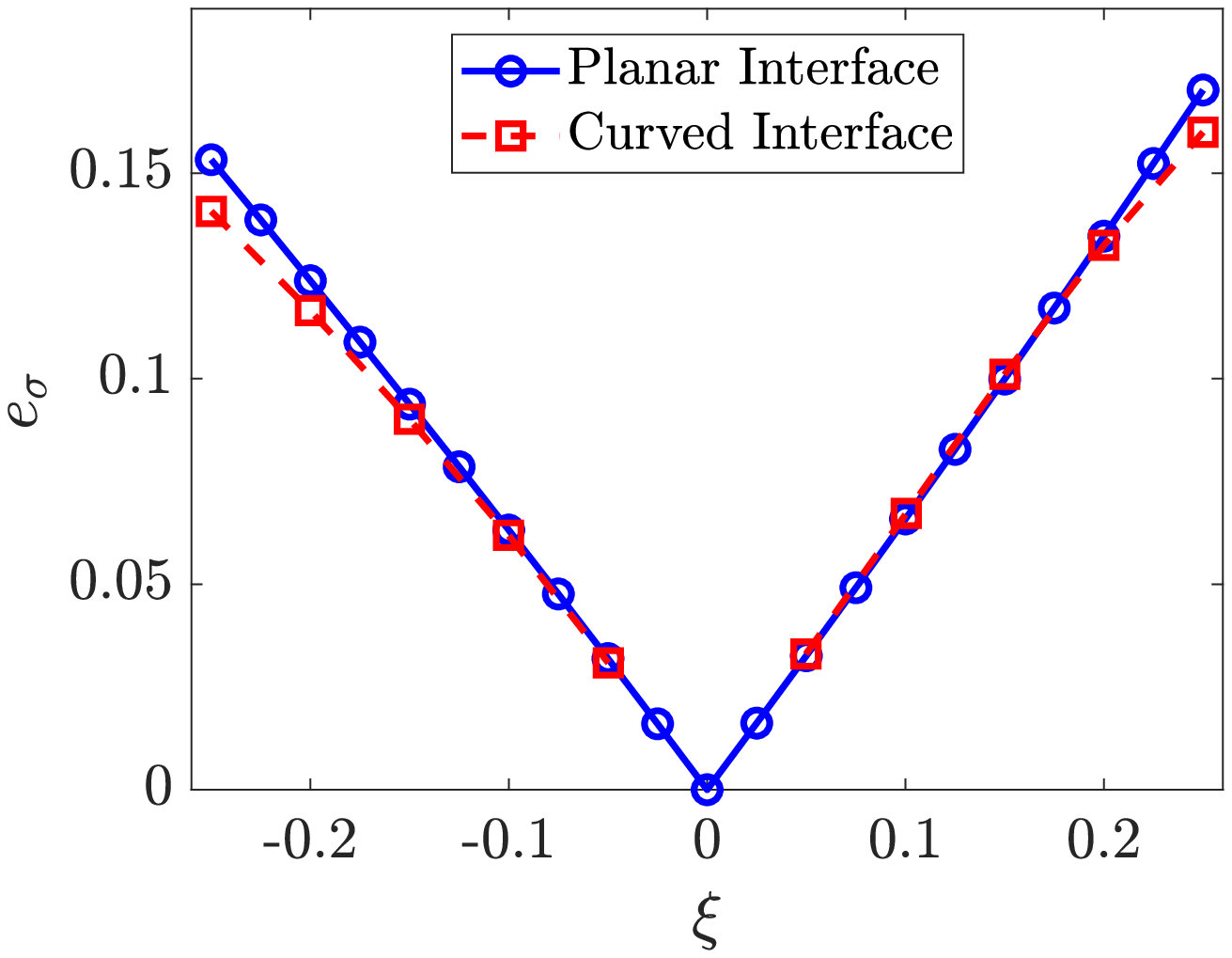}
    		\caption*{\hspace{30pt}(b)}
    	\end{minipage}
    	\caption{Dependence of the interface errors on the convective distortion parameter $\xi$: (a) the relative interface thickness error $e_{\varepsilon}$, and (b) the relative surface tension force error $e_{\sigma}$. } 
    	\label{m1}
    \end{figure}

 The case at $\xi=0.1$ is used to illustrate the simulation results. The contour of the order parameter from $t=0$ to $t=0.75$ are shown in Figs.~\ref{ag} (a)-(d). The time history of $\tilde{\varepsilon}_d$ on $\theta=0$ and $\theta=\pi/2$, which are the bottom boundary and the left boundary respectively, are shown in Fig.~\ref{ag} (e). It can be observed that due to the convective distortion, the interface is expanded at $\theta=0$ and compressed at $\theta=\pi/2$. The distortion increases during the convection of the curved interface. On the other hand, as the interface deviates further from the equilibrium interface profile, the effect of the free energy minimization increases. When the free energy minimization opposes the convective distortion, the deviation will stop increasing, which leads to a constant $\tilde{\varepsilon}_d$.
 Following similar definitions of the errors in Eqs.~(\ref{thkerror}), (\ref{surftenserror}) and similar calculation techniques, the relative interface thickness error and the relative surface tension force error are calculated at $\theta=0$ and $\theta=\pi/2$. The results are shown in Fig.~\ref{m1} (a)-(b) respectively. 

\subsection{Relationship between the errors and the convective distortion parameter}
\begin{figure}[h]
	
	\begin{minipage}[b]{0.5\textwidth}
		\centering
		\includegraphics[scale=0.6,trim= 0 0 0 0,clip]{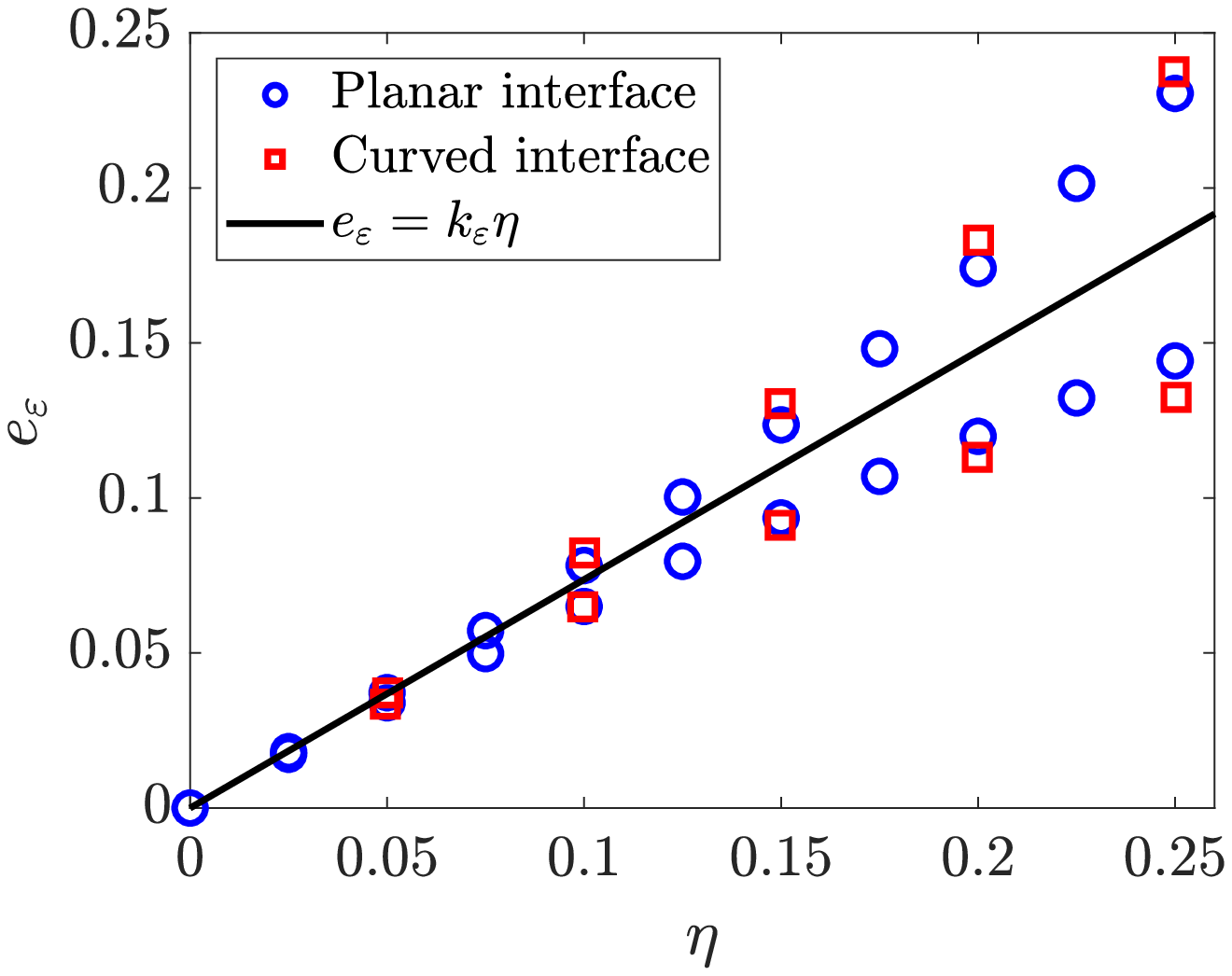}
		\caption*{\hspace{30pt}(a)}
	\end{minipage}
	\hspace{0pt}
	\begin{minipage}[b]{0.5\textwidth}
		\centering
		\includegraphics[scale=0.6,trim= 0 0 0 0,clip]{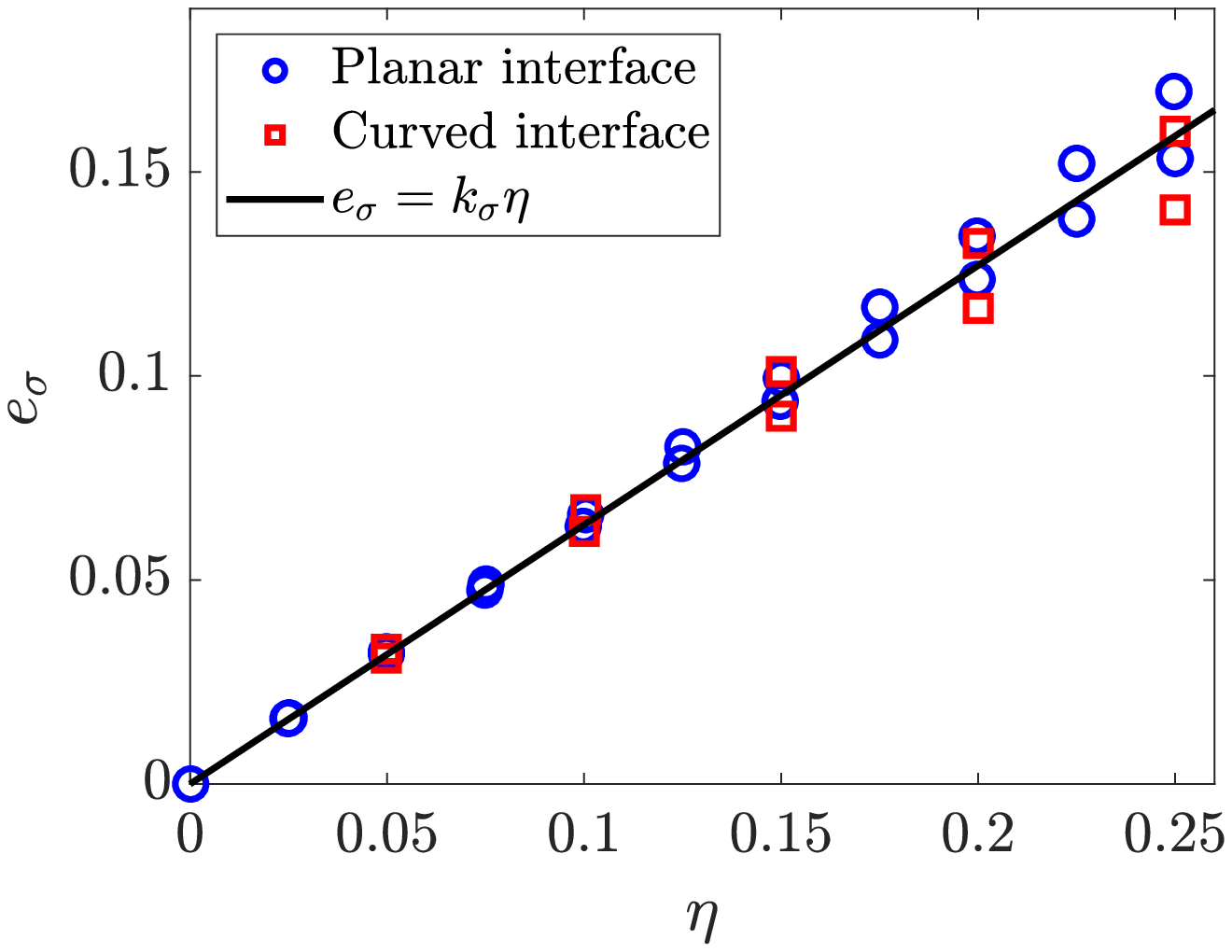}
		\caption*{\hspace{30pt}(b)}
	\end{minipage}
	\caption{Correlations of the interface errors as a function of RMS convective distortion parameter $\eta$: (a) the relative interface thickness error $e_{\varepsilon}$, and (b) the relative surface tension force error $e_{\sigma}$. } 
	\label{fitting}
\end{figure}

 In Fig.~(\ref{m1}), the relative interface thickness error and the relative surface tension force error are plotted as a function of $\xi$ for the convection of the planar and the curved interfaces. It is worth mentioning that the errors vary continuously with respect to $\xi$, while the singularities at $\xi=0$ is merely a result of taking the absolute value of the errors. When the effect of the volume-preserving mean curvature flow is minimized by reducing $\varepsilon/R$ in the curved cases, the dependency of the errors on the convective distortion parameter $\xi$ is almost identical for both the cases. It shows that the relationship is consistent in one and two dimensions with different problem setup. Considering the numerical results and the analysis of the interface-preserving condition~(\ref{convective distortion parameter}), we can further infer that the relationship between the errors and a constant $\xi$ is general. As shown in the figure, the errors decrease with the decrease in $|\xi|$. For the current cases with a constant $\xi$, at the discussed location, the RMS convective distortion parameter $\eta=|\xi|$.  Hence a small $\eta$ can be used to reduce the error due to the convective distortion and to improve the interface-preserving capability. In Fig.~\ref{m1}, the interface errors $e_{\varepsilon}$ and $e_{\sigma}$ can be fitted as functions which are proportional to $\eta$ with the corresponding coefficients $k_{\varepsilon}$ and $k_{\sigma}$:
 \begin{equation}\label{correlation}
 e_{\varepsilon}=k_{\varepsilon}\eta,\ e_{\sigma}=k_{\sigma}\eta,
 \end{equation}
 where the coefficients $k_{\varepsilon}=0.7371$ and $k_{\sigma}=0.6352$. The simple correlations are shown in Fig.~\ref{fitting}. When diffuse interface is subjected to complex motions, there can be the loss of accuracy or the unresolved gradients due to the extensional and compressional distortions. In that case, there is a need to minimize the interface thickness error $e_{\varepsilon}$. Furthermore, the control of $e_{\sigma}$ is critical when the surface tension force and the interface dynamics play an important role. The RMS convective distortion parameter $\eta$ can be selected as:
 $\eta=\mathrm{min}\left\{\frac{e_{\varepsilon}}{k_{\varepsilon}},\frac{e_{\sigma}}{k_{\sigma}}\right\}$     
for a desired value of the relative interface thickness and surface tension force errors.

 
\section{Rising bubble problem}
After analyzing the selection of the RMS convective distortion parameter, we assess the effectiveness of our Navier-Stokes Allen-Cahn system with the time-dependent mobility model by simulating 2D and 3D rising bubble benchmark cases. A systematic convergence study is performed for the volume-preserving mean curvature flow and the interface-preserving capability. 
We compare the accuracy of the diffuse interface formulation with the corresponding sharp interface method. We will show that the interface-preserving capability is necessary to guarantee an accurate solution from the diffuse interface formulation. 

\subsection{Two-dimensional rising bubble benchmark}
\begin{figure}

  \begin{minipage}[b]{0.5\textwidth}
  \hspace*{0.05\linewidth}
  \centering
    \begin{tikzpicture}[xscale=3.2, yscale=3.2,baseline={(0,-0.6)},line width=0.6]
    
 \draw [fill=gray!20](0,0) rectangle (1,2);

    \draw[black,fill=white](0.5,0.5) circle [radius=0.25];

\node [above] at (0.51,0.52) { $\Omega_2$};
\node [below] at (0.51,0.5) {($\rho_2,\mu_2$)};
\node [above] at (0.5,1.4) {$\Omega_1$};
\node [below] at (0.5,1.4) {($\rho_1,\mu_1$)};
\draw [fill=black] (0.5,0.5) circle (0.01);

\draw [-stealth] (0.5,0.5) -- (0.27,0.6);
\node [left] at (0.27,0.62) {$R$};

\draw [-stealth,thick] (0,0) -- (0,0.2);
\node [left] at (0,0.2) {Y};
\draw [-stealth,thick] (0,0) -- (0.2,0);
\node [below] at (0.2,0) {X};

\draw [dashed] (0,2) -- (0,2.15);
\draw [dashed] (1,2) -- (1,2.15);
\draw [dashed] (1,0) -- (1.15,0);
\draw [dashed] (1,2) -- (1.15,2);
\draw [stealth-stealth] (0,2.15) -- (1,2.15);
\draw [stealth-stealth] (1.15,0) -- (1.15,2);
\node [above] at (0.5,2.15) {1};
\node [right] at (1.15,1) {2};
\draw [dashed](0.6,0)--(0.7,-0.05);
\node [below] at (0.7,-0.05) {$u=v=0$};
\draw [dashed](0,1.1)--(-0.05,1);
\node [left] at (-0.05,1) {$u=0$};
\draw [dashed](1,1.1)--(0.95,1);
\node [left] at (0.95,1) {$u=0$};
\draw [dashed](0.4,2)--(0.5,1.95);
\node [below] at (0.5,1.95) {$u=v=0$};

  \end{tikzpicture}
  \caption*{(a)}
  \end{minipage}
  \hspace{-1cm}
  \begin{minipage}[b]{0.495\textwidth}
    \centering
    \includegraphics[trim=1 25 1 1 ,clip,scale=0.42]{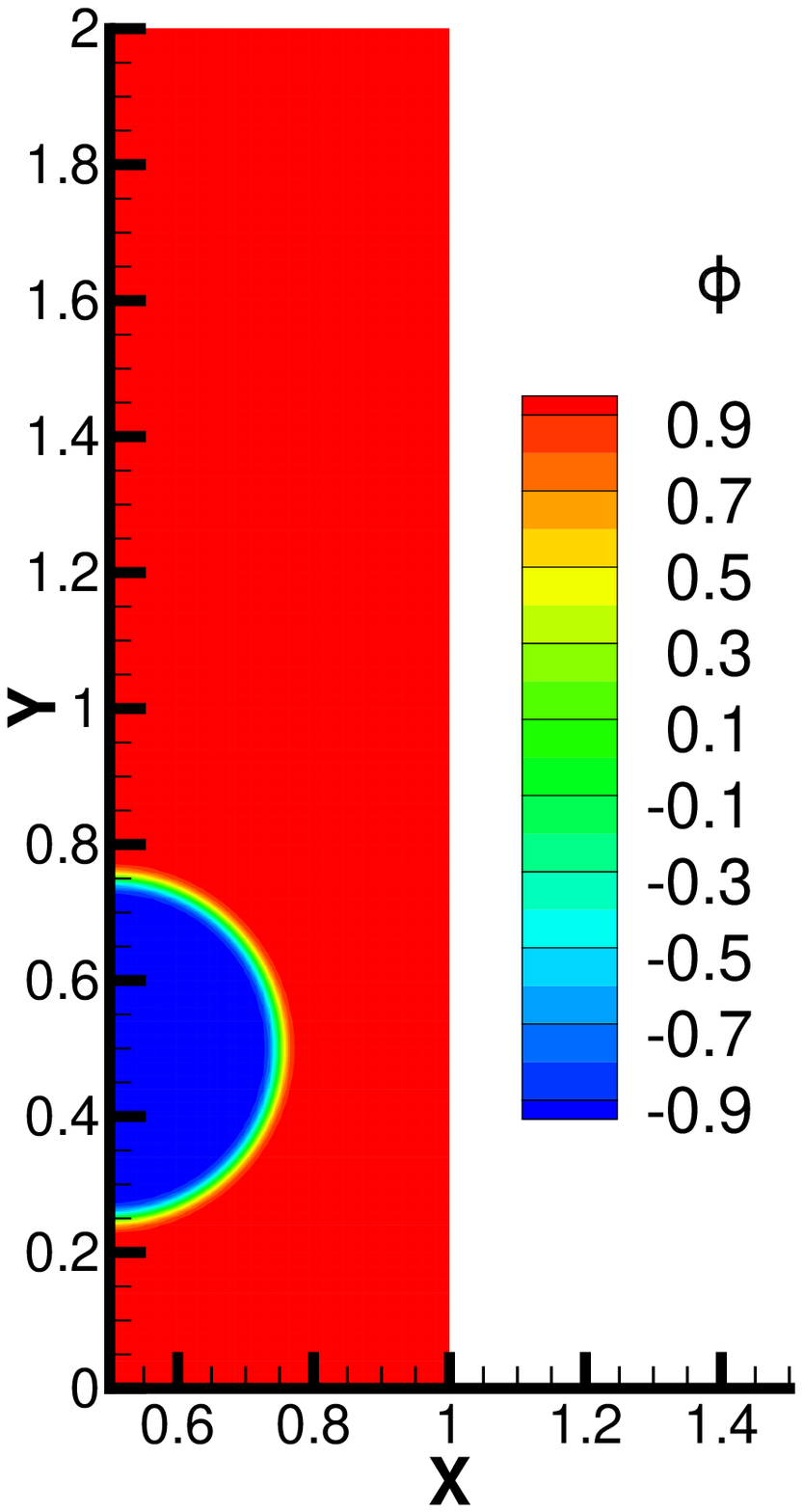}
    \caption*{(b)}
  \end{minipage}

  \caption{Two-dimensional rising bubble problem: (a) schematic diagram of the computational domain, and (b) contour of the order parameter at $t=0$.}
  \label{RBI}

\end{figure}

In this subsection, we present a well-known two-dimensional rising bubble benchmark case for assessing the role of the time-dependent mobility model on the accurate surface tension dynamics.
The benchmark case considers the rising and deforming of an initially circular bubble immersed in a quiescent fluid. A rectangular computational domain $[0, 1]\times [0,2]$ is considered for the benchmark case. The initial condition for the circular bubble is given by:
\begin{equation}
        \phi(x,y,0)=-\tanh\left(\frac{R-\sqrt{(x-x_c)^2+(y-y_c)^2}}{\sqrt{2}\varepsilon}\right),
\end{equation}
where $R=0.25$ is the radius of the bubble with its center at  $(x_c,y_c)=(0.5,0.5)$. The quiescent fluid is initialized with zero velocity and pressure. The slip boundary condition is satisfied on the left and the right boundaries, while the no-slip boundary condition is prescribed on the top and the bottom boundaries. A zero flux Neumann boundary condition is imposed on all the boundaries for the order parameter. The density and the dynamic viscosity of the fluid and the bubble are selected as $\rho_1=1000$, $\rho_2=100$ and $\mu_1=10$, $\mu_2=1$. The surface tension coefficient is set to be $\sigma=24.5$. The gravitational acceleration is taken as $\boldsymbol{g}=(0,-0.98)$. The problem setup is illustrated in Fig.~\ref{RBI} (a). The benchmark problem has been studied by several research groups employing various numerical techniques in \cite{hysing2009quantitative}. In the current study, we consider the data from the first group in \cite{hysing2009quantitative} for comparison purposes, which employs the finite element method for spatial discretization and the sharp level-set method for the interface capturing.
Because the problem has a symmetric axis $x=0.5$, we conduct the simulation with the right half of the computational domain. A symmetric boundary condition is imposed on the axis of symmetry. The contour of the order parameter in the computational domain at $t=0$ is shown in Fig.~\ref{RBI} (b). The computational domain is discretized with a uniform structured mesh of grid size $\Delta x=\Delta y=h$. The grid size is selected according to the mesh resolution at the interface as $\varepsilon/h=1$. The time step size is chosen as $\Delta t =0.00125$. 
 
To quantify the mass conservation and the rising bubble dynamics, the mass of the order parameter $m$, the circularity of the bubble $\slashed{c}$, the rise velocity of the bubble $V_b$ and the center of mass of the bubble $Y_b$ are defined as follows:
\begin{align}
	m&=\int_{\Omega} \phi \mathrm{d}\Omega,\nonumber\\
	\slashed{c}&=P_a/P_b,\nonumber\\
    V_b&=\frac{\int_{\Omega_2}v \mathrm{d}\Omega}{\int_{\Omega_2} 1 \mathrm{d}\Omega},\nonumber\\
    Y_b&=\frac{\int_{\Omega_2}y \mathrm{d}\Omega}{\int_{\Omega_2} 1 \mathrm{d}\Omega},\nonumber
\end{align}
 where $P_a$ is the perimeter of the circle which has the same area as the deformed bubble, $P_b$ denotes the perimeter of the bubble, $v$ is the velocity in the Y direction and $y$ is the Y coordinate. The defined variables and the bubble shape at $t=3$ are compared in the convergence study.

\subsubsection{Convergence of the volume-preserving mean curvature flow}
\begin{figure}[h]
    \begin{minipage}[b]{0.5\textwidth}
   \captionsetup{skip=1pt}
  \centering
\includegraphics[scale=0.57]{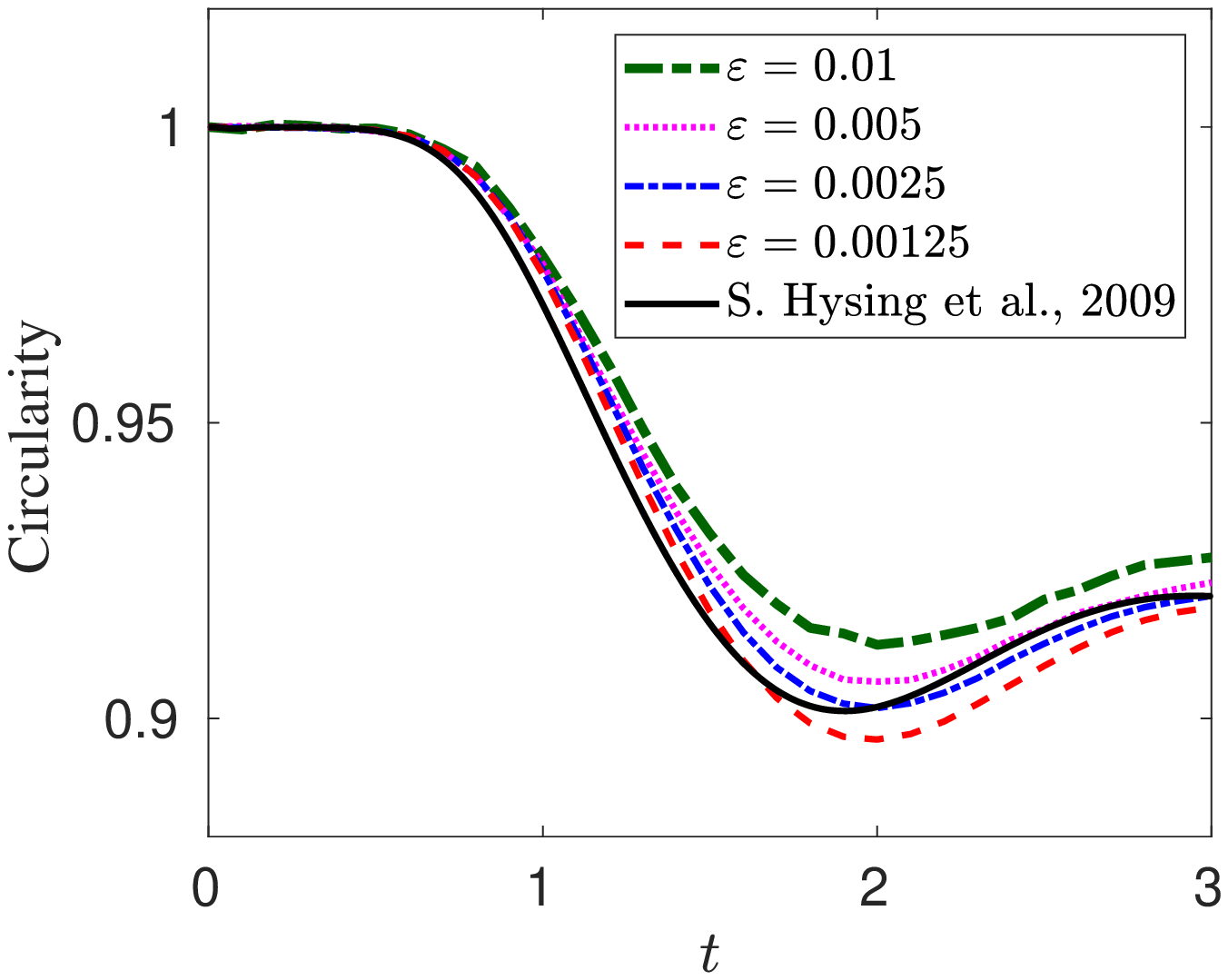}
    \caption*{\hspace{35pt}(a)}
  \end{minipage}
\hspace{5pt}
  \begin{minipage}[b]{0.5\textwidth}
  	\captionsetup{skip=1pt}
  \centering
\includegraphics[scale=0.57]{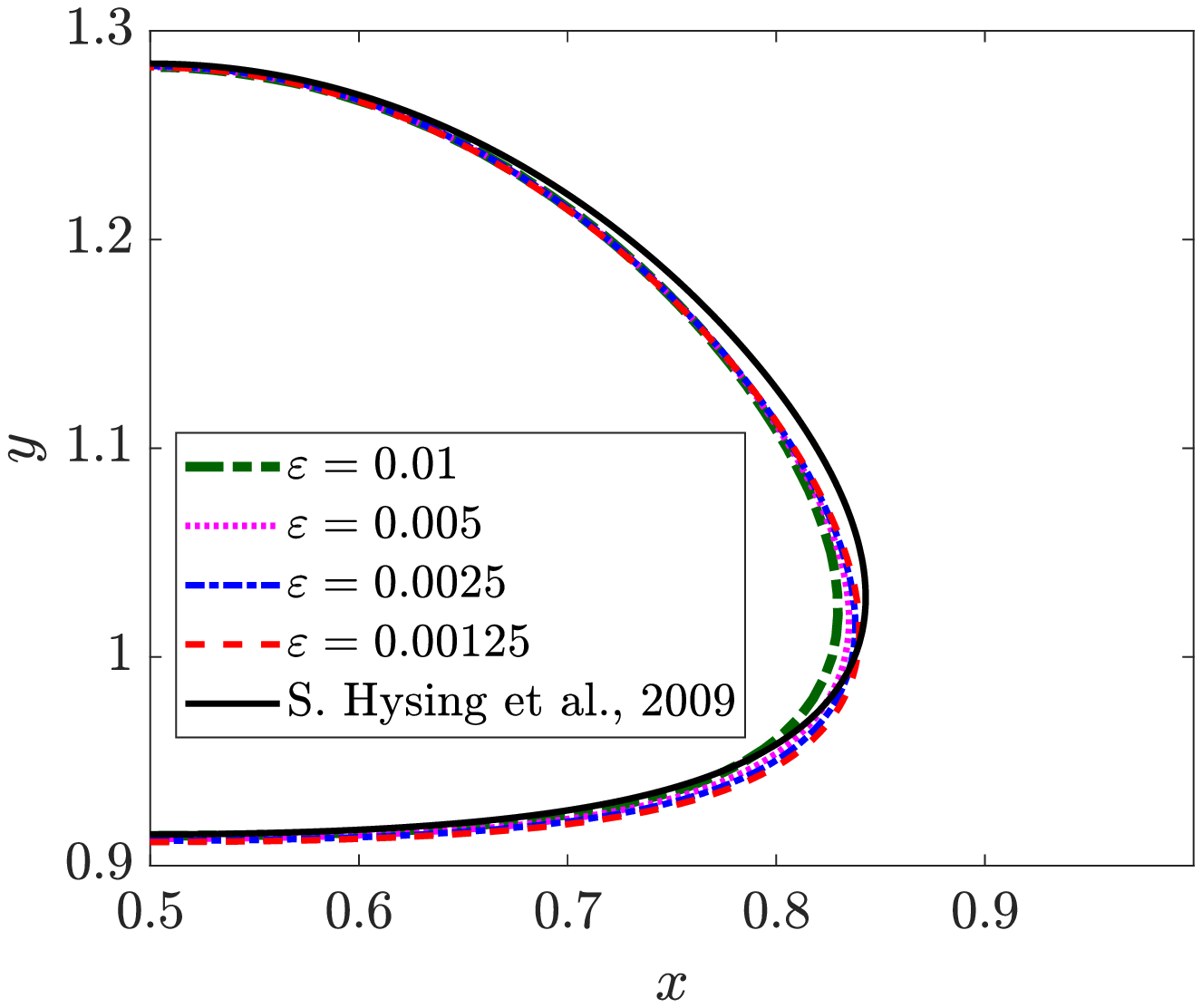}
    \caption*{\hspace{20pt}(b)}
  \end{minipage}
  
  \begin{minipage}[b]{0.5\textwidth}
  	\captionsetup{skip=1pt}
  \centering
\includegraphics[scale=0.57]{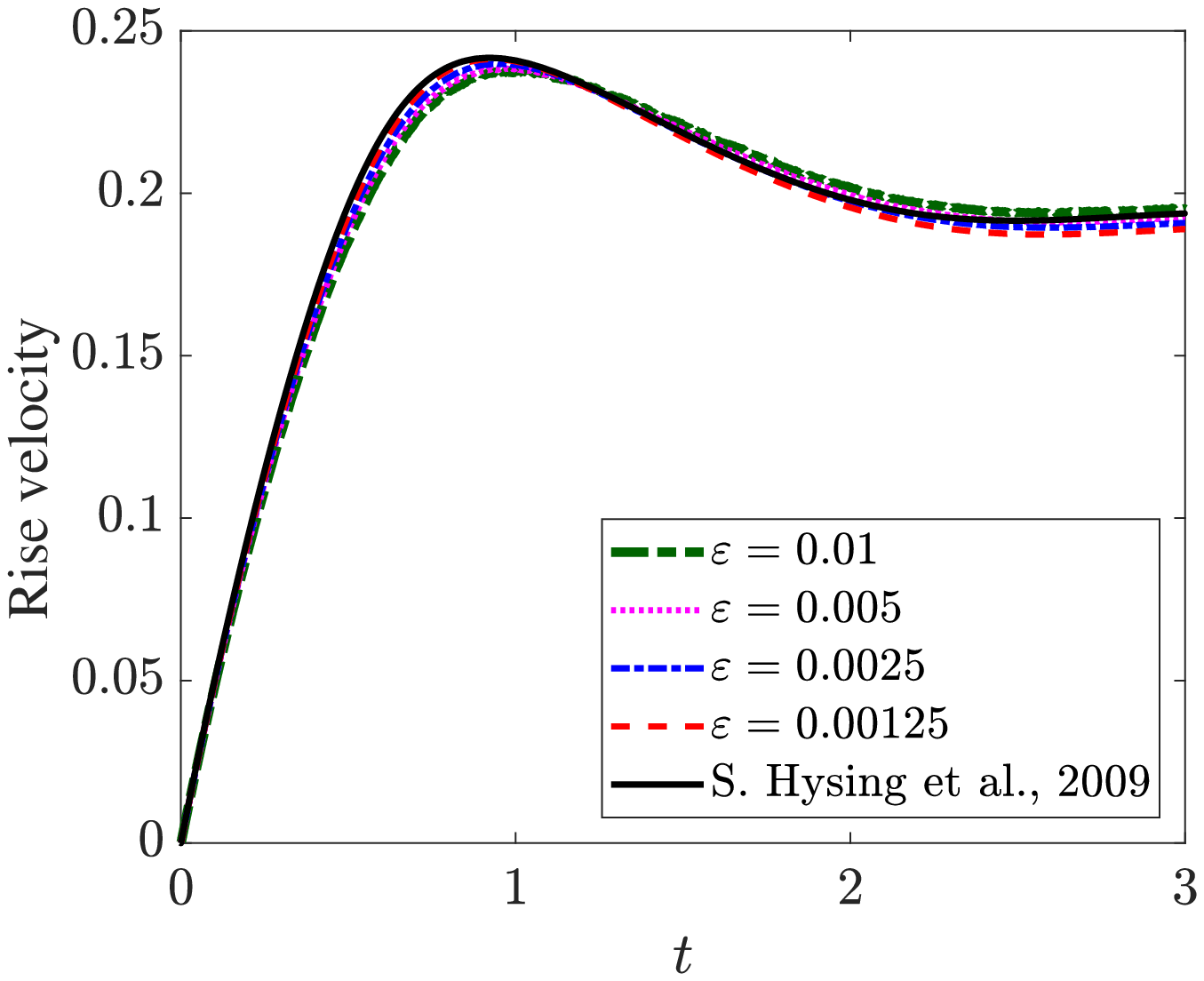}
    \caption*{\hspace{35pt}(c)}
  \end{minipage}
\hspace{0pt}
  \begin{minipage}[b]{0.5\textwidth}
  	\captionsetup{skip=1pt}
  \centering
\includegraphics[scale=0.57]{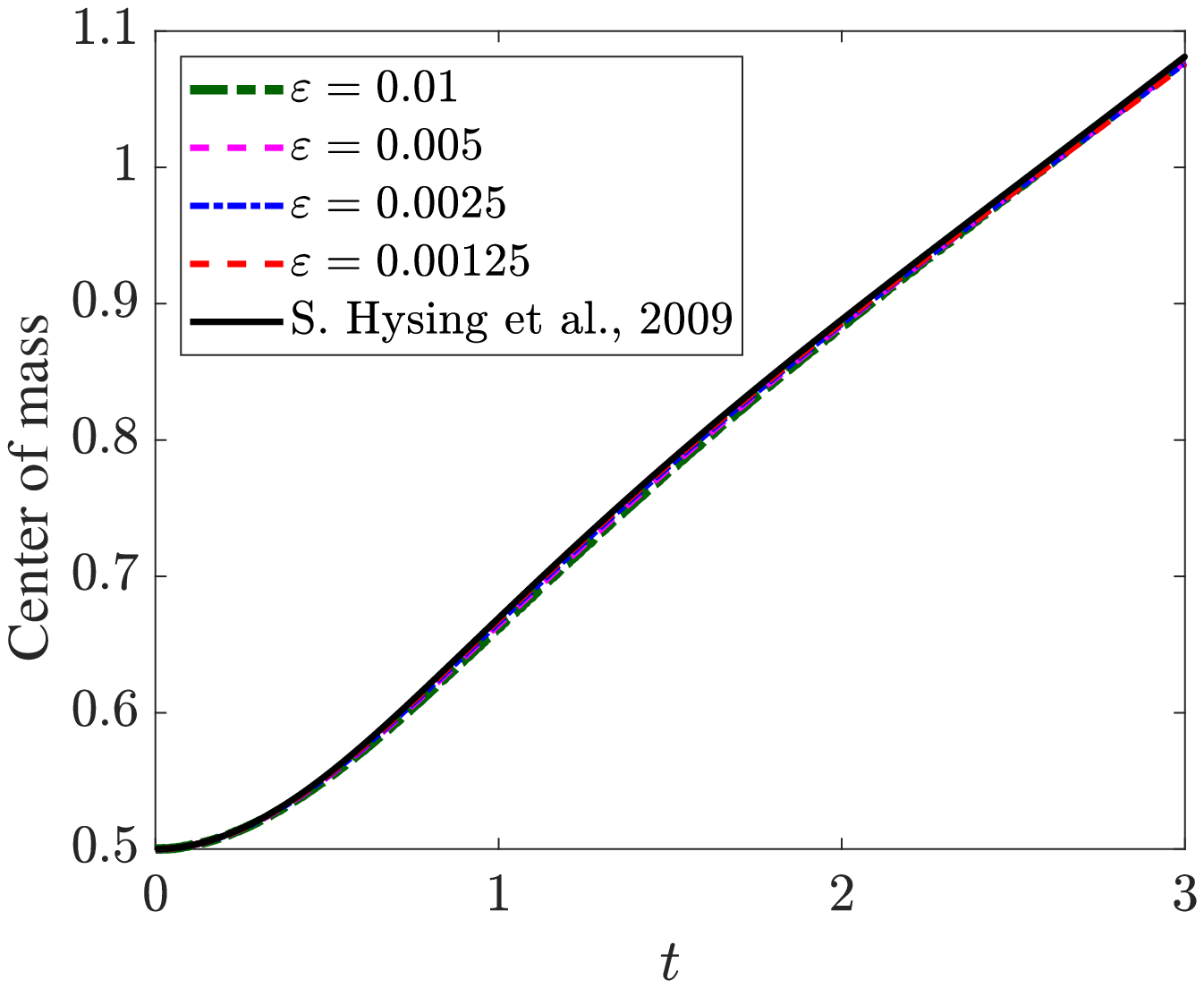}
    \caption*{\hspace{30pt}(d)}
  \end{minipage}
  
  	\captionsetup{skip=1pt}
  
    \caption{Volume-preserving mean curvature flow convergence study for 2D rising bubble benchmark case: (a) circularity of the bubble, (b) interface shape at $t=3$, (c) rise velocity, and (d) center of mass.} 
    \label{2dmc}
\end{figure}

The volume-preserving mean curvature flow velocity disturbs the convection of the two-phase interface according to the fluid flow velocity. Therefore we perform a convergence study to make sure that the disturbance is negligible. With the time-dependent mobility, the velocity is given by:
 \begin{align}
    \boldsymbol{v}(\boldsymbol{x},t)&=\gamma(t)\varepsilon^2\left(\kappa(\boldsymbol{x},t)-\frac{1}{|\Gamma_I^{\phi}(t)|}\int_{\Gamma_I^{\phi}}\kappa(\boldsymbol{x},t) ds\right)\boldsymbol{n}^{\phi}_L(\boldsymbol{x},t)\nonumber\\
    &=\frac{1}{\eta}\mathcal{F}(|\zeta(\boldsymbol{x},t)|)\varepsilon^2\left(\kappa(\boldsymbol{x},t)-\frac{1}{|\Gamma_I^{\phi}(t)|}\int_{\Gamma_I^{\phi}}\kappa(\boldsymbol{x},t) ds\right)\boldsymbol{n}^{\phi}_L(\boldsymbol{x},t),\quad \boldsymbol{x}\in\Gamma_I^{\phi}(t).\label{mc_with_vg}
\end{align}
We refer to the quantity $\gamma(t)\varepsilon^2$ as the scaling factor of the volume-preserving mean curvature flow. According to Eq.~(\ref{mc_with_vg}), the scaling factor is affected by the user-defined parameters $\eta$ and $\varepsilon$. In the present convergence study, we decrease $\varepsilon$ by a factor of $2$ from $\varepsilon=0.01$ to $\varepsilon=0.00125$, while keeping $\eta=0.1$. Thus, the volume-preserving mean curvature flow is reduced while the RMS convective distortion parameter is kept the same.  The resulting scaling factor $\gamma(t)\varepsilon^2$ at $t=3$ at various $\varepsilon$ is summarized in Table \ref{mcc errors}. The circularity $\slashed{c}$, the bubble shape $\phi=0$ at $t=3$ , the rise velocity $V_b$, the center of mass $Y_c$, and their comparison with the data in \cite{hysing2009quantitative} are shown in Fig.~\ref{2dmc}.

To further quantify the results, we define the mass conservation error and the convergence errors as follows:
\begin{align}
   e_m&=\left|\frac{m_{t=0}-m_{t=3}}{m_{t=0}}\right|,\nonumber\\
   e_{\slashed{c}}&=\frac{||\slashed{c}-\slashed{c}_{\mathrm{ref}} ||_2}{||\slashed{c}_{\mathrm{ref}}||_2},\nonumber\\
    e_{V_b}&=\frac{||V_b-V_{b,\mathrm{ref}} ||_2}{||V_{b,\mathrm{ref}}||_2},\nonumber\\
    e_{Y_b}&=\frac{||Y_b-Y_{b,\mathrm{ref}} ||_2}{||Y_{b,\mathrm{ref}}||_2},\nonumber
\end{align}
where $|\cdot|$ denotes the absolute value and $||\cdot||_2$ denotes the $L^2$ norm of the vector. The simulation results from the case with $\varepsilon=0.00125$ are taken as the reference. The errors including the relative interface thickness and surface tension force errors estimated by Eq.~(\ref{correlation}) are summarized in Table \ref{mcc errors}.

\begin{table}[h]
    \centering
        \caption{Quantification of the errors for 2D rising bubble case: convergence with respect to $\varepsilon$ at a constant $\eta$.}
    \begin{tabular}{cccccccccccc}
    \hline
        $\eta$&$\varepsilon$&$e_{\varepsilon}$&$e_{\sigma}$&$\gamma(t)\varepsilon^2\ (t=3)$&$e_m$ &$e_{\slashed{c}}$&$e_{V_b}$&$e_{Y_b}$\\
         \hline
         $0.1$&$0.01$&$0.074$&$0.064$&$1.1\times10^{-3}$&0.0082& 0.0105& 0.0279& 0.0048 \\

         $0.1$&$0.005$&$0.074$&$0.064$&$3.1\times10^{-4}$&0.0047& 0.0061& 0.0172& 0.0029\\

         $0.1$&$0.0025$&$0.074$&$0.064$&$9.1\times10^{-5}$&0.0025& 0.0034& 0.0089& 0.0014\\
      
         $0.1$&$0.00125$&$0.074$&$0.064$&$2.7\times10^{-5}$&0.0013& -& -& -\\
         \hline
    \end{tabular}

    \label{mcc errors}
\end{table}

It can be observed from Table \ref{mcc errors} that the mass conservation error is below $1\%$ and it decreases with the reduction of $\varepsilon$. The total mass is well conserved. The decrease in the scaling factor $\gamma(t)\varepsilon^2$ at $t=3$ along with the reduction in $\varepsilon$ reflects the diminishing of the volume-preserving mean curvature flow. Since the flow directly affects the topology of the bubble, its influence is shown clearly in the convergence of the circularity and the bubble shape. As shown in Fig.~\ref{2dmc} (a), the circularity is reduced with the reduction of $\varepsilon$. This indicates that the bubble experiences a relatively larger deformation and a higher curvature due to the reduction of the volume-preserving mean curvature flow. This is consistently observed in the convergence of the bubble shape shown in Fig.~\ref{2dmc} (b). However, the bubble shape deviates from the data in \cite{hysing2009quantitative}. The deviation can be observed in the rise velocity plotted in Fig.~\ref{2dmc} (c) as well. This deviation is due to the error in the surface tension force calculation introduced by the convective distortion. To conclude, the decreasing of the volume-preserving mean curvature flow with the decreasing of $\varepsilon$ cannot guarantee the accuracy of the solution, which leads to the investigation of the interface-preserving property.

\subsubsection{Convergence of the interface-preserving property}
\begin{figure}[h]
    \begin{minipage}[b]{0.5\textwidth}
	\captionsetup{skip=1pt}
	\centering
	\includegraphics[scale=0.57]{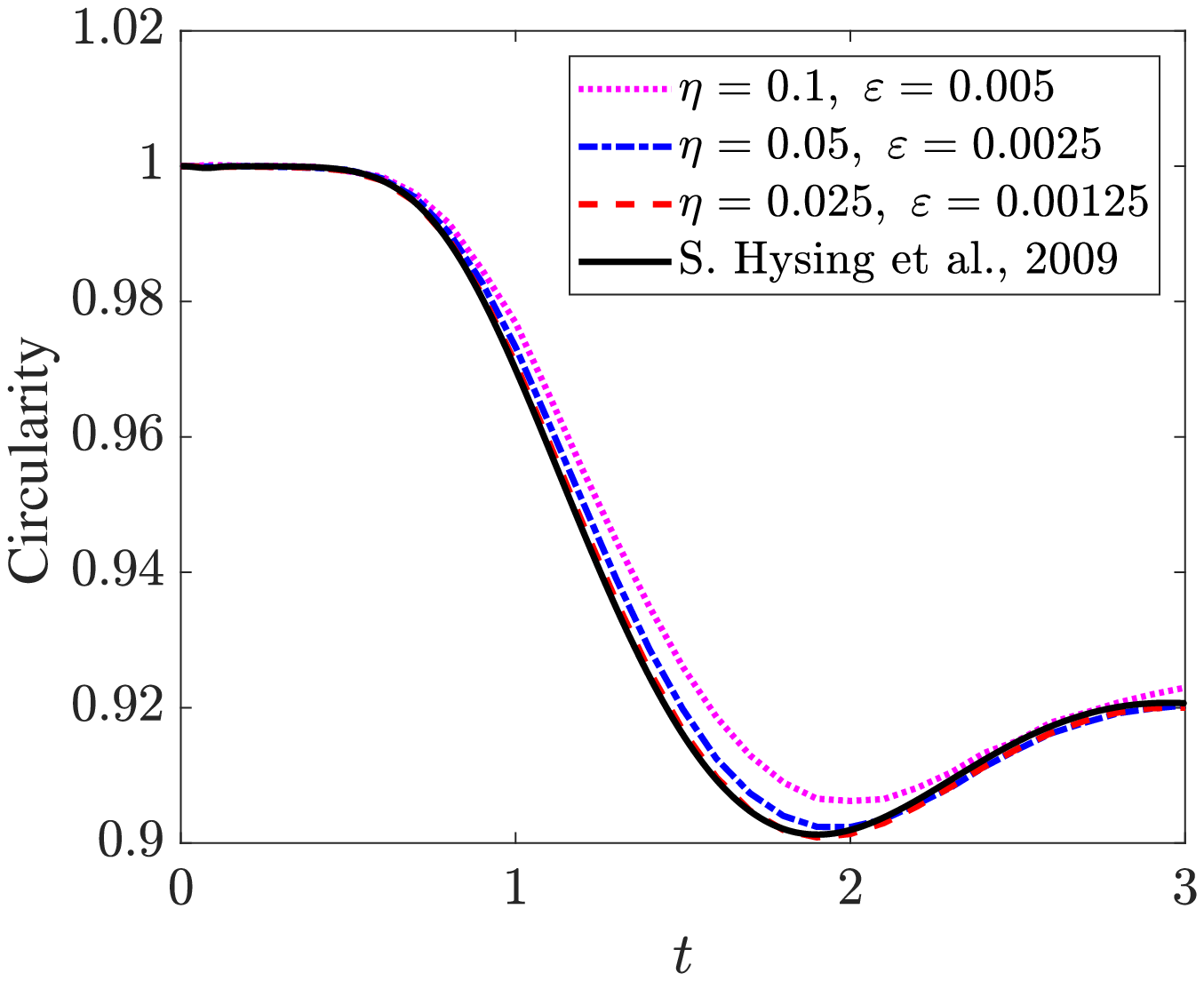}
	\caption*{\hspace{35pt}(a)}
\end{minipage}
\hspace{5pt}
\begin{minipage}[b]{0.5\textwidth}
	\captionsetup{skip=1pt}
	\centering
	\includegraphics[scale=0.57]{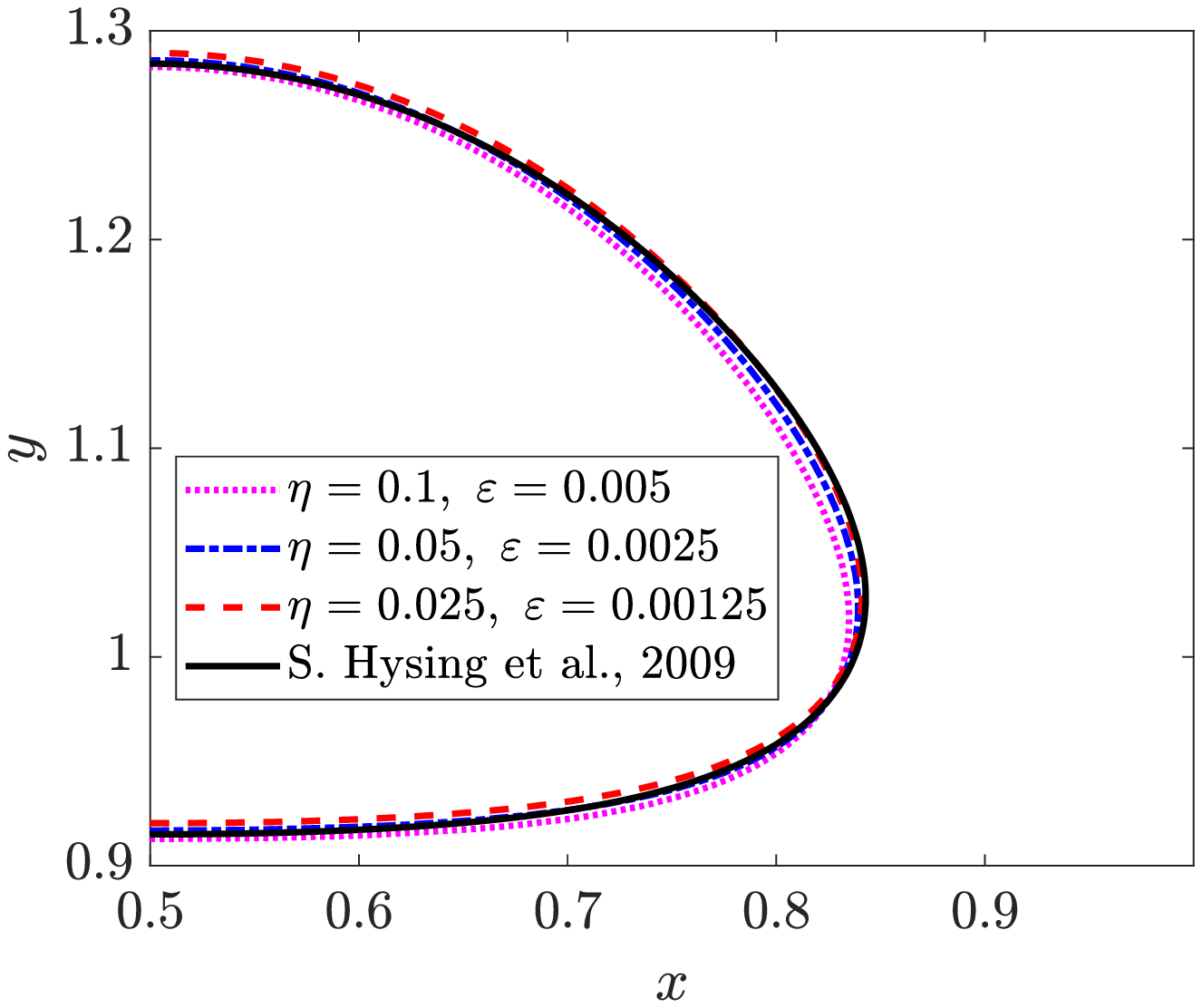}
	\caption*{\hspace{20pt}(b)}
\end{minipage}

\begin{minipage}[b]{0.5\textwidth}
	\captionsetup{skip=1pt}
	\centering
	\includegraphics[scale=0.57]{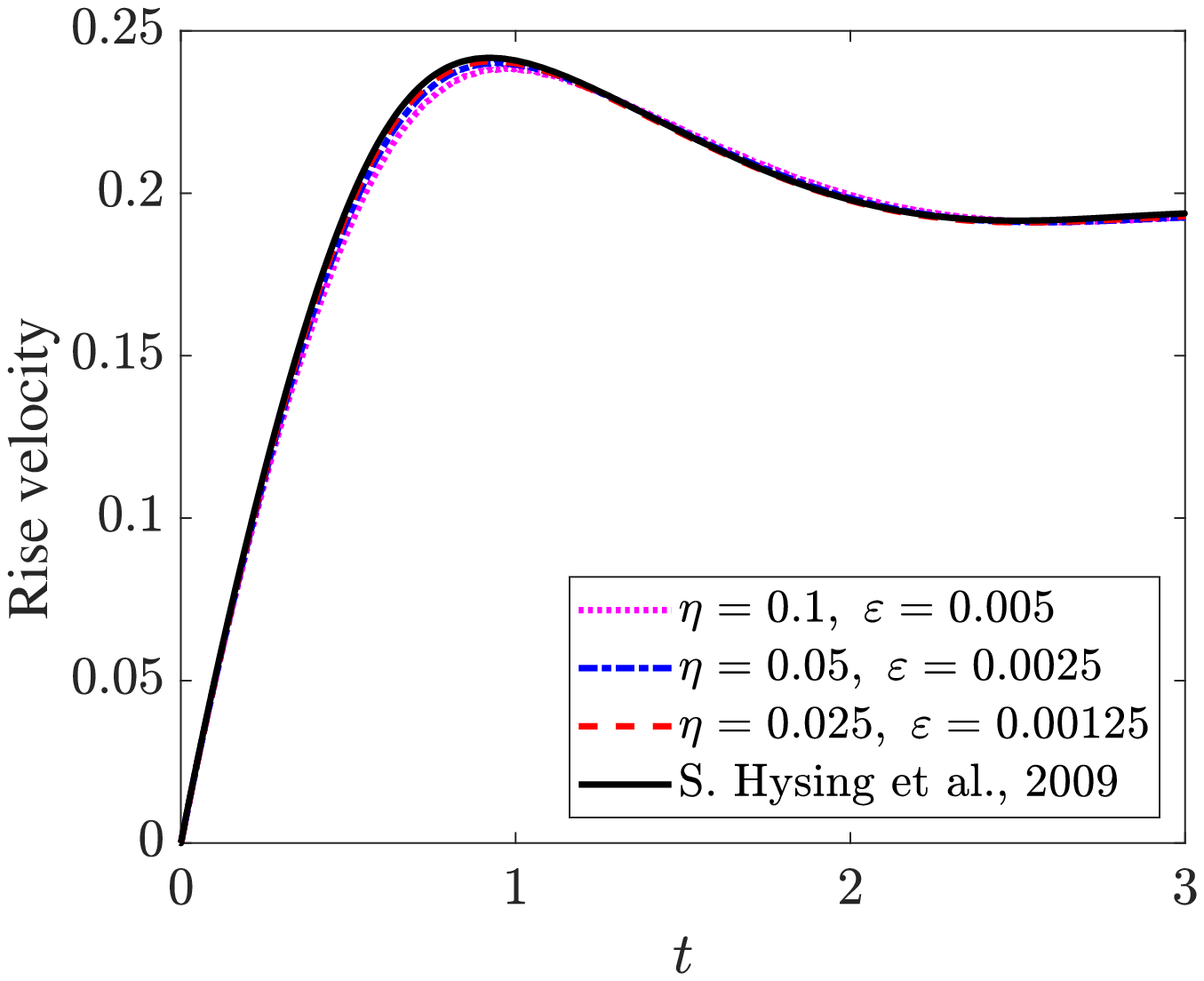}
	\caption*{\hspace{35pt}(c)}
\end{minipage}
\hspace{0pt}
\begin{minipage}[b]{0.5\textwidth}
	\captionsetup{skip=1pt}
	\centering
	\includegraphics[scale=0.57]{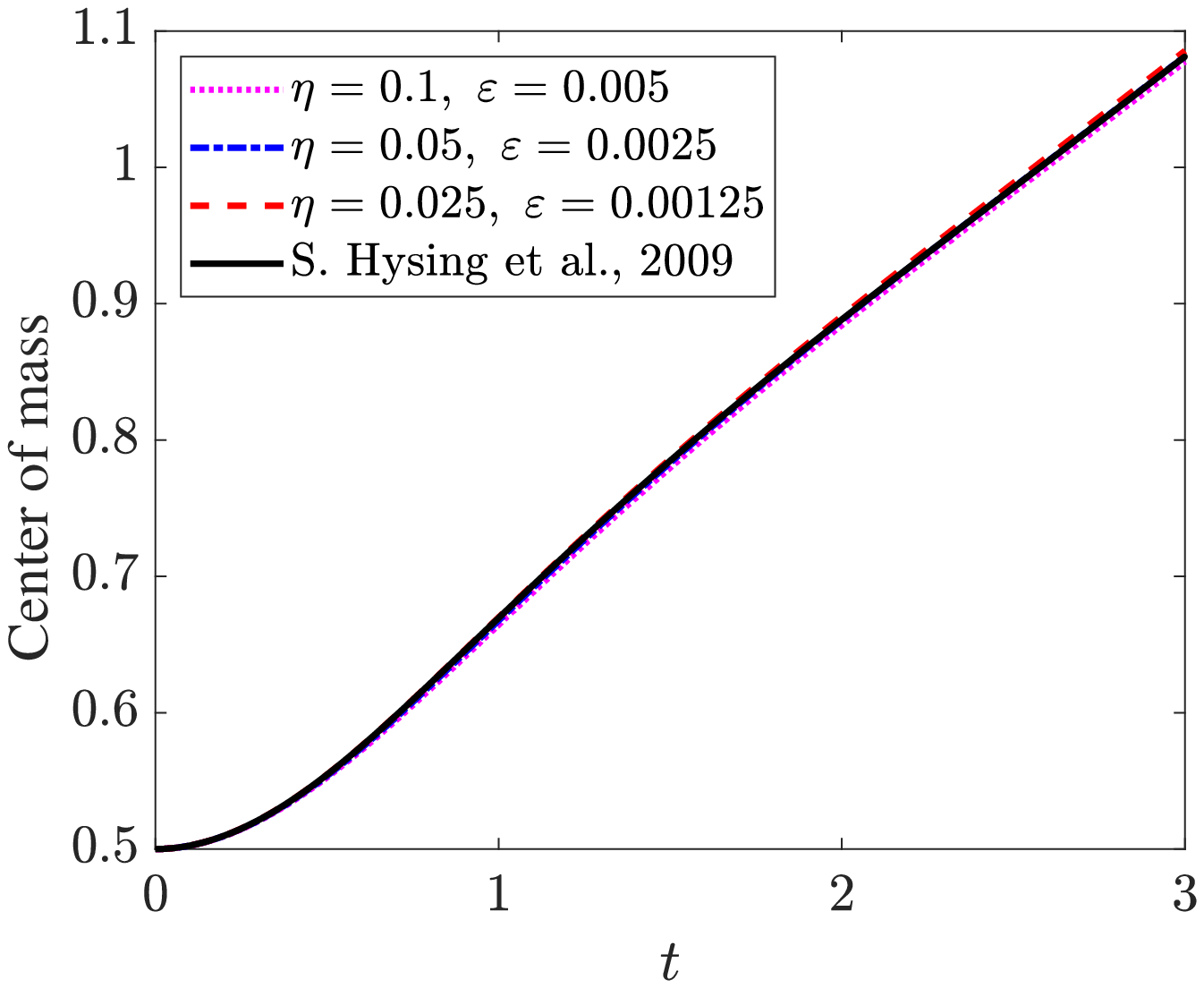}
	\caption*{\hspace{30pt}(d)}
\end{minipage}

\captionsetup{skip=1pt}

    \caption{The interface-preserving capability convergence study for 2D rising bubble benchmark case: (a) circularity of the bubble (b) interface shape at $t=3$ (c) rise velocity, and (d) center of mass.} 
    \label{2dgc}
\end{figure}

Following the convergence study of the volume-preserving mean curvature flow, we study the convergence of the interface-preserving capability. From the discussion in Section 4, we know that by decreasing the RMS convective distortion parameter $\eta$, the convective distortion error will decrease. However, as shown in Eq.~(\ref{mc_with_vg}), a decrease in $\eta$ will increase the volume-preserving mean curvature flow. Thus we need to decrease both $\eta$ and $\varepsilon$ to improve the interface-preserving capability and decrease the volume-preserving mean curvature flow simultaneously. Considering this, we conduct studies with the following parameter combinations: $\eta=0.1,\varepsilon=0.005$; $\eta=0.05,\varepsilon=0.0025$ and $\eta=0.025,\varepsilon=0.00125$. The scaling factor of the volume-preserving mean curvature flow $\gamma(t)\varepsilon^2$ at $t=3$ for each combination is shown in Table \ref{vp errors}. The circularity $\slashed{c}$, the bubble shape at $t=3$, the rise velocity $V_b$, the center of mass of the bubble $Y_b$ and the comparison with the data in \cite{hysing2009quantitative} are shown in Fig.~\ref{2dgc}. Taking $\eta=0.025,\varepsilon=0.00125$ as the reference case, the errors including the relative interface thickness and surface tension force errors are shown in Table \ref{vp errors}.

It can be observed from Table \ref{vp errors} that the mass is well conserved. The scaling factor of the volume-preserving mean curvature flow decreases with the decreasing of $\varepsilon$ and $\eta$ in the test cases. The decreasing of the RMS convective distortion parameter $\eta$ reduces the errors from the convective distortion and improves the interface preservation property. As shown in Fig.~\ref{2dgc}, with the decrease in the volume-preserving mean curvature flow and the enforcement of the interface preservation capability, the simulation results converge and match well with the data in \cite{hysing2009quantitative}.

\begin{table}[h]
    \centering
        \caption{Quantification of the errors for 2D rising bubble case: variation of $\varepsilon$ and $\eta$}
    \begin{tabular}{ccccccccccc}
    \hline
        $\eta$&$\varepsilon $&$e_{\varepsilon}$&$e_{\sigma}$&$\gamma(t)\varepsilon^2\ (t=3)$&$e_m$&$e_{\slashed{c}}$& $e_{V_b}$&$e_{Y_b}$\\
         \hline
         $0.1$&$0.005$&$0.074$&$0.064$&$3.1\times10^{-4}$&$0.0047$& 0.0051& 0.0130& 0.0070\\

         $0.05$&$0.0025$&$0.037$&$0.032$&$1.8\times10^{-4}$&0.0027& 0.0017& 0.0047& 0.0033\\
         
         $0.025$&$0.00125$&$0.018$&$0.016$&$1.2\times10^{-4}$&$0.0016$& -&  -& -\\
         \hline
    \end{tabular}

    \label{vp errors}
\end{table}

\subsubsection{Error due to insufficient interface-preserving capability}
We further emphasize the importance of the interface-preserving capability by comparing the results of the 2D rising bubble benchmark case simulated by the constant mobility model with $\gamma=1$ and the time-dependent mobility with $\eta=0.1$. The complete computational domain $[0,1]\times[0,2]$ is used in the simulations. The interface thickness parameter is set to be $\varepsilon=0.01$ and a uniform structured mesh $\Delta x = \Delta y =h$ of grid size $\varepsilon/h=1$ is employed for the spatial discretization via linear finite elements. The time step size is taken as $\Delta t = 0.005$.

\begin{figure*}

	\begin{minipage}[b]{0.45\textwidth}
		\centering
		\includegraphics[scale=0.5,trim=3 3 3 3,clip]{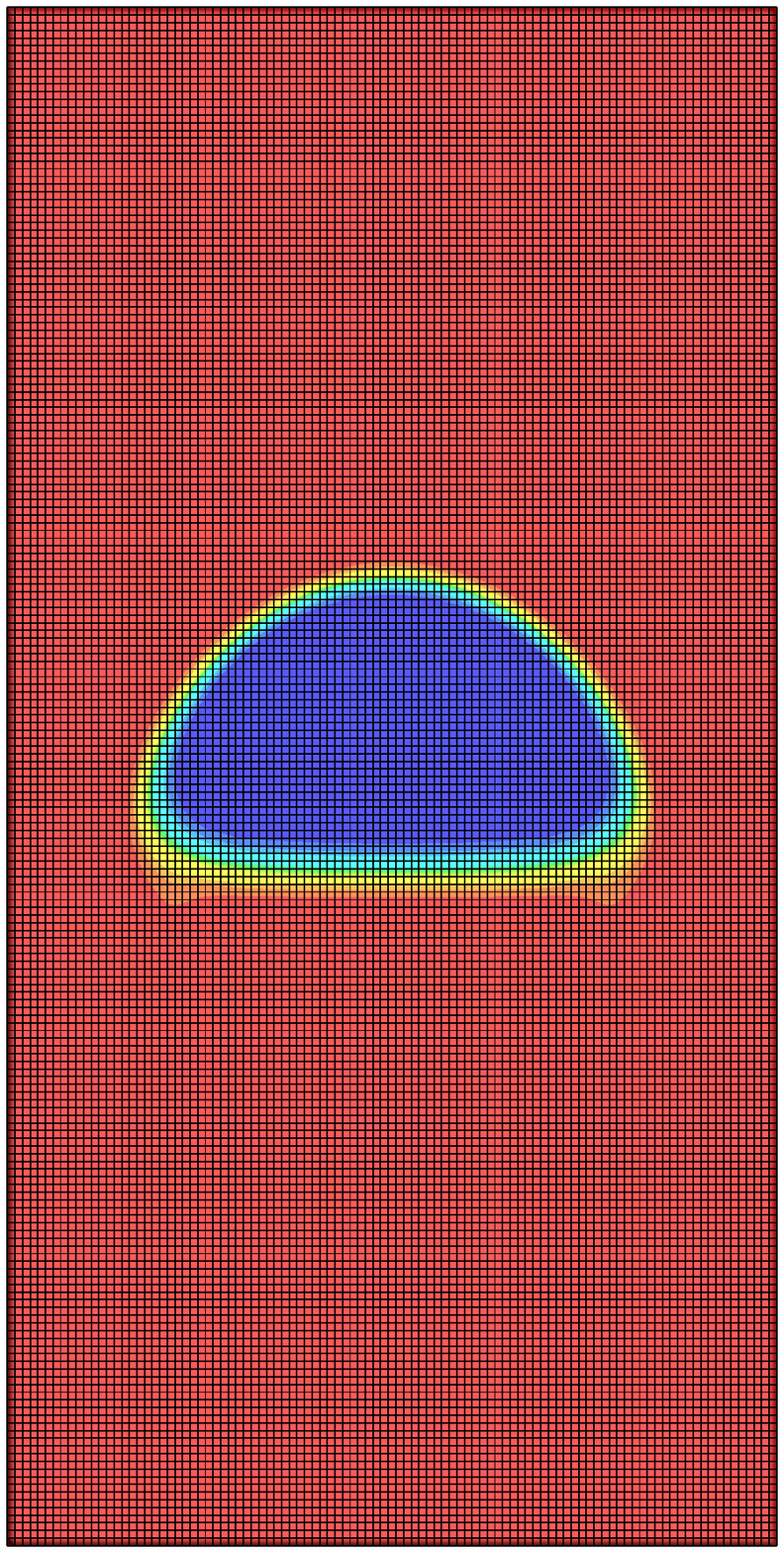}
		\caption*{(a)}
	\end{minipage}
\hspace{-1cm}
	\begin{minipage}[b]{0.45\textwidth}
		\centering
		\includegraphics[scale=0.5,trim=3 3 3 3,clip]{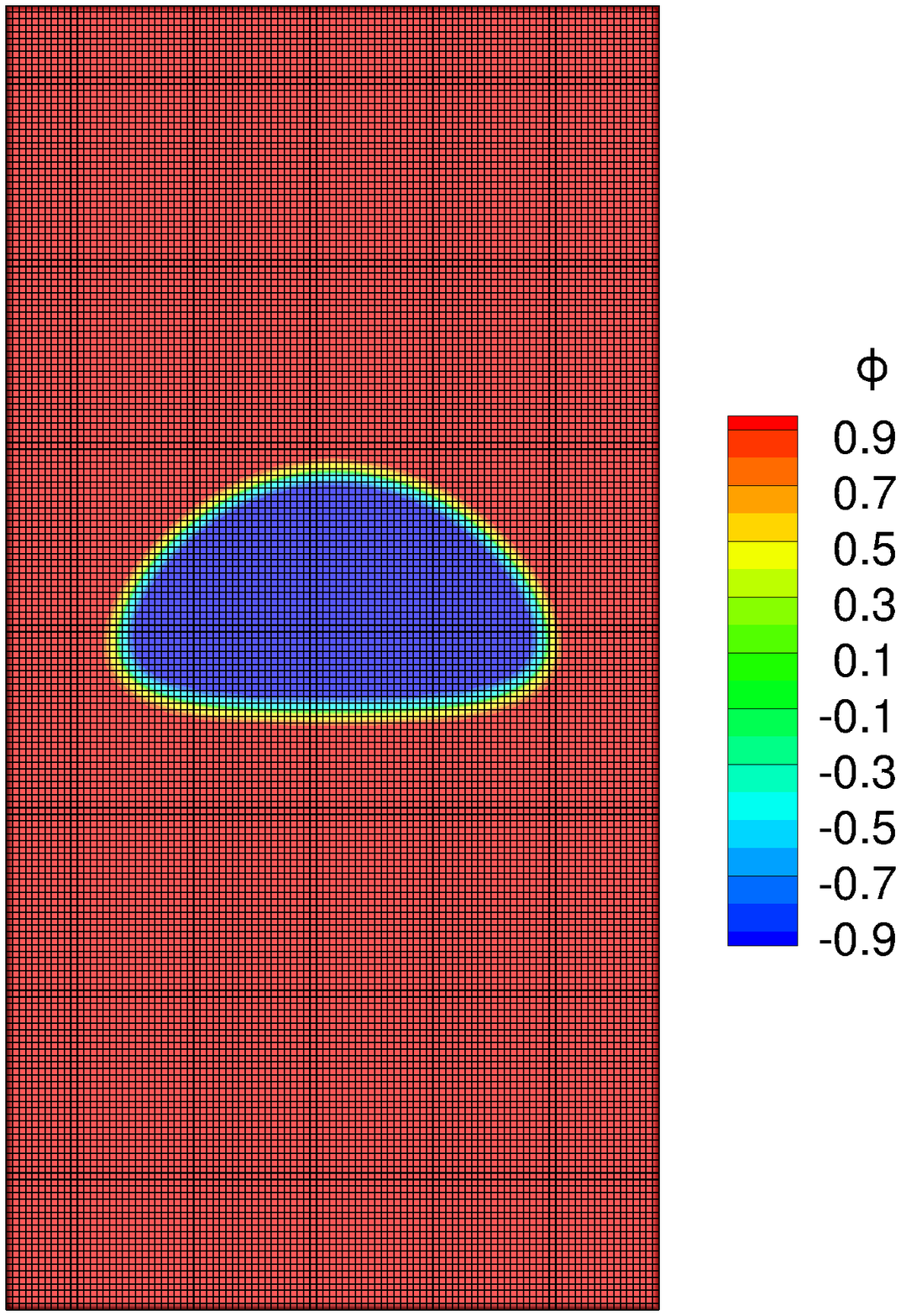}
		\caption*{\hspace{3cm}(b)}
	\end{minipage}

\hspace{-1cm}
	\begin{minipage}[b]{0.5\textwidth}
		\centering
	\includegraphics[scale=0.57]{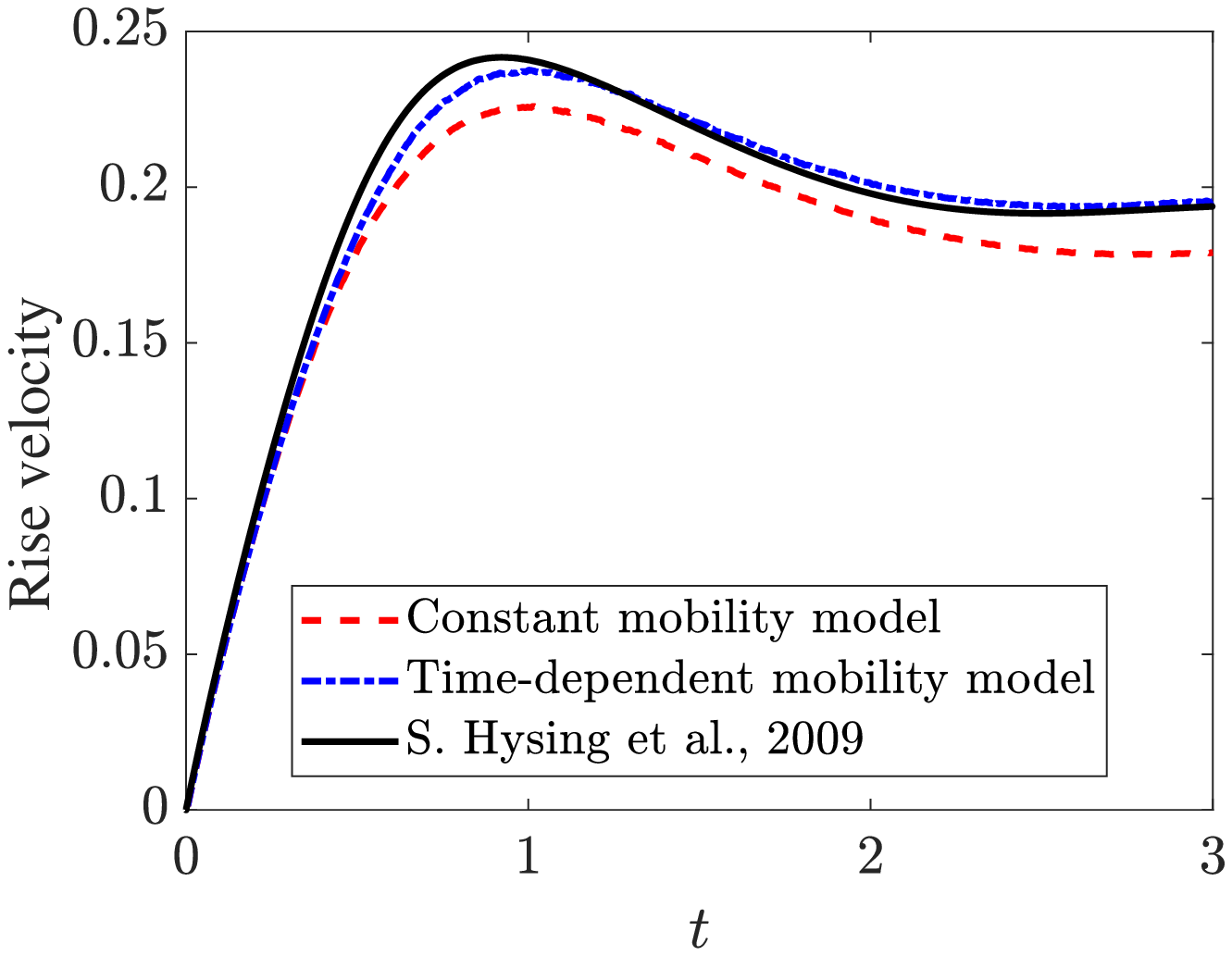}
	\caption*{\hspace{1.2cm}(c)}
	\end{minipage}
\hspace{-0.3cm}
	\begin{minipage}[b]{0.5\textwidth}
		\centering
	\includegraphics[scale=0.57]{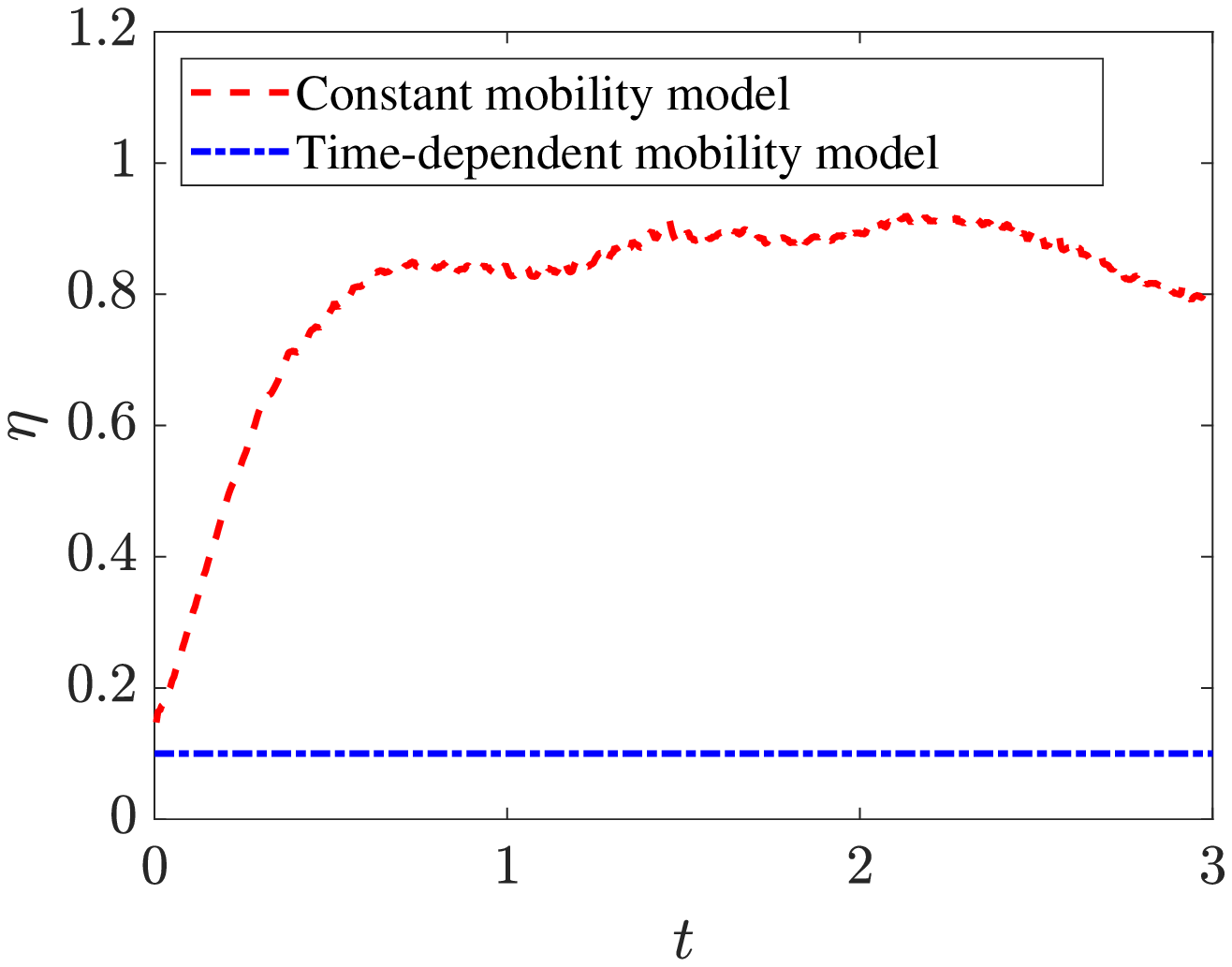}
	\caption*{\hspace{0.95cm}(d)}
	\end{minipage}
	\caption{Comparison of the constant and the time-dependent mobility model in 2D rising bubble case: contour of the order parameter at $t=3$ simulated with (a) constant mobility coefficient $\gamma=1$, (b) time-dependent mobility model at $\eta=0.1$, and the difference in (c) rise velocity, and (d) the time history of the RMS convective distortion parameter $\eta$.}
	\label{infc}
\end{figure*}

The contour of the order parameter at $t=3$ superimposed on the mesh simulated by the constant and the time-dependent mobility are shown in Figs. \ref{infc} (a) and (b),  respectively. As observed in Fig.~\ref{infc} (a), when $\gamma=1$, the interface-preserving capability is not sufficient to keep the interface profile against the convective distortion. Therefore, at the bottom of the bubble, the interface is subjected to an observable extensional distortion, which leads to an excessively low Laplace pressure. This changes the shape of the bubble, decreases the buoyancy force, and further reduces the rise velocity as shown in Fig.~\ref{infc} (c). On the contrary, when the time-dependent mobility model with $\eta=0.1$ is used, the interface profile is preserved well as shown in Fig.~\ref{infc} (b), which gives a correct bubble shape and the surface tension force. To further justify the above statements, we quantify the convective distortion by calculating the RMS convective distortion parameter as $\eta=\mathcal{F}(|\zeta(\boldsymbol{x},t)|)/\gamma$. As shown in Fig .\ref{infc} (d), a larger $\eta$ representing insufficient interface-preserving capability is observed in the simulation with the constant mobility compared to the time-dependent mobility. The comparison shows that the proposed time-dependent model provides an approach to estimate as well as control the convective distortion.

\subsection{Three-dimensional rising bubble benchmark}
\begin{figure}
  \centering
    \begin{minipage}[b]{0.43\textwidth}
      \centering
    \begin{tikzpicture}[xscale=3.3, yscale=3.3,line width=0.6]

\draw[top color=gray!20,bottom color=gray!50, opacity = 0.5,draw=none]  (0,0) -- (0,2)-- (0.2,2.2)-- (1.2,2.2)-- (1.2,0.2)--(1,0)--(0,0);
 \draw [dashed] (0.2,0.2) -- (0.2,2.2);
 \draw [dashed] (0.2,0.2) -- (1.2,0.2);
 \draw [dashed] (0,0) -- (0.2,0.2);
 \draw (0.2,2.2) -- (1.2,2.2);
 \draw (1.2,0.2) -- (1.2,2.2);
 \draw (1,0)--(1.2,0.2);
\draw (0,2)--(0.2,2.2);
\draw (1,2)--(1.2,2.2);
\draw (0,0) rectangle (1,2);
\draw (0.6,0.6) circle [radius=0.25];

\draw [-stealth,thick] (0,0) -- (0.1,0.1);
\node [right] at (0.1,0.1) {X};
\draw[-stealth,thick] (0,0) -- (0.141,0);
\node [below] at (0.141,0) {Z};
\draw[-stealth,thick] (0,0) -- (0,0.141);
\node [left] at (0,0.141) {Y};

  \draw[fill=white,opacity=0.75] (0.6,0.6) circle (0.25);
  \draw (0.35,0.6) arc (180:360:0.25 and 0.1);
  \draw[dashed] (0.85,0.6) arc (0:180:0.25 and 0.1);
  \fill[fill=black] (0.6,0.6) circle (0.5pt);

\node at (0.7,1) {$\Omega_2$($\rho_2,\mu_2$)};
\draw (0.7 ,0.9)--(0.65,0.75);
\node [above] at (0.6,1.6) {$\Omega_1$};
\node [below] at (0.6,1.6) {($\rho_1,\mu_1$)};
\draw [-stealth] (0.6,0.6)--(0.4,0.73);
\node [above left] at (0.38,0.73) {$R$};

\node [above] at (0.7,2.3) {1};
\node [right] at (1.3,1.2) {2};
\draw [stealth-stealth] (1.3,0.2) -- (1.3,2.2);
 \draw [dotted,line width = 0.9] (1,0) -- (1.1,0);
\draw [stealth-stealth] (0.2,2.3) -- (1.2,2.3);
\draw [stealth-stealth] (1.1,0) -- (1.3,0.2);
\node [below right] at (1.2,0.1) {1};

\draw [dotted,line width = 0.9] (0.2,2.2) -- (0.2,2.3);
\draw [dotted,line width = 0.9] (1.2,2.2) -- (1.2,2.3);
\draw [dotted,line width = 0.9] (1.2,0.2) -- (1.3,0.2);
\draw [dotted,line width = 0.9] (1.2,2.2) -- (1.3,2.2);



  \end{tikzpicture}
   \caption*{(a)}
  \end{minipage}
  \hspace{0pt}
  \begin{minipage}[b]{0.495\textwidth}
  \centering
    \includegraphics[trim=5 5 5 5,scale=0.43,clip]{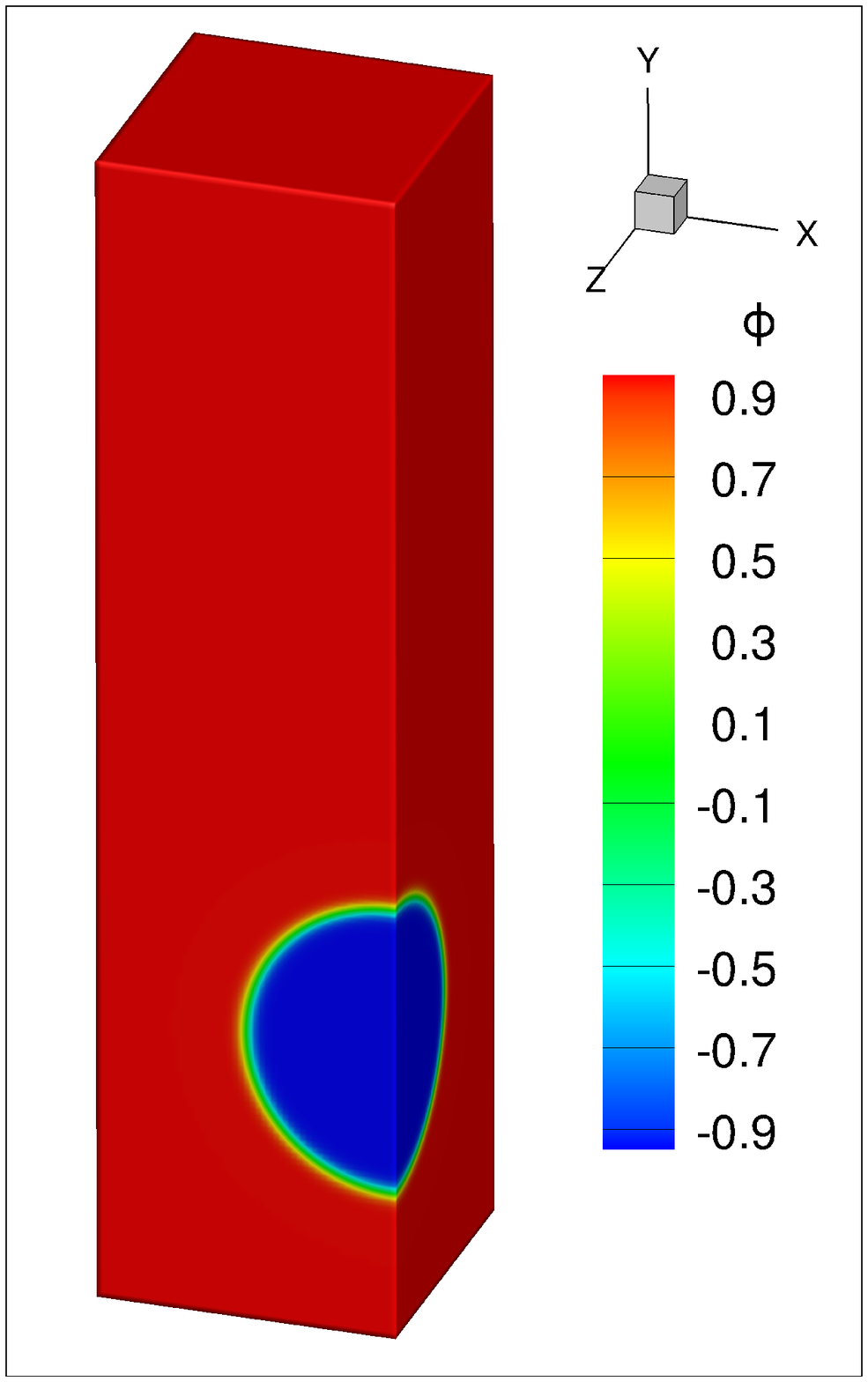}
    \caption*{(b)}
  \end{minipage}
  \caption{Three-dimensional rising bubble problem: (a) schematic diagram of the computational domain, and (b) contour of the order parameter at $t=0$.}
  \label{3Dil}
 \end{figure}

 The 3D rising bubble benchmark case is a generalization of the 2D rising bubble case with increasing complexity and practicality due to the 3D topology and motion of the bubble. We use the benchmark case to further assess the Navier-Stokes Allen-Cahn CSF system with the time-dependent mobility model. The benchmark case considers the rising and deforming of an initially spherical bubble in a cuboid tank occupying the spatial domain $\Omega\in[0,1]\times[0,2]\times[0,1]$. The phase-field function describing the bubble is initialized as:
 \begin{align}
 \phi(x,y,z,0)=-\tanh\left(\frac{R-\sqrt{(x-x_c)^2+(y-y_c)^2+(z-z_c)^2}}{\sqrt{2}\varepsilon}\right),
 \end{align}
 where $R=0.25$ is the radius of the bubble with its center at $(x_c,y_c,z_c)=(0.5,0.5,0.5)$. The no-slip boundary condition and the zero flux Neumann boundary condition are imposed on all the boundaries for the velocity and the order parameter respectively. The density and the viscosity of the fluid and the bubble are taken as $\rho_1=1000$, $\rho_2=100$, $\mu_1=10$, $\mu_2=1$. The surface tension coefficient is taken as $\sigma=24.5$.  The gravitational acceleration is set to be $\boldsymbol{g}=(0,-0.98,0)$. The problem setup is illustrated in Fig.~\ref{3Dil} (a). The 3D rising bubble benchmark case is investigated in \cite{adelsberger20143d} by several research groups. We consider the data from the second group in \cite{adelsberger20143d} for comparison purposes, in which the finite difference method and the sharp level-set method are employed for the discretization and the interface capturing respectively.
 As the case is symmetric with respect to planes $x=0.5$ and $z=0.5$, we simulate one quarter of the computational domain, as shown in Fig.~\ref{3Dil} (b) with the symmetric boundary condition imposed on the symmetric planes. The computational domain is discretized with a uniform structured mesh of grid size $\Delta x=\Delta y=\Delta z=h$. The mesh resolution at the interface is selected as $\varepsilon/h=1$, while the time step is taken as $\Delta t=0.005$.  

 We define the following variables to assess the simulation results quantitatively: the total mass of the order parameter m, the sphericity of the bubble $\slashed{s}$, the diameter of the bubble in X direction $D_x$ and Y direction $D_y$, the rise velocity of the bubble $V_b$ and the center of mass of the bubble $Y_b$, which are given by:
\begin{align}
	m&=\int_{\Omega}\phi d\Omega,\nonumber\\
   \slashed{s}&=A_a/A_b,\nonumber\\
  D_x&=\max(\{x|x\in\Omega_2\})-\min(\{x|x\in\Omega_2\}),\nonumber\\
   D_y&=\max(\{y|y\in\Omega_2\})-\min(\{y|y\in\Omega_2\}),\nonumber\\
    V_{b}&=\frac{\int_{\Omega_2} v d\Omega}{\int_{\Omega_2} 1 d\Omega},\nonumber\\
    Y_b&=\frac{\int_{\Omega_2} y d\Omega}{\int_{\Omega_2} 1 d\Omega},\nonumber
\end{align}
 where  $A_a$ is the area of the sphere which has the same volume as the deformed bubble, $A_b$ denotes the surface area of the bubble, $x$ and $y$ are the coordinates in X and Y directions respectively, and $v$ is the velocity in the Y direction. The defined variables are compared in the convergence study.

\begin{figure}[h]
	\begin{minipage}[b]{0.5\textwidth}
		\captionsetup{skip=1pt}
		\centering
		\includegraphics[scale=0.55]{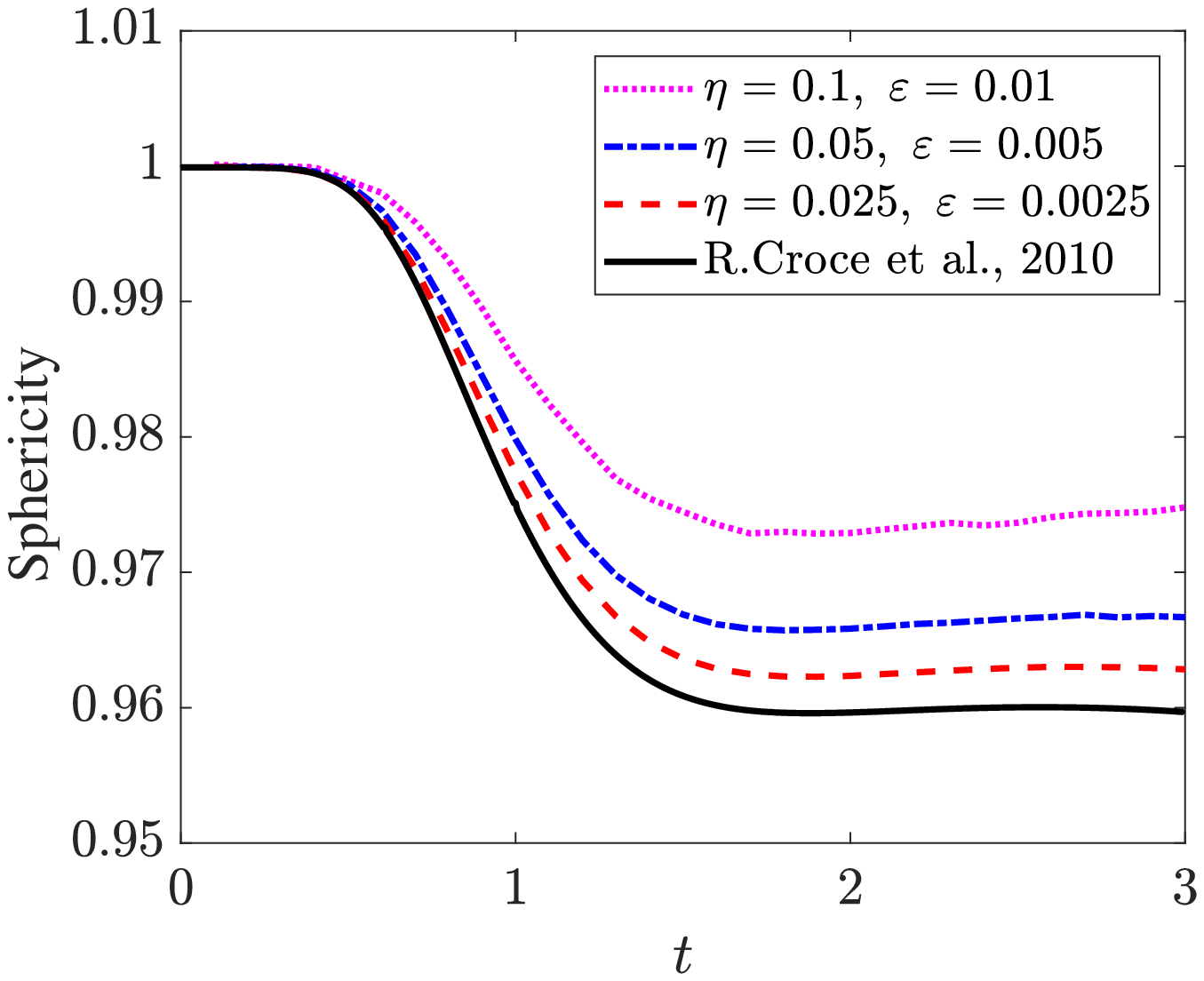}
		\caption*{\hspace{30pt}(a)}
	\end{minipage}
	\hspace{-5pt}
	\begin{minipage}[b]{0.5\textwidth}
		\captionsetup{skip=1pt}
		\centering
		\includegraphics[scale=0.55]{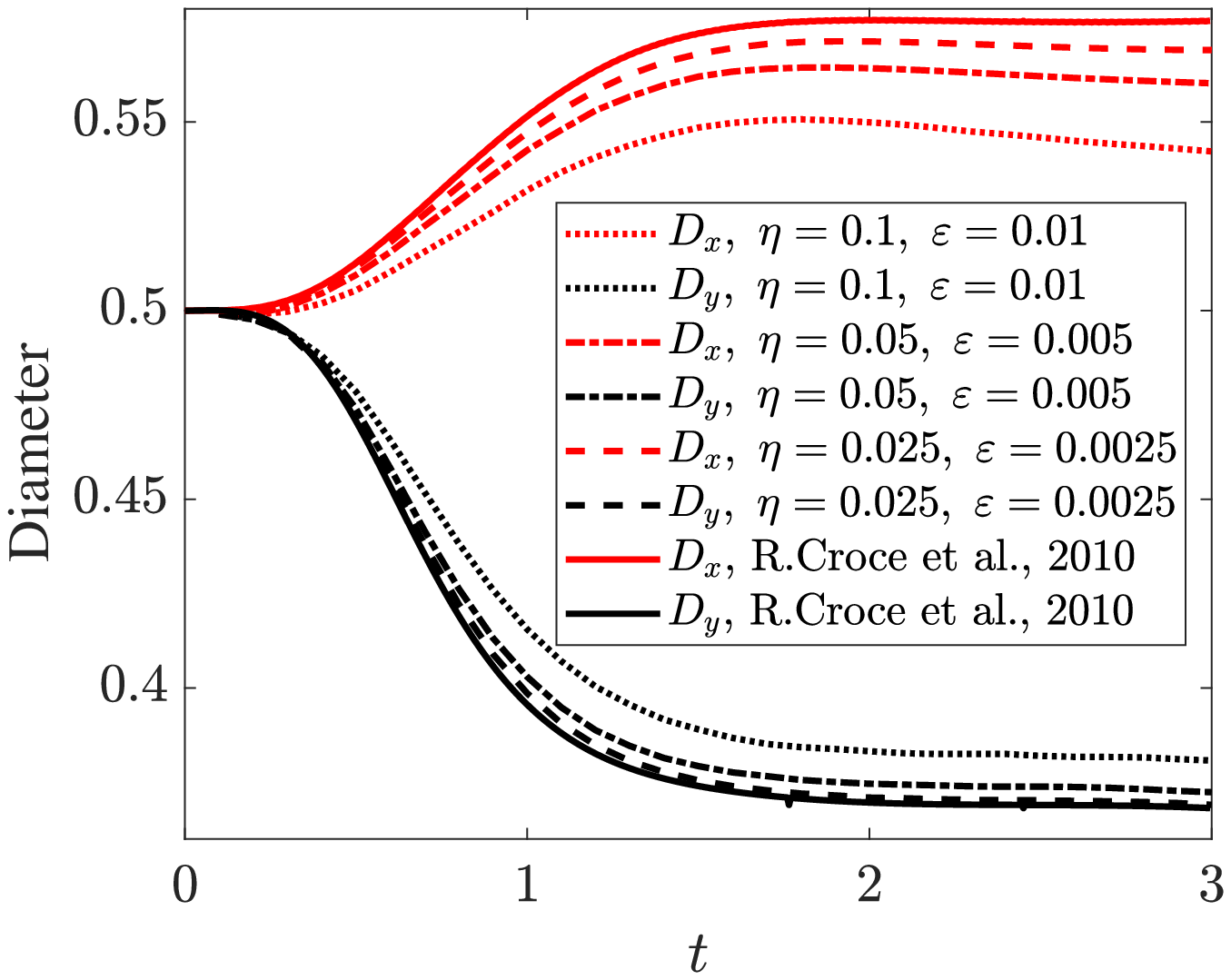}
		\caption*{\hspace{30pt}(b)}
	\end{minipage}
	
	\begin{minipage}[b]{0.5\textwidth}
		\captionsetup{skip=1pt}
		\centering
		\includegraphics[scale=0.55]{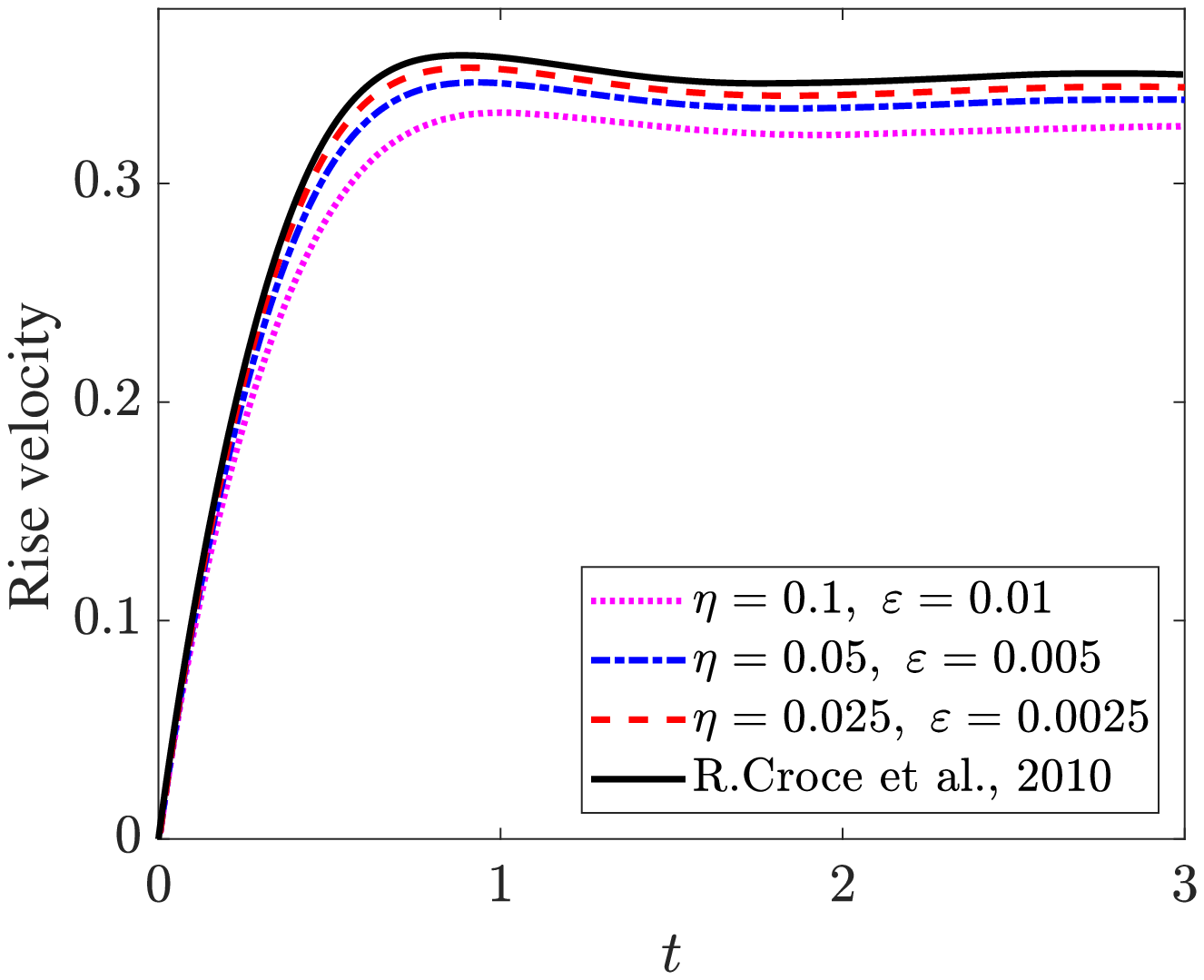}
		\caption*{\hspace{25pt}(c)}
	\end{minipage}
	\hspace{0pt}
	\begin{minipage}[b]{0.5\textwidth}
		\captionsetup{skip=1pt}
		\centering
		\includegraphics[scale=0.55]{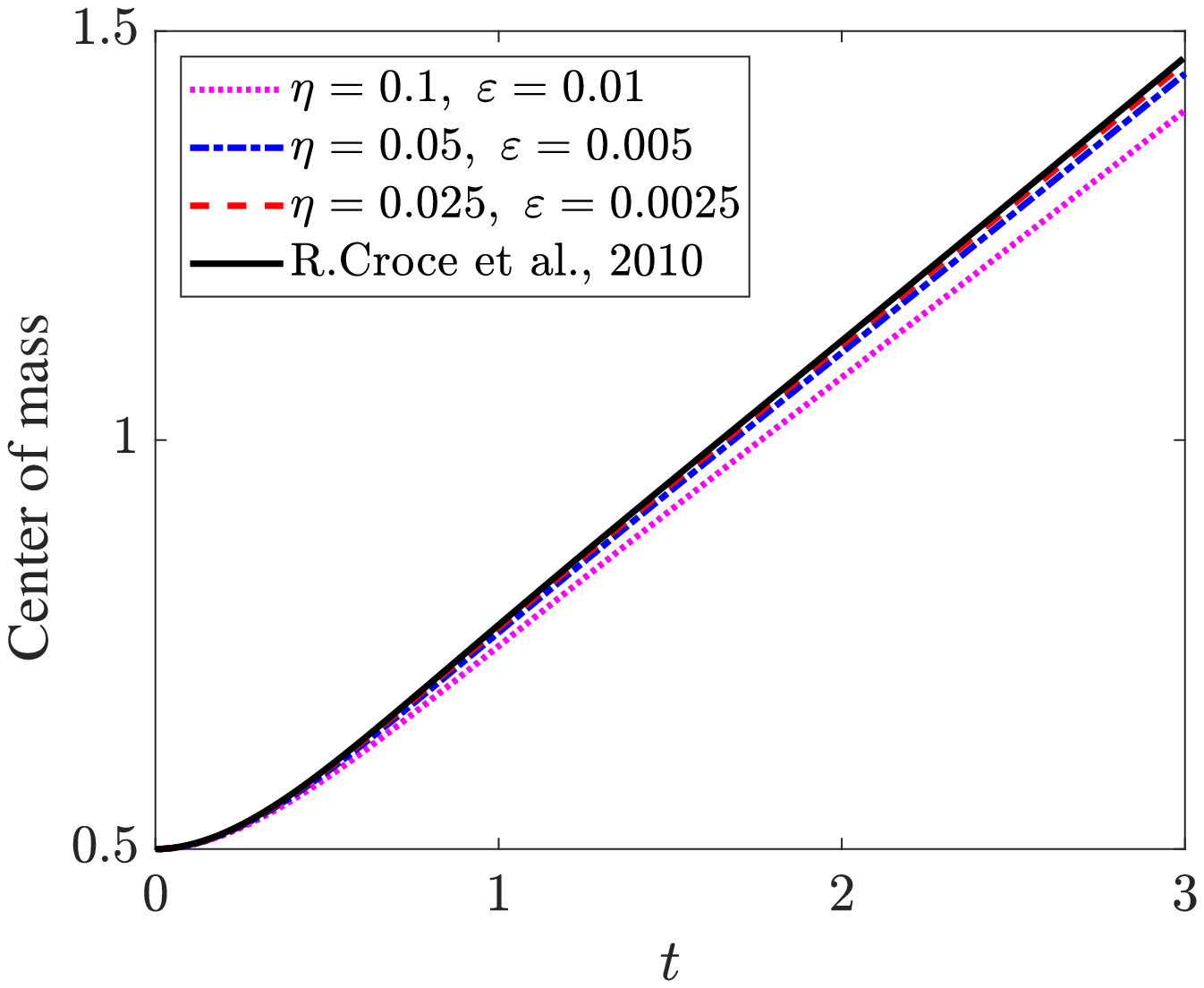}
		\caption*{\hspace{25pt}(d)}
	\end{minipage}
	
    \caption{Convergence study for the 3D bubble rising problem: (a) sphericity of the bubble, (b) bubble diameters in X and Y directions,
 (c) rise velocity, and (d) center of mass.} 
\label{3dgc}
\end{figure}

Similar to the convergence study of the 2D rising bubble case, we decrease $\eta$ and $\varepsilon$ to improve the interface preservation capability and decrease the volume-preserving mean curvature flow simultaneously.The following combinations are tested: $\eta=0.1,\varepsilon=0.01$; $\eta=0.05,\varepsilon=0.005$ and $\eta=0.025,\varepsilon=0.0025$. The scaling factor of the volume-preserving mean curvature flow $\gamma(t)\varepsilon^2$ at $t=3$ for each combination is tabulated in Table \ref{3Dip errors}. The defined variables and their comparison to the data in \cite{adelsberger20143d} are shown in Fig.~\ref{3dgc}.

To further quantify the results, we define the mass conservation error and convergence errors as:
\begin{align}
e_m&=\left|\frac{m_{t=0}-m_{t=3}}{m_{t=0}}\right|,\nonumber\\
e_{\slashed{s}}&=\frac{||\slashed{s}-\slashed{s}_{\mathrm{ref}}||_2}{||\slashed{s}_{\mathrm{ref}}||_2},\nonumber\\
e_{D_x}&=\frac{||D_x-D_{x,\mathrm{ref}}||_2}{||D_{x,\mathrm{ref}}||_2},\nonumber\\
e_{D_y}&=\frac{||D_y-D_{y,\mathrm{ref}}||_2}{||D_{y,\mathrm{ref}}||_2},\nonumber\\
e_{V_b}&=\frac{||V_b-V_{b,\mathrm{ref}}||_2}{||V_{b,\mathrm{ref}}||_2},\nonumber\\
e_{Y_b}&=\frac{||Y_b-Y_{b,\mathrm{ref}}||_2}{||Y_{b,\mathrm{ref}}||_2}.\nonumber
\end{align}
The simulation results of $\eta=0.025$, $\varepsilon=0.0025$ are taken as the reference. The errors including the relative interface thickness and surface tension force errors are tabulated in Table \ref{3Dip errors}.

\begin{table}[h]
    \centering
        \caption{Quantification of the errors for the 3D rising bubble benchmark}
    \begin{tabular}{cccccccccccc}
    \hline
        $\eta$&$\varepsilon $&$e_{\varepsilon}$&$e_{\sigma}$&$\gamma(t)\varepsilon^2\ (t=3)$&$e_m$&$e_{\slashed{s}}$& $e_{D_x}$&$e_{D_y}$&$e_{V_b}$&$e_{Y_b}$\\
         \hline
         $0.1$&$0.01$&$0.074$&$0.064$&$1.5\times10^{-3}$&$0.0063$ &0.0093& 0.0340& 0.0302& 0.0614&0.0348\\

         $0.05$&$0.005$&$0.037$&$0.032$&$8.8\times10^{-4}$&$0.0034$& 0.0030& 0.0110& 0.0085& 0.0197&0.0083\\
         
         $0.025$&$0.0025$&$0.018$&$0.016$&$5.0\times10^{-4}$&$0.0016$& -& -& -& -&-\\
         \hline
    \end{tabular}
    \label{3Dip errors}
\end{table}

It can be observed that the total mass is well conserved with less than $1\%$ relative error. The relative interface thickness and surface tension force errors decreases with the decrease in the RMS convective distortion parameter $\eta$. The reduction of the volume-preserving mean curvature flow is reflected by the reduction of its scaling factor $\gamma(t)\varepsilon^2$ at $t=3$ in current cases. As a result, lower sphericity representing larger deformation and higher curvature is observed in Fig.~\ref{3dgc} (a). This is further confirmed in Fig.~\ref{3dgc} (b), where a larger $D_x$ and a smaller $D_y$ representing more deviation from the spherical shape are observed. With both the improvement of the interface-preserving capability and the decrease in the volume-preserving mean curvature flow, the converged simulation results agree well with the data in \cite{adelsberger20143d}.

\section{Two rising bubbles merging with a free surface}
\begin{figure}
  \begin{minipage}{0.5\textwidth}
  \centering

    \begin{tikzpicture}[xscale=4, yscale=4,line width=0.6]

\draw [dashed] (0,1.5) -- (0.2,1.7);
\draw [dashed] (0.2,1.7) -- (1.2,1.7);
\draw (1.2,1.7) -- (1,1.5);
\draw (1,1.5) -- (0,1.5);
\draw[top color=gray!20,bottom color=gray!50, opacity = 0.5,draw=none]  (0,0) -- (0,1.5)-- (0.2,1.7)-- (1.2,1.7)-- (1.2,0.2)--(1,0)--(0,0);

 \draw [dashed] (0.2,0.2) -- (0.2,2.2);
 \draw [dashed] (0.2,0.2) -- (1.2,0.2);
 \draw [dashed] (0,0) -- (0.2,0.2);
 \draw (0.2,2.2) -- (1.2,2.2);
 \draw (1.2,0.2) -- (1.2,2.2);
 \draw (1,0)--(1.2,0.2);
 \draw (0,2)--(0.2,2.2);
 \draw (1,2)--(1.2,2.2);
 \draw (0,0) rectangle (1,2);
 \draw (0.6,0.6) circle [radius=0.25];
 \draw (0.6,1.1) circle [radius=0.2];



\node at (0.6,1.38) {$\Omega_1$($\rho_1,\mu_1$)};

\node [above] at (0.6,1.83) {$\Omega_2$};
\node [below] at (0.6,1.85) {($\rho_2,\mu_2$)};

\draw (0.55,1.6)--(0.65,1.6);
\draw(0.57,1.57)--(0.63,1.57);
\draw (0.6,1.6)--(0.63,1.65)--(0.57,1.65)--(0.6,1.6);






 \draw[fill = white,opacity=0.75] (0.6,0.6) circle (0.25);
 \draw [-stealth] (0.6,0.6)--(0.39,0.73);
 \node[above left] at (0.42,0.73) {$R_l$};
\draw (0.35,0.6) arc (180:360:0.25 and 0.1);
\draw[dashed] (0.85,0.6) arc (0:180:0.25 and 0.1);
\fill[fill=black] (0.6,0.6) circle (0.5pt);

 \draw[fill = white,opacity=0.75] (0.6,1.1) circle (0.2);
 \draw [-stealth] (0.6,1.1)--(0.42,1.20);
  \node[above left] at (0.42,1.20) {$R_u$};
\draw (0.4,1.1) arc (180:360:0.2 and 0.1);
\draw[dashed] (0.8,1.1) arc (0:180:0.2 and 0.1);
\fill[fill=black] (0.6,1.1) circle (0.5pt);
\draw [-stealth,thick] (0,0) -- (0.1,0.1);
\node [right] at (0.1,0.1) {X};
\draw[-stealth, thick] (0,0) -- (0.141,0);
\node [below] at (0.141,0) {Z};
\draw[-stealth,thick] (0,0) -- (0,0.141);
\node [left] at (0,0.141) {Y};
  \node [right] at (0.62,1.1) {$\Omega_2$};
\node [right] at (0.64,0.6) {$\Omega_2$};
\draw[stealth-stealth] (1.1,0.1)--(1.1,1.6);
\node[right] at (1.23,0.85) {$y_{wl}$};
\draw [stealth-stealth] (1.45,0.2)--(1.45,2.2);
\draw[dotted,line width=0.9] (1.2,0.2)--(1.45,0.2);
\draw[dotted,line width=0.9] (1.2,2.2)--(1.45,2.2);
\node[right] at (1.5,1.2) {2};
\draw[dotted,line width=0.9] (1.0,0)--(1.25,0);
\draw [stealth-stealth] (1.25,0)--(1.45,0.2);
\node [right] at (1.4,0.06) {1};
\draw[dotted,line width=0.9] (0.2,2.2)--(0.2,2.4);
\draw[dotted,line width=0.9] (1.2,2.2)--(1.2,2.4);
\draw [stealth-stealth] (0.2,2.4)--(1.2,2.4);
\node [above] at (0.7,2.4) {1};

  \end{tikzpicture}
  \caption*{(a)}
    \end{minipage}
\hfill
  \begin{minipage}{0.5\textwidth}
	\centering
	\includegraphics[scale=0.48,trim=1 1 1 1,clip]{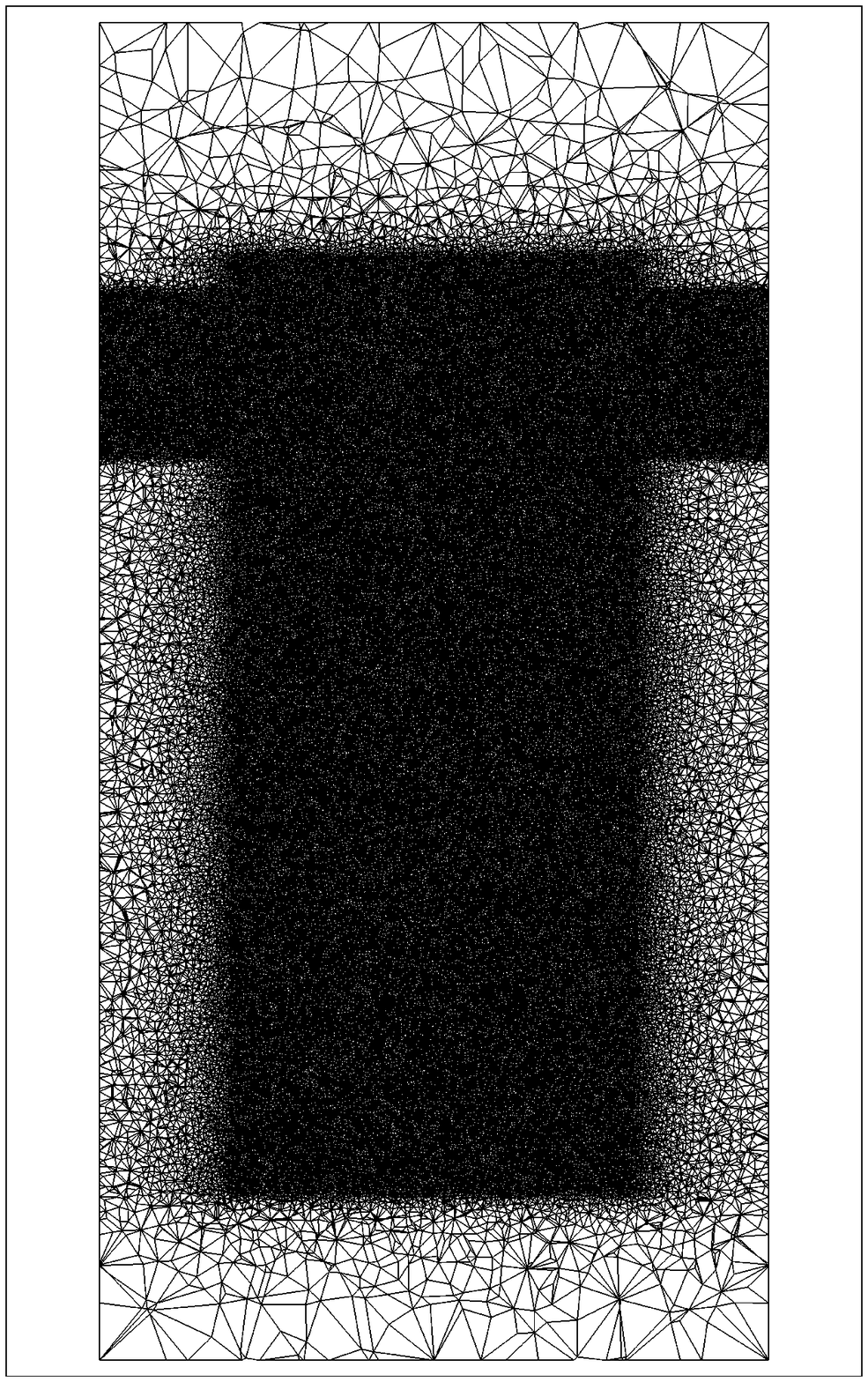}
	  \caption*{(b)}
	 \end{minipage}
\caption{Two rising bubbles merging with a free surface problem: (a) schematic diagram showing the computational domain, and (b) unstructured finite element mesh at the plane $x=0.5$.}
\label{2brf}
\end{figure}

\begin{figure}
  \begin{minipage}[b]{0.32\textwidth}
  \centering
    \includegraphics[scale=0.27,trim=50 2 20 2,clip]{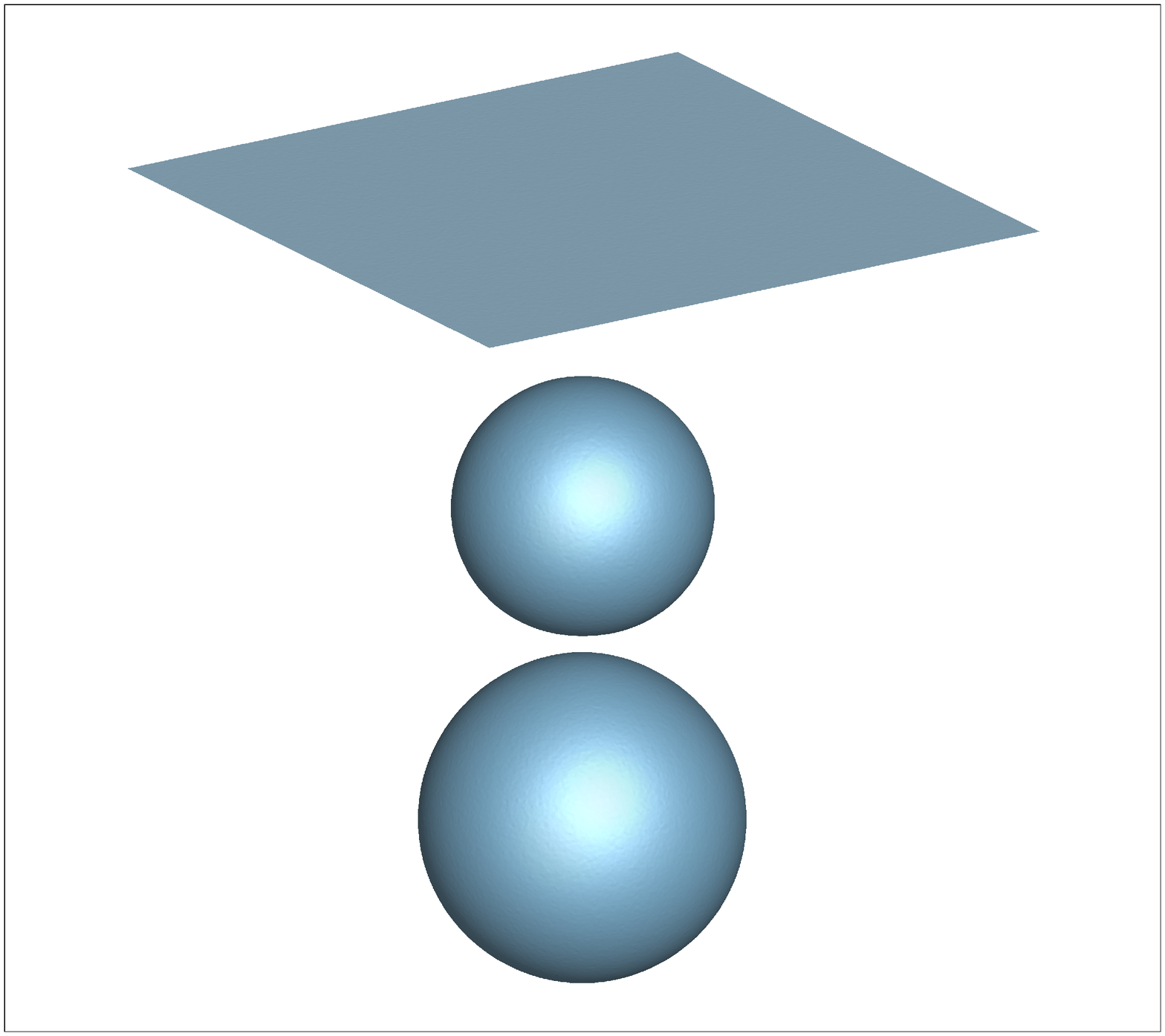}
    \caption*{(a)}
    \end{minipage}
  \begin{minipage}[b]{0.32\textwidth}
    \centering
    \includegraphics[scale=0.27,trim=50 2 20 2,clip]{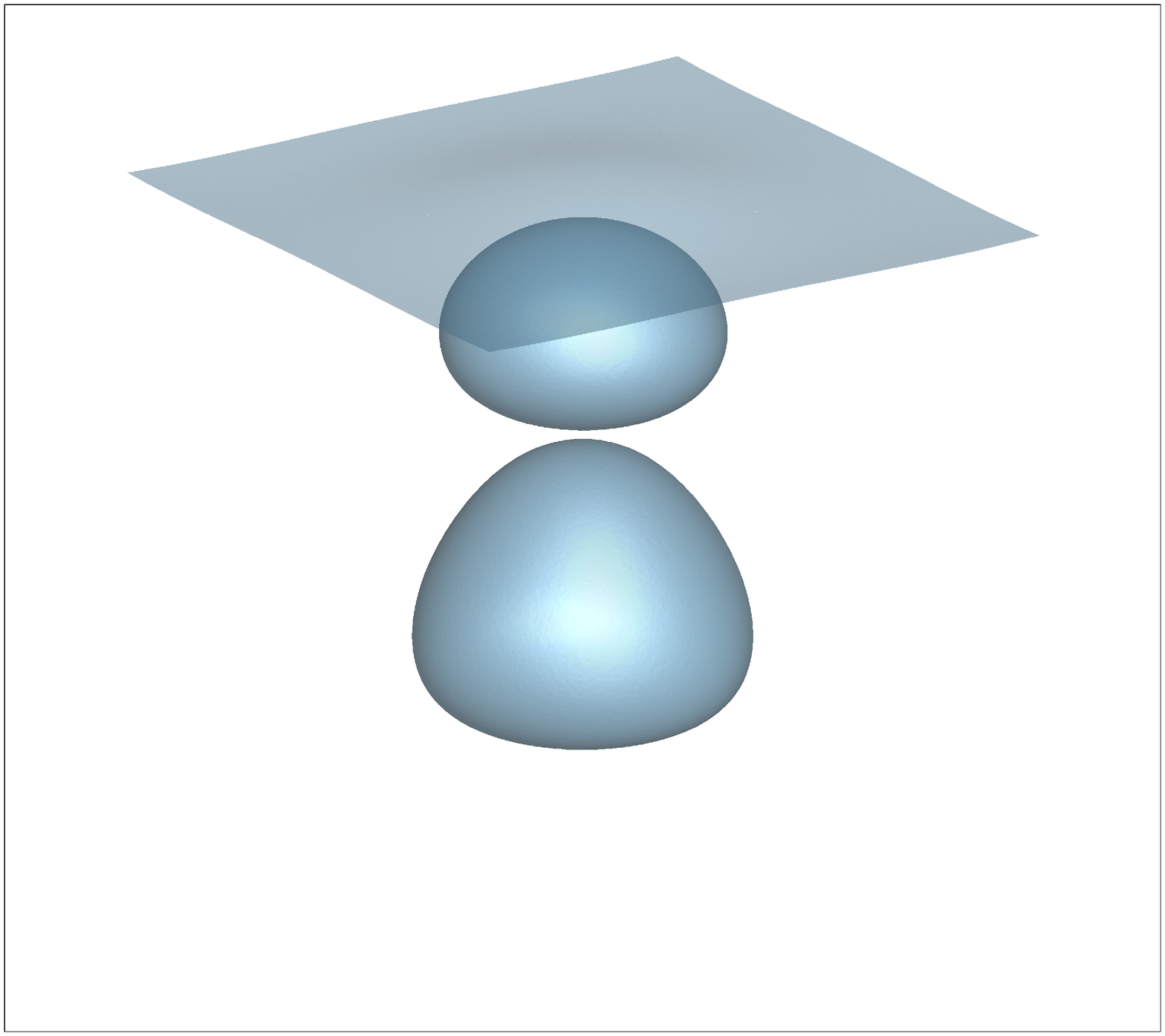}
    \caption*{(b)}
        \end{minipage}
  \begin{minipage}[b]{0.32\textwidth}
    \centering
    \includegraphics[scale=0.27,trim=50 2 20 2,clip]{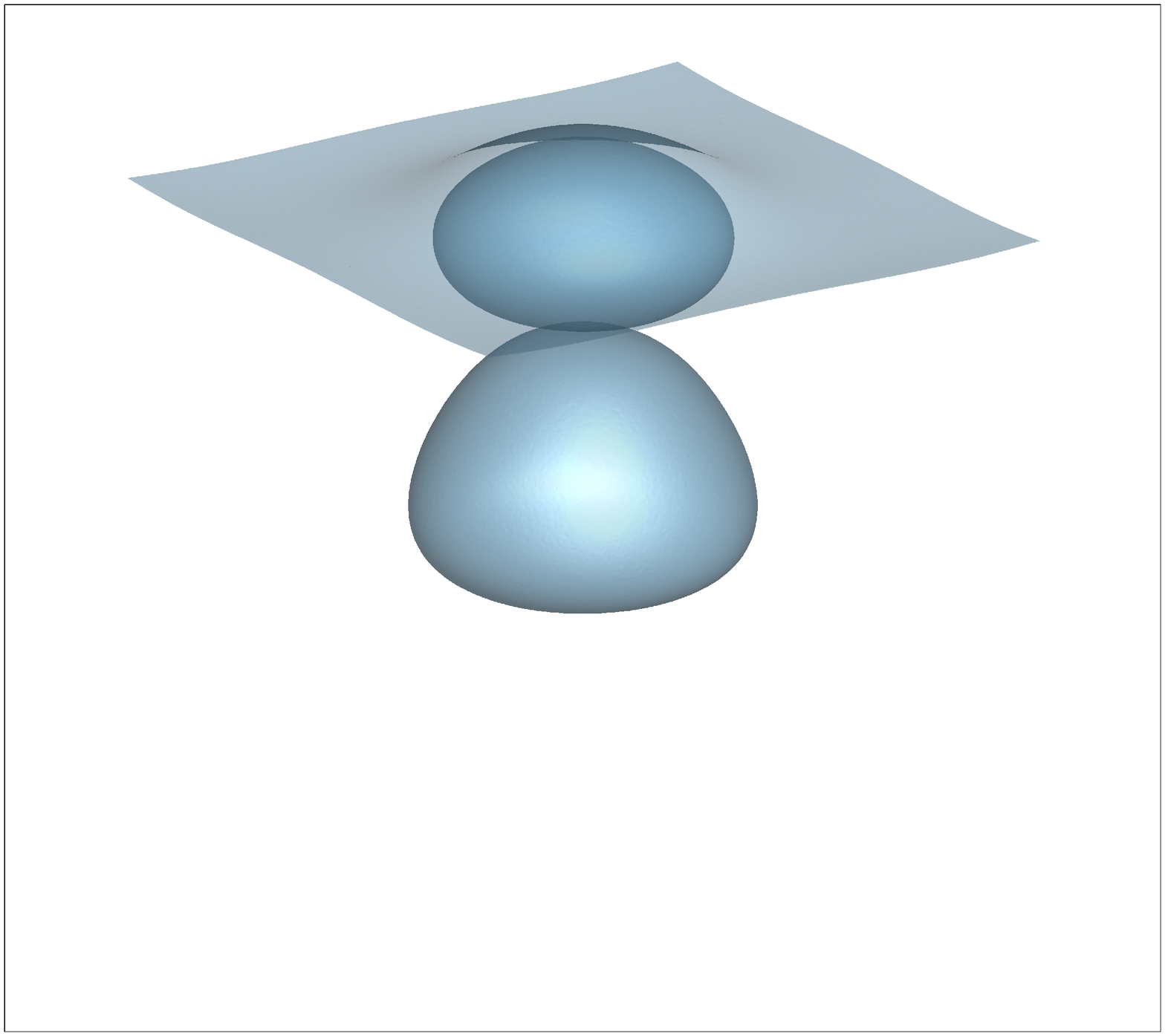}
    \caption*{(c)}
            \end{minipage}

 \begin{minipage}[b]{0.32\textwidth}
  \centering
    \includegraphics[scale=0.27,trim=50 220 20 2,clip]{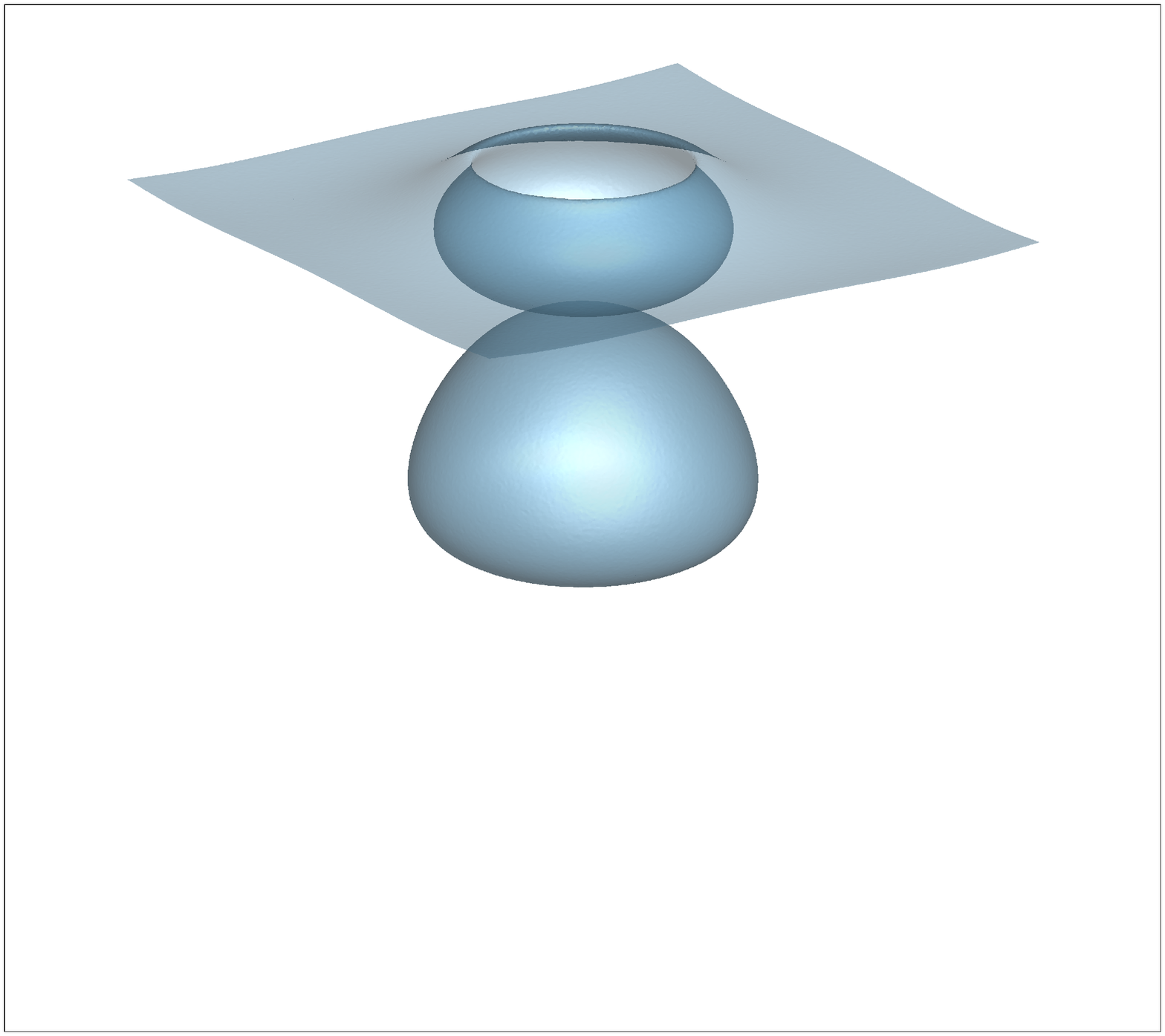}
    \caption*{(d)}
    \end{minipage}
  \begin{minipage}[b]{0.32\textwidth}
    \centering
    \includegraphics[scale=0.27,trim=50 220 20 2,clip]{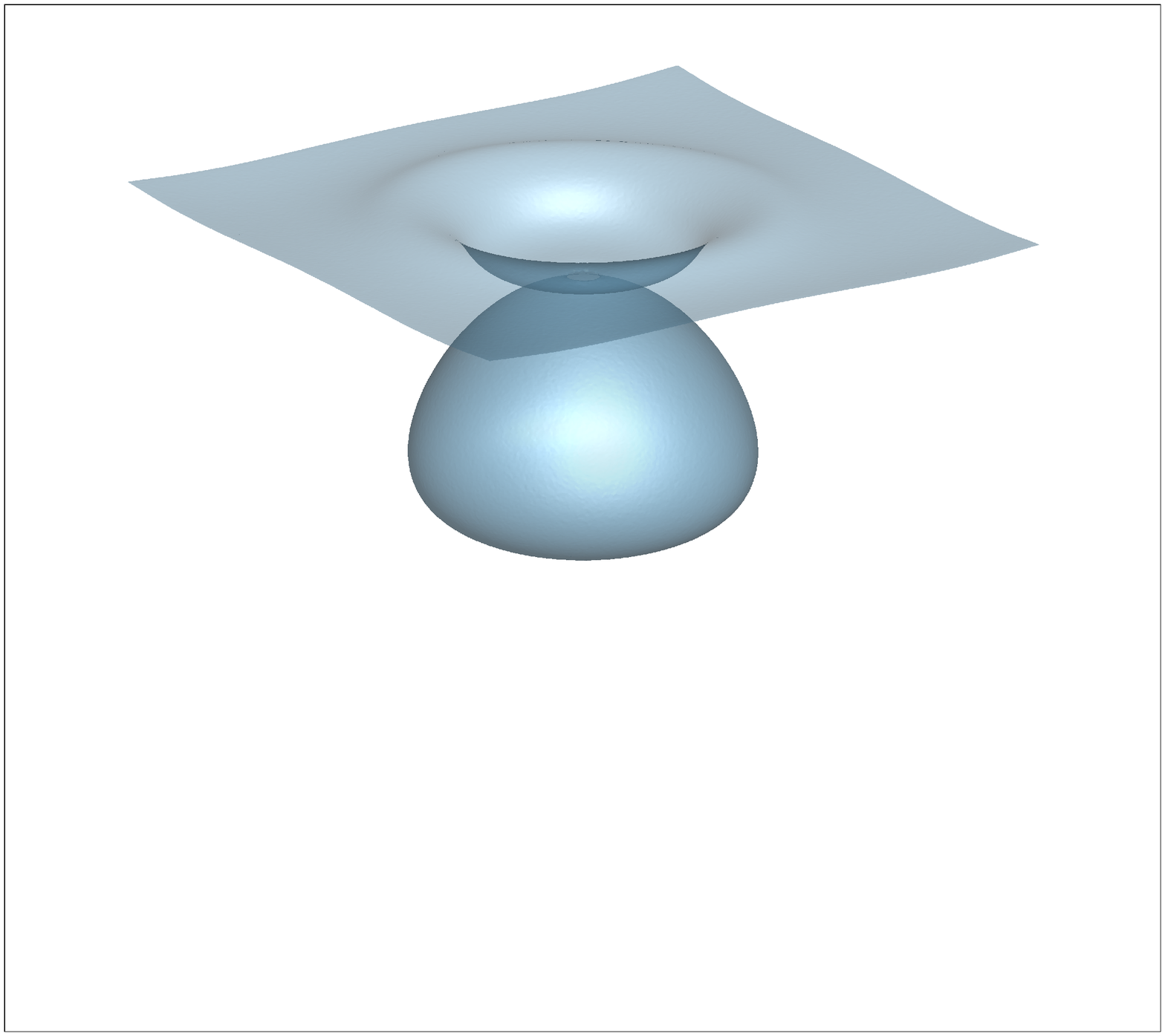}
    \caption*{(e)}
        \end{minipage}
  \begin{minipage}[b]{0.32\textwidth}
    \centering
    \includegraphics[scale=0.27,trim=50 220 20 2,clip]{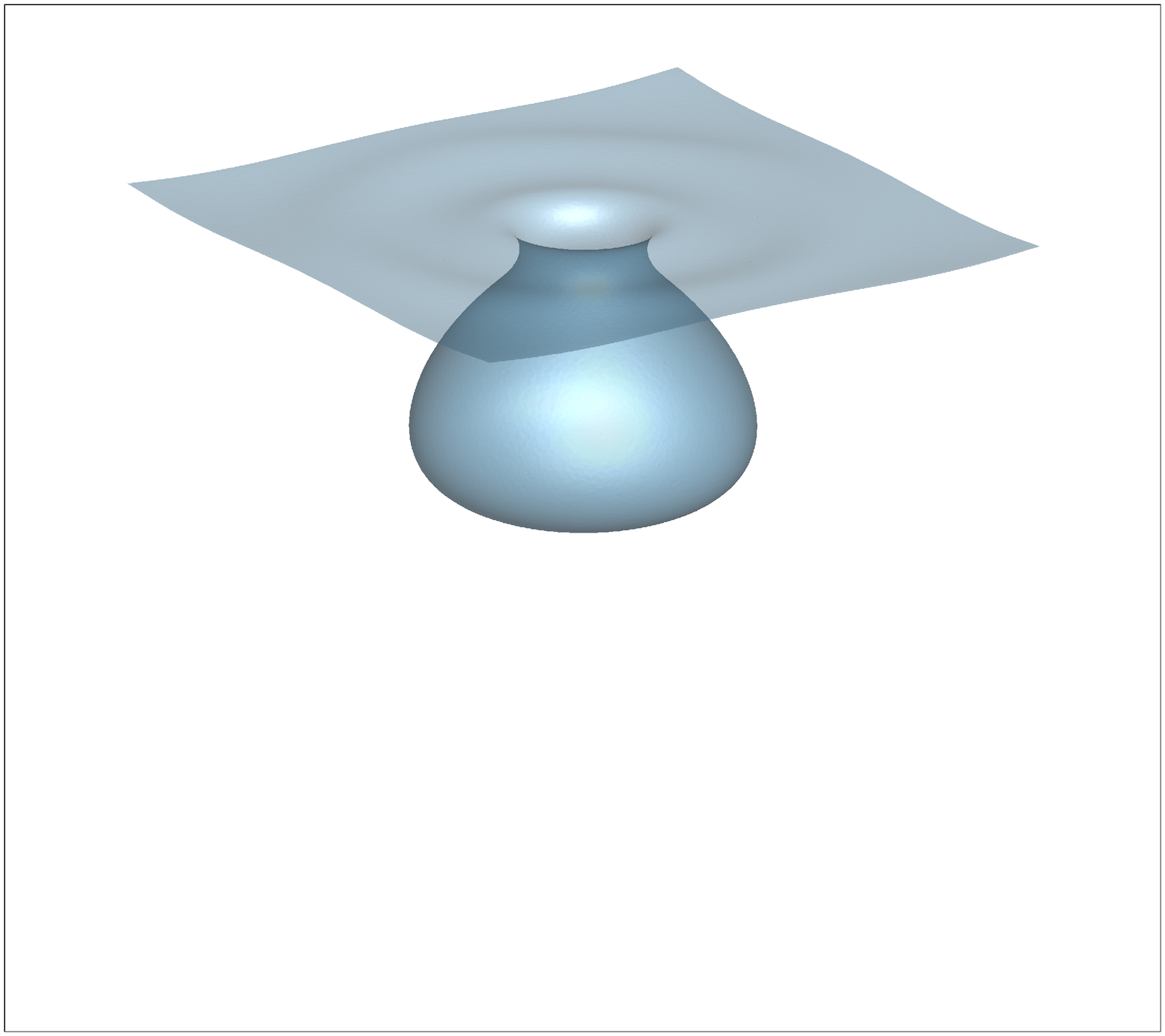}
    \caption*{(f)}
            \end{minipage}

         \begin{minipage}[b]{0.32\textwidth}
  \centering
    \includegraphics[scale=0.27,trim=50 300 20 2,clip]{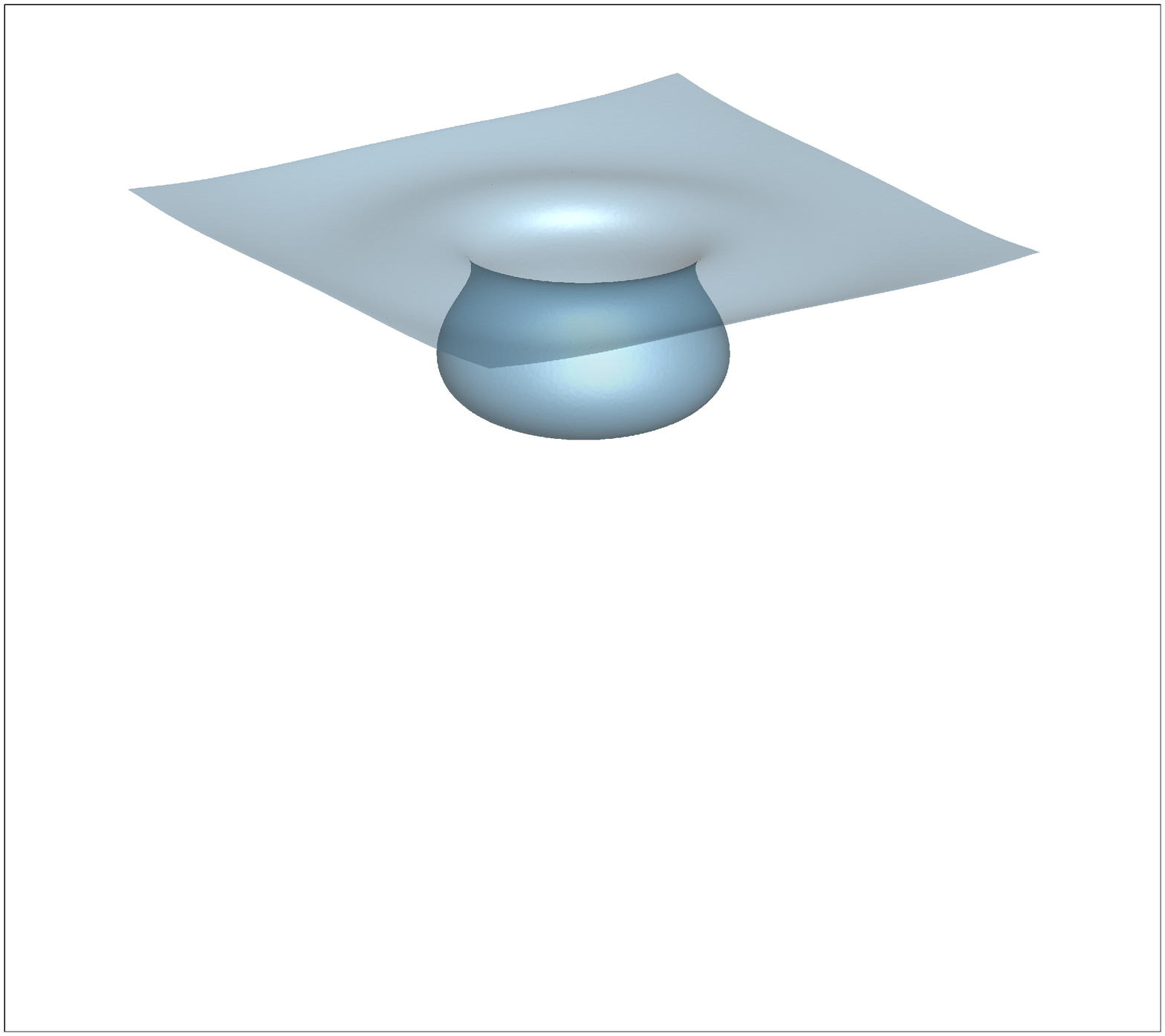}
    \caption*{(g)}
    \end{minipage}
  \begin{minipage}[b]{0.32\textwidth}
    \centering
    \includegraphics[scale=0.27,trim=50 300 20 2,clip]{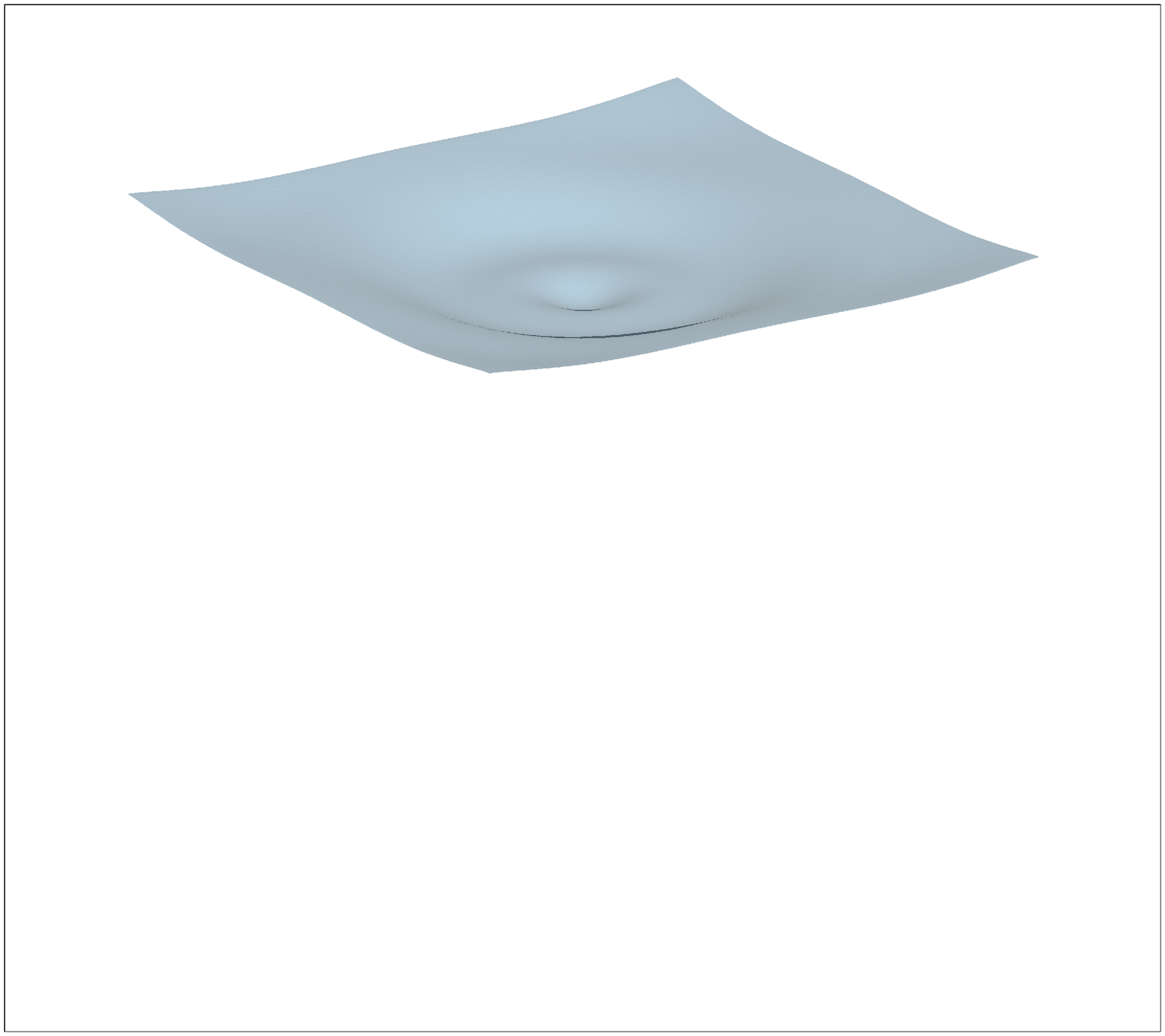}
    \caption*{(h)}
        \end{minipage}
  \begin{minipage}[b]{0.32\textwidth}
    \centering
    \includegraphics[scale=0.27,trim=50 300 20 2,clip]{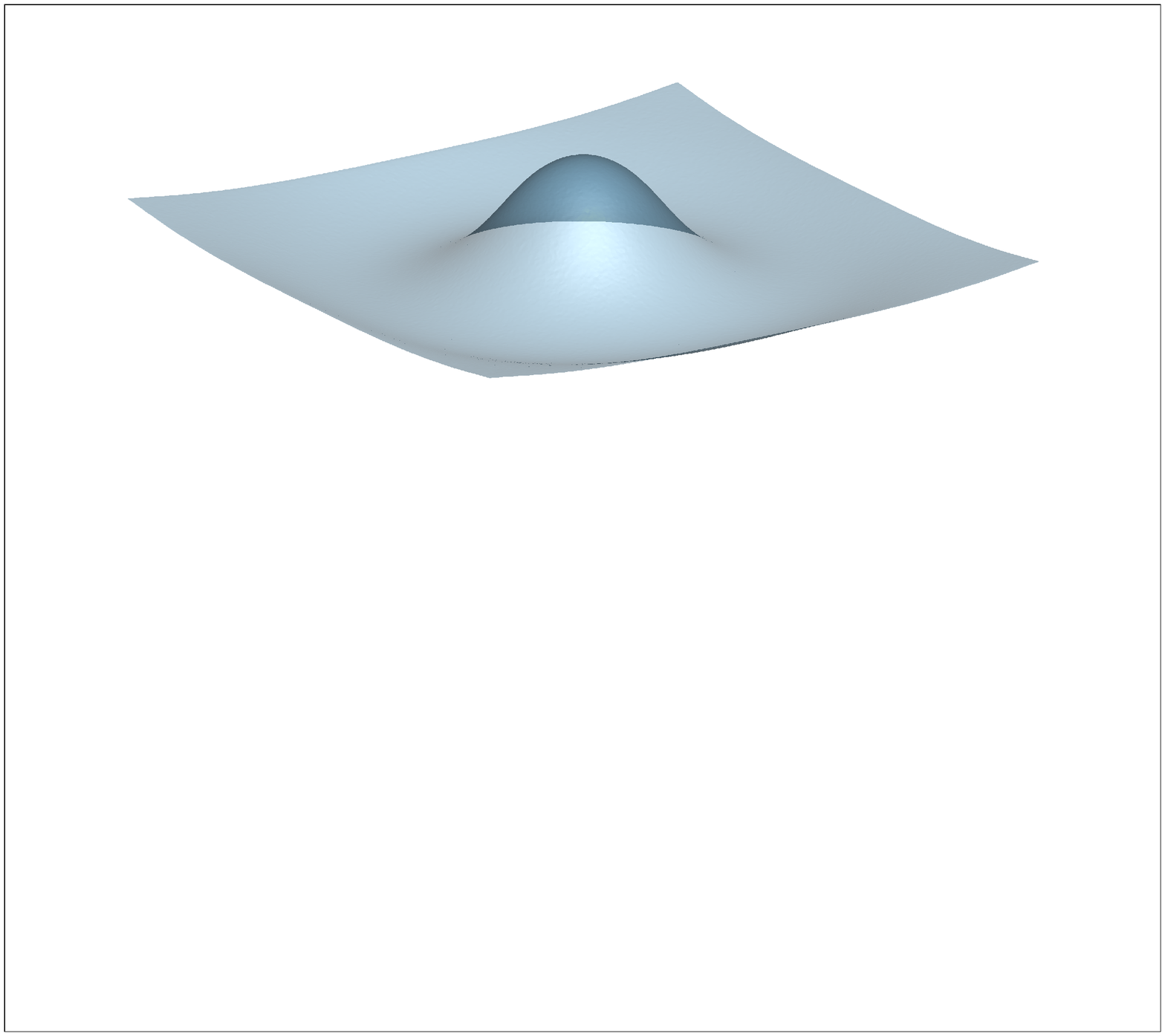}
    \caption*{(i)}
            \end{minipage}

\caption{Two rising bubbles merging with a free surface: the evolution of the interface $\phi=0$ at different time instants $t=$  (a) 0.1, (b) 1.1, (c) 1.6, (d) 1.7, 
(e) 1.8, (f)1.9, (g) 2.2, (h) 2.4, and (i) 2.7.}
\label{of2rbf}
\end{figure}

In this section, we demonstrate the applicability of the proposed model in the case of two rising bubbles merging with a free surface, in which complicated topological changes of the interface and dynamics of the bubble-bubble and bubble-free surface interaction occur in an unstructured finite element mesh. The case considers the rising of two vertically aligned spherical bubbles driven by gravitational force and the merging of bubbles with the free surface in a cuboid tank. The computational domain is taken as $\Omega\in[0,1]\times[0,2]\times[0,1]$. The no-slip boundary condition and the zero flux Neumann boundary condition are applied on all the boundaries for the velocity and the order parameter, respectively. The order parameter is initialized for the bubbles and the free surface as:
\begin{align}
\phi(x,y,z,0)=
&-\tanh\left(\frac{R_{l}-\sqrt{\left(x-x_{lc}\right)^2+\left(y-y_{lc}\right)^2+\left(z-z_{lc}\right)^2}}{\sqrt{2}\varepsilon}\right)\nonumber\\
&-\tanh\left(\frac{R_{u}-\sqrt{\left(x-x_{uc}\right)^2+\left(y-y_{uc}\right)^2+\left(z-z_{uc}\right)^2}}{\sqrt{2}\varepsilon}\right)\nonumber\\
&-\tanh\left(\frac{y-y_{wl}}{\sqrt{2}\varepsilon}\right)-2,
\end{align}
where $R_{l}=0.25$ is the radius of the lower bubble with its center at $(x_{lc},y_{lc},z_{lc})=(0.5,0.5,0.5)$, $R_{u}=0.2$ is the radius of the upper bubble with its center at $(x_{uc},y_{uc},z_{uc})=(0.5,1,0.5)$, $y_{wl}=1.5$ is the water level of the free surface.
The density and viscosity of the fluid and the bubbles are taken as $\rho_1=1000$, $\rho_2=100$,  $\mu_1=10$, $\mu_2=1$. The  surface tension coefficient is chosen as $\sigma=24.5$. The gravitational acceleration is set to be $\boldsymbol{g}=(0,-0.98,0)$. The problem definition is illustrated in Fig.~\ref{2brf} (a). The interface thickness parameter is selected as $\varepsilon=0.005$ to reduce the volume-preserving mean curvature flow. In the time-dependent mobility model, the RMS convective distortion parameter $\eta=0.05$ is used to get an accurate surface tension force calculation. The above combinations of $\varepsilon$ and $\gamma$ has been proven to be accurate in the simulation of the 3D rising bubble case, which uses the same physical parameters. A non-uniform unstructured mesh consisting of 7,077,043 nodes and 45,078,392 tetrahedrons is employed for the spatial discretization.  The mesh at the plane $x=0.5$ is shown in Fig.~\ref{2brf} (b). The time step size is selected as $\Delta t = 0.0025$. The evolution of the interface $\phi=0$ is shown in Fig.~\ref{of2rbf}. The complex topological changes including the rising of the bubbles, the merging of the bubbles with the free surface and the wave formation at the free surface in the merging process can be observed clearly using our 3D phase-field Navier-Stokes solver with the unstructured mesh.

\section{Conclusion}
In the present work, a variational interface-preserving Allen-Cahn phase-field formulation relying on a novel time-dependent mobility model has been developed for accurate surface tension force calculation. By writing the convective Allen-Cahn equation in a non-dimensional moving orthogonal curvilinear coordinate system, we have derived the governing equation for the interface profile. We have identified the convective distortion term and the effective parameter determining the deviation of the diffuse interface profile from the hyperbolic tangent profile in the governing equation. A time-dependent mobility model to control the convective distortion parameter in the diffuse interface region and to preserve the hyperbolic tangent profile has been constructed accordingly. Following the verification of our implicit PPV-based steady-state Allen-Cahn solver for a generic bistable convection-reaction-diffusion system, we established the correlation between the convective distortion parameter and the interface errors through numerical simulations of the planar and curved interface convection problems. We then assessed the solutions of the proposed model in two- and three-dimensional rising bubble benchmark cases against the sharp interface counterparts. Through the assessment, it has been shown  that the interface preservation achieved by the proposed model and the minimization of the volume-preserving mean curvature flow realized by decreasing the interface thickness parameter are essential for the accurate surface tension dynamics. Finally, by simulating two rising bubbles merging with a free surface, we have shown that the proposed technique is applicable in a practical problem involving complex topology changes of the interface in an unstructured mesh and complex dynamics involving bubble-bubble and bubble-free surface interactions. 

\section*{Acknowledgement}
The authors would like to acknowledge the Natural Sciences and Engineering Research Council of Canada (NSERC) for the funding. This research was supported in part through computational resources and services provided by Advanced Research Computing at the University of British Columbia.

\appendix
\setcounter{equation}{0} 
\setcounter{figure}{0}
\section{Quantification of the convective distortion intensity}\label{appdixA}
In this appendix, we demonstrate the quantification of the intensity of the convective distortion by the normal velocity gradient in the normal direction $\zeta$. Consider two level sets of $\phi=\phi_1$ and $\phi=\phi_2$. Due to the finite thickness of the region, the convective velocity of the two level sets can be different. The tangential velocities are ignored since they do not affect the propagation of level sets. The normal velocity of the level sets are denoted as $u_n(\boldsymbol{x}_1,t)$ and $u_n(\boldsymbol{x}_2,t)$, where $\boldsymbol{x}_1$ and $\boldsymbol{x}_2$ are in the same normal axis with identical tangential coordinates.  Suppose $\phi$ is continuous and differentiable with respect to $\boldsymbol{x}\in(\boldsymbol{x}_1,\boldsymbol{x}_2)$, we want to quantify the convective distortion intensity at $\boldsymbol{x}$.

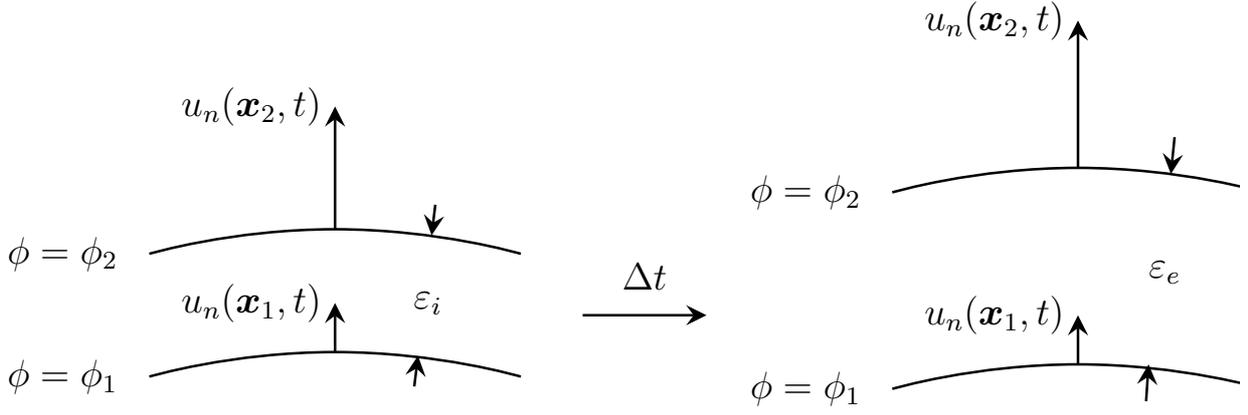
\begin{figure}
	\centering
	\trimbox{20 100 0 100}{
		\begin{tikzpicture}[scale=2.4,every node/.style={scale=0.5}]
		
		\begin{axis}[ 
		ymin=-0.5,
		ymax=2.5,
		xmax=4.8,
		xmin=-5.3,
		xticklabel=\empty,
		yticklabel=\empty,
		ytick style={draw=none},
		xtick style={draw=none},
		axis line style = {draw=none},
		minor tick num=0,
		axis lines = middle,
		label style = {at={(ticklabel cs:1.1)}},
		axis equal
		]
		\def\mu{0.5}
		\draw plot [smooth, tension=1] coordinates { (axis cs: -4,0+\mu) (-2.5,0.2+\mu) (-1,0+\mu)};
		\draw [-stealth] (-2.5,0.2+\mu)--(-2.5,1.2+\mu);
		\node [left] at (axis cs: -2.5,1.2+\mu) {$u_n(\boldsymbol{x}_2,t)$};
		
		\draw plot [smooth, tension=1] coordinates { (axis cs: -4,-1+\mu) (-2.5,-0.8+\mu) (-1,-1+\mu)};
		\draw [-stealth] (-2.5,-0.8+\mu)--(-2.5,-0.4+\mu);
		\node [left] at (axis cs: -2.5,-0.4+\mu) {$u_n(\boldsymbol{x}_1,t)$};
		
		\node at (axis cs: -4.7,0+\mu) {$\phi=\phi_2$};
		\node at (axis cs: -4.7,-1+\mu) {$\phi=\phi_1$};
		\draw[-stealth] (axis cs: -1.69,0.4+\mu)--(axis cs: -1.72,0.15+\mu);
		\draw[-stealth] (axis cs: -1.86,-1.08+\mu)--(axis cs: -1.83,-0.83+\mu);
		\node at (axis cs: -1.75,-0.4+\mu) {$\varepsilon_i$};
		\draw[-stealth] (axis cs: -0.5,0)--(axis cs: 0.5,0);
		\node [above] at (axis cs: 0,0.1 ) {$\Delta t$};
		
		\draw plot [smooth, tension=1] coordinates { (axis cs: 2,1) (3.5,1.2) (5,1)};
		\draw [-stealth] (3.5,1.2)--(3.5,2.4);
		\node [left] at (axis cs: 3.5,2.4) {$u_n(\boldsymbol{x}_2,t)$};

		\draw plot [smooth, tension=1] coordinates { (axis cs: 2,-0.6) (3.5,-0.4) (5,-0.6)};
		\draw [-stealth] (3.5,-0.4)--(3.5,0);
		\node [left] at (axis cs: 3.5,0) {$u_n(\boldsymbol{x}_1,t)$};

		\node at (axis cs: 1.3,1) {$\phi=\phi_2$};
		\node at (axis cs: 1.3,-0.6) {$\phi=\phi_1$};
		
		\draw[-stealth] (axis cs: 4.28,1.45)--(axis cs: 4.25,1.15);
		\draw[-stealth] (axis cs: 4.05,-0.7)--(axis cs: 4.07,-0.4);
		\node at (axis cs: 4.2,0.35) {$\varepsilon_e$};
		
		\end{axis}
		\end{tikzpicture}}
	\caption{Schematic diagram of the convective distortion of the thickness between the level sets of $\phi=\phi_1$ and $\phi=\phi_2$ due to different normal velocities.}
	\label{zeta}
\end{figure}

Consider the initial distance between the two level sets at the discussed normal axis as $\varepsilon_i=||\boldsymbol{x}_2-\boldsymbol{x}_1||_2$. After a small time increment $\Delta t$, the thickness is distorted due to the difference in the normal velocity, which can be linearly approximated as:
\begin{equation}
 \varepsilon_e=\varepsilon_{i}+(u_n(\boldsymbol{x}_2,t)-u_n(\boldsymbol{x}_1,t)) \Delta t,
 \end{equation} 
 where $\varepsilon_e$ represents the distance between the two level sets at the end of the time increment, as shown in Fig.~\ref{zeta}. 
 The intensity of the convective distortion at $\boldsymbol{x}\in (\boldsymbol{x}_1,\boldsymbol{x_2})$ can be considered as the relative change of the thickness per unit time due to the difference in the convective velocity:
\begin{equation}
\lim\limits_{\substack{ \varepsilon_i\to 0 \\ \Delta t\to 0 }}\frac{\varepsilon_e-\varepsilon_i}{\varepsilon_i\Delta t}=\lim\limits_{\substack{ \varepsilon_i\to 0 \\ \Delta t\to 0 }}\frac{u_n(\boldsymbol{x}_2,t)-u_n(\boldsymbol{x}_1,t)}{\varepsilon_i}=\frac{\partial u_n}{\partial n}(\boldsymbol{x},t)=\zeta(\boldsymbol{x},t)
\end{equation}

\section{Frame independent form of time-dependent mobility}\label{appdixB}
The normal velocity gradient in the normal direction $\zeta(\boldsymbol{x},t)$ in Eq.~(\ref{time dependent mobility}) can be written in a frame independent form:
\begin{equation}
   \zeta(\boldsymbol{x},t)=\nabla(\boldsymbol{u}\cdot\boldsymbol{n}^{\phi}_L)\cdot\boldsymbol{n}^{\phi}_L . 
\end{equation}
To simplify the equation, we expand the normal velocity gradient in the normal direction as:
\begin{align}
    \nabla(\boldsymbol{u}\cdot\boldsymbol{n}^{\phi}_L)\cdot\boldsymbol{n}^{\phi}_L=\boldsymbol{n}^{\phi}_L\cdot\nabla\boldsymbol{u}\cdot\boldsymbol{n}^{\phi}_L+\boldsymbol{u}\cdot\nabla\boldsymbol{n}^{\phi}_L\cdot\boldsymbol{n}^{\phi}_L+\boldsymbol{u}\times(\nabla\times\boldsymbol{n}^{\phi}_L)\cdot\boldsymbol{n}^{\phi}_L+\boldsymbol{n}^{\phi}_L\times(\nabla\times\boldsymbol{u})\cdot\boldsymbol{n}^{\phi}_L \nonumber
\end{align}
where the second, the third and the fourth terms are zero. For example, the second term:
$ \boldsymbol{u}\cdot\nabla\boldsymbol{n}^{\phi}_L\cdot\boldsymbol{n}^{\phi}_L=\boldsymbol{u}\cdot(\nabla\boldsymbol{n}^{\phi}_L\cdot\boldsymbol{n}^{\phi}_L)=\boldsymbol{u}\cdot\frac{1}{2}\nabla(\boldsymbol{n}^{\phi}_L\cdot\boldsymbol{n}^{\phi}_L)=0$
and  the third term $ \nabla\times\boldsymbol{n}^{\phi}_L=\nabla\times\left(\frac{\nabla\phi}{|\nabla\phi|}\right)=\nabla\left(\frac{1}{|\nabla\phi|}\right)\times\nabla\phi+\frac{1}{|\nabla\phi|}\nabla\times\nabla\phi$.
Notice that with the assumption of uniform interface profile along the interface, $1/|\nabla\phi|$ is a constant on the level sets of $\phi$. Thus both $\nabla(1/|\nabla\phi|)$ and $\nabla \phi$ are normal to the level sets of $\phi$,  which leads to $\nabla(1/|\nabla\phi|)\times\nabla\phi=0$. The curl of gradient of a scalar field is identically zero. As a result, the third term vanishes. The last term equals to zero since for arbitrary vectors $\boldsymbol{a}$ and $\boldsymbol{b}$, $(\boldsymbol{a}\times\boldsymbol{b})\cdot\boldsymbol{a}=0$.
As a result, the normal velocity gradient in the normal direction can be simplified as:
\begin{equation}
   \zeta(\boldsymbol{x},t)=\boldsymbol{n}^{\phi}_L\cdot\nabla\boldsymbol{u}\cdot\boldsymbol{n}^{\phi}_L . 
\end{equation}
and by substituting $\boldsymbol{n}^{\phi}_L=\nabla\phi/|\nabla\phi|$, we get the frame-independent form of the normal velocity gradient in the normal direction:
\begin{equation}\label{NVND frame independent}
   \zeta(\boldsymbol{x},t)=\frac{\nabla\phi\cdot\nabla\boldsymbol{u}\cdot\nabla\phi}{|\nabla\phi|^2} . 
\end{equation}
Finally, the frame independent form of the time-dependent mobility model is given by:
\begin{align}
 \gamma(t)&=\frac{1}{\eta}\mathcal{F}\left(\left|\frac{\nabla\phi\cdot\nabla\boldsymbol{u}\cdot\nabla\phi}{|\nabla\phi|^2}\right|\right),
\end{align}
where  $\mathcal{F}(\varphi(\boldsymbol{x},t))=\sqrt{\frac{\int(\varphi(\boldsymbol{x},t))^2d\Omega}{\int 1 d\Omega}},\ \boldsymbol{x}\in\Gamma^{\phi}_{DI}(t)$.
The frame independent form facilitates its numerical implementation in Cartesian coordinate system.

\bibliography{bibfile}
\end{document}